\documentclass[useAMS,usenatbib]{mn2e}

\usepackage{subfigure}
\usepackage[dvips]{color}
\usepackage{rotating}  
\usepackage{amssymb}

\newcommand{\zs}{$z_{s}$ }

\newcommand{\zd}{$z_{d}$}

\newcommand{\evec}{{\bf{e}}}
\newcommand{\gvec}{{\bf{g}}}
\newcommand{\gammavec}{{\mbox{\boldmath$\gamma$}}}

\newcommand{\beq}{\begin{equation}}
\newcommand{\eeq}{\end{equation}}
\newcommand{\bea}{\begin{eqnarray}}
\newcommand{\eea}{\end{eqnarray}}
\newcommand{\lx}{$L_{{\rm X} (0.1-2.4 {\rm keV})}$}

\newcommand{\MegaPrime}{\texttt{MegaPrime}\space}

\title[Weak lensing mass estimates of galaxy groups]{Weak lensing mass
  estimates of galaxy groups and the line-of-sight contamination}

\author[P. F. Spinelli et
al.]{P. F. Spinelli$^{1}$\thanks{E-mail:pat@usm.uni-muenchen.de},
  S. Seitz$^{1,2}$, M. Lerchster$^{1,2}$, F. Brimioulle$^{1}$ \&
  A. Finoguenov$^{2,3}$\\
  $^{1}$Universit\"{a}tssternwarte M\"{u}nchen, Scheinerstr. 1, 81679 M\"{u}nchen, Germany\\
  $^{2}$Max Planck Institute for Extraterrestrial Physics,
  Giessenbachstrasse, 85748 Garching, Germany\\
  $^{3}$Center for Space Science Technology, University of Maryland
  Baltimore County, 1000 Hilltop Circle, Baltimore, MD 21250, USA}

\begin{document}

\date{Accepted 2$^{\rm nd}$ November 2011. Received 17$^{\rm th}$ April 2011}

\pagerange{\pageref{firstpage}--\pageref{lastpage}} \pubyear{2002}

\maketitle

\label{firstpage}

\begin{abstract}
  Weak lensing is an important technique to determine the masses of
  galaxy groups. However, the distortion imprint on the shape of the
  background galaxies is not only affected by the gravitational field
  of the main group but by all the mass content along the
  line-of-sight. Using COSMOS shear mock data we study the shear
  profile around 165 groups and investigate the level at which the
  neighbouring groups can enhance or suppress the shear signal from
  the main halo. The mock data are based on CFHT and Subaru
  observations, which are used to obtain the photometric redshifts of
  galaxies in the field and a realistic galaxy density, given by the
  weak lensing distortion analysis of the observed data. We further
  use information on the galaxy groups (having a median mass and
  redshift of $M_{200}=3.1\times 10^{13}$ M$_\odot$ and $z=0.68$) from
  the COSMOS X-ray catalogue of extended sources. The expected
  gravitational shear field of these groups is calculated assuming
  that the haloes are described by NFW density profiles, and the total
  shear is computed by summing the shear over all the lenses. We
  conclude that, on average, the signal-to-noise for a detection of
  the main halo is affected by $\thickapprox 15\% \times
  \sqrt{n_{gal}/30}$ with respect to the signal-to-noise the same halo
  would have if it was isolated in the sky. Groups with neighbours
  that are close in projected distance ($\lesssim1^{\prime}$) are the
  most affected, but haloes located at larger angular distances also
  cause a measurable shear signal. These (angular) distant groups can be
  interpreted as uncorrelated large-scale structure. The average bias
  in the mass excess estimate of individual groups that is introduced
  by the external haloes is zero with an rms of $\sim6-72\%$,
  depending on the aperture size used. One way to eliminate this bias
  is by stacking the density profile of several groups. The shear
  signal introduced by large-scale structure acts as an external
  source of noise. The averaged uncertainty introduced is
  $\sigma_{\gamma_{\rm t}}^{\rm LSS}\sim0.006$ per component for an
  aperture size of $\theta\sim5^{\prime}$, which corresponds to
  $\sim1.8\%$ of the one-component intrinsic ellipticity value. This
  large-scale structure noise error becomes equal to intrinsic
  ellipticity noise if there are measurements for $\sim3000$ galaxies
  within a certain aperture, a number that is already achieved by
  current deep surveys such as COSMOS and, therefore, that should not
  be ignored.
\end{abstract}


\section{Introduction}

Weak gravitational lensing is sensitive to both dark matter and dark
energy, making it a valuable tool to map the matter content of the
universe and its evolution with time. In fact, weak lensing has been
identified by the report of the Dark Energy Task Force \citep{de-task}
as one of the most promising tools to understand the nature of dark
energy.

Weak gravitational lensing is also an attractive technique to study
groups and clusters of galaxies. Since the signal does not depend on
the dynamical or evolutionary state of systems under investigation, it
has advantage upon other techniques, such as X-Rays
\citep[e.g.][]{boehringer00} or Sunyaev-Zeldovich effect
\citep[e.g.][]{birkinshaw99,carlstrom02} for which mass estimates
assume the hydrostatic equilibrium of the intra-cluster gas. However,
the current systems analysed using the weak lensing technique are
biased towards galaxy clusters ($\gtrsim10^{14}$ M$_{\odot}$), for
which the lensing signal is stronger and therefore not so affected by
the intrinsic shape noise. The weak lensing analysis of individual
galaxy groups requires a much higher density of galaxies in order to
eliminate this noise. Nevertheless, the shear measurements of such
systems are of great interest, since galaxy groups constitute the most
common association of galaxies and can be found in abundance at the
redshift range used to discriminate between cosmological models.

In practise, weak lensing is rather challenging. The induced
gravitational shear changes the intrinsic ellipticity of galaxies by a
very tiny amount. For instance, galaxies have an average intrinsic
ellipticity of the order of $\evec^{s}\sim0.4\pm0.4$, whereas the
change introduced by gravitation is of the order of
$\gamma\sim0.03$. To overcome this problem, the average distortion
within an area is measured. This holds because the orientation of the
intrinsic ellipticity of galaxies has no preferred direction, being
randomly distributed. There are, however, other sources of uncertainty
which are often ignored and which limit the precision of the
measurements, such as the shear signal introduced by the large-scale
structure (LSS) and the possible presence of multiple haloes along the
line-of-sight. In case the shear signal is affected by such external
contributors, the calculated physical parameters of the halo are not
reliable.

In a consideration of the first problem, \cite{hoekstra01b,hoekstra03}
estimated analytically the contribution of the uncorrelated LSS to the
mass estimates of clusters of galaxies via weak
lensing. \cite{hoekstra01b,hoekstra03} found that the large-scale
structure does not bias the mass estimate itself but it does introduce
uncertainty in the measurement that can not be ignored. These findings
were confirmed in a recent work \citep{hoekstra10} using N-body
simulations.

With regard to the second problem, \citet{brainerd10} made a study of
the frequency and the effect of multiple deflections in galaxy-galaxy
lensing. Deflections by multiple lenses included all foreground lenses
apart from the nearest lens to the source (in projected
distance). \citet{brainerd10} concluded that if the observed shear is
used to constrain fundamental parameters associated with the galaxy
halo it is crucial to take the multiple lens calculations into
account.

For massive galaxy clusters ($M\sim10^{15}$\,M$_{\odot}$), there is a
small probability that two or more clusters can be aligned along the
line-of-sight. Therefore, the distortion on the shape of a background
galaxy induced by any other deflector along the line-of-sight is not
comparable to the magnitude of the distortion that a massive object
such as a galaxy cluster induces. This statement does not hold for
less massive haloes such as galaxy groups, for which the shape of a
background source galaxy can be equally distorted by other groups
along the line-of-sight, given that there is realistic probability of
finding such a configuration. When this is the case, the total
distortion measured can not be associated to an unique galaxy group.
 
The line-of-sight and LSS contamination of weak lensing measurements
can be studied via simulations. The distortion induced by a foreground
mass on the shape of a background source galaxy depends on the mass
distribution of the foreground lens and on the ratio of the distance
of the lens relative to the background source over the distance of the
source. Hence, if the foreground mass distribution is known and the
positions and redshifts of background galaxies are available, the
expected shear field along the line-of-sight can be computed. The
total shear is obtained by adding the contribution of all systems
acting as lenses.

A robust way to set up such simulations is by using observational
data. The COSMOS field \citep{cosmos07} is an ideal data set for this
purpose due to the broad wavelength coverage with which the field has
been observed. The XMM-Newton and Chandra data provide information on
the galaxy group and cluster distribution over the redshift range
$z\sim0.07-1.8$. Field galaxies, for which reliable shapes can be
measured, were observed with three different telescopes: CFHT, Subaru
and HST. Multi-wavelength imaging and spectroscopy of galaxies in the
field allow to estimate the photometric redshifts of galaxies over the
redshift range $z\sim 0.01-2.5$. In this work, we have used the
available information on the COSMOS field to create realistic shear
mock catalogues of this patch of the sky. We compare how the
gravitational shear changes for the case where lenses are considered
as isolated systems in the sky and when they lie embedded in their
environment. We investigate if the difference in the shear magnitude
affects the likelihood of a system being detected by its weak lensing
signal as well as how much its density contrast profile is affected.

This paper is organised as follows. In Section \ref{cats} we provide
details on the data set used and how we obtain the necessary
catalogues: photometric, photometric redshifts and shear catalogues
(Sections \ref{cats:phot}, \ref{cats:photo-z} and \ref{cats:shear}
respectively). The reader who is not interested in this part can jump
directly to Section \ref{cats:x-ray}, where we give an outlook on the
COSMOS X-ray catalogue and how we select the galaxy groups
investigated in this paper. In section \ref{shear-sim} we use all the
information compiled in Section \ref{cats} to create shear mock
catalogues of the COSMOS field. Current techniques used in weak
lensing analysis are applied to isolated and multiple lensing mock
catalogues and are discussed in Section \ref{resul}.  A summary of the
paper and the conclusions are described in Section \ref{conc}. In the
appendices we include the information on the CFHT and Subaru COSMOS
data reduction, used to derive the shear catalogues.

Throughout this paper we adopt WMAP5 $\Lambda$CMD cosmology with
$\Omega_{m}=0.258$, $\Omega_{\Lambda}=0.742$ and $H_{0}=72.0$ km
s$^{-1}$ Mpc$^{-1}$  \citep{wmap5}. We follow the standard lensing
notation for distances, where $D_d$, $D_s$, $D_{ds}$ \,stands, respectively,
for the angular diameter distances between the observer and the lens,
the observer and the source and the lens and the source. The notation
for the redshift also follows the same convention: \zd \,is the
redshift of the lens and \zs \,the redshift of the
source. \texttt{MegaPrime}/CFHT and \texttt{Suprime-Cam}/Subaru
filters are differentiated by adding a prime (CFHT), e.g. {\it i'}, and
a cross (Subaru), e.g. {\it i$^+$}, in the filter name.

\section{Data}\label{cats}

\begin{table*}
 \centering
 \begin{minipage}{130mm}
    \caption{Summary of the data used in the Lensing Analysis.}
    \begin{tabular}{ccccccc}
      \hline
      \hline
      Telescope -- Filter & $\rm EXP\_TIME$ & ${\rm mag_{\,lim}}^{\rm c}$ & Seeing &
      Astrometry$^{\rm d}$ & Area & Pixel Scale \\
      \hline
      CFHT -- {\it i'} band & $\sim32.5^{\rm a}$ h & 26.9 & $0.71^{\prime\prime}$  & $0.14^{\prime\prime}$ & 1.00 degrees$^2$ & $0.186^{\prime\prime}$ \\ 
      Subaru -- {\it i$^+$} band & $\sim0.7^{\rm b}$ h  & 26.0 & $0.60^{\prime\prime}$  & $0.22^{\prime\prime}$ & 0.55 degrees$^2$ & $0.200^{\prime\prime}$  \\
      \hline
    \end{tabular}
    \label{tab:data}
    \begin{flushleft}
    \scriptsize
      $^{\rm a}$ For the lensing analysis we only stack exposures
      taken during CFHT \MegaPrime phase three. \\
      $^{\rm b}$ We only stack individual exposures taken with the same
      camera orientation angle and offset between different exposures less
      then $3^{\prime}$. \\
      $^{\rm c}$ The $5\sigma$ limiting magnitude within $2^{\prime\prime}$
      diameter aperture. \\
      $^{\rm d}$ With respect to SDSS-R6 catalogue.
   \end{flushleft}
\end{minipage}
\end{table*}

The COSMOS field ($\alpha$=10:00:28.6, $\delta$=+02:12:21.0) is the
largest contiguous area imaged deeply with Hubble Space Telescope
(HST) using the Advanced Camera for Surveys (ACS). The field covers
approximately 1.64 degrees$^2$ and has also been imaged with many
other telescopes.  The wavelength coverage spans from X-rays to
radio. In this work we use public CFHT {\it u$^*$, g', r', i'} and
{\it z'} bands, {\it H} band obtained with CAHA telescope and COSMOS
public {\it K$_s$} band data to derive multi-colour catalogues and
photometric redshifts. CFHT {\it i'} and Subaru {\it i$^+$} bands are
used in the gravitational shear analysis. X-ray data observed with
XMM-Newton and Chandra are used to obtain information on the galaxy
groups and clusters present in the field.

\begin{figure}
\center
\includegraphics[scale=.40]{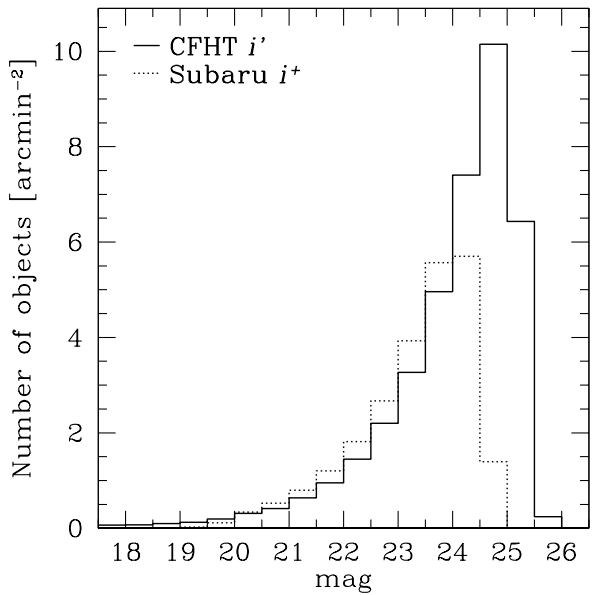}
\caption{The $5\sigma$ galaxy magnitude distribution in {\it i'}
  (CFHT) and {\it i$^+$} (Subaru) bands within $2^{\prime\prime}$
  diameter aperture. For this plot we have used 32 h of CFHT and
  0.7 h of Subaru data. See the text for further
  details.\label{mag}}
\end{figure}

The CFHT data cover an area of 1.0 degrees$^2$, which corresponds to
the \MegaPrime instrument field-of-view (FOV). For the Subaru data, we
only use individual exposures taken with the same camera orientation
angle and offset $<3^{\prime}$ in the final stack. This yields an
image coverage of 0.55 degrees$^2$. This is done because by stacking
all the exposures, the resulting PSF pattern could not be corrected to
the level required in the lensing analysis. Therefore, for the
CFHT-like mock data, which will be introduced in Section
\ref{shear-sim}, we restrict ourselves to 1.0 degrees$^2$ of data. For
the Subaru-like mocks, we are restricted to 0.55 degrees$^2$. Details
about the CFHT and Subaru data acquisition and reduction used in the
derivation of shear and photometric redshifts catalogues (hereafter
called photo-z) are described in the Appendix \ref{app:data}. Table
\ref{tab:data} provides a summary of the data used in the lensing
analysis, while Fig. \ref{mag} shows the magnitude distribution of
detected objects using both CFHT {\it i'} and Subaru {\it i$^+$} band
images. Note that the detection of the field objects is done prior to
the lensing analysis.

In the following subsections we describe the creation of the
photometric, photo-z and shear catalogues and the halo selection obtained
from the X-ray data.  An observational final catalogue, containing the
position, redshift, shear, magnitude and other properties of each
galaxy, is created by combining the photo-z and shear catalogues. This
information was subsequently used to create the shear mock
catalogues. The reader who is not interested in this part can jump
directly to Section \ref{cats:x-ray}, where we explain the halo
selection from the X-ray catalogue.

\subsection{Photometric Catalogues}\label{cats:phot}

From the CFHT reduced images, we create photometric catalogues that are
used to estimate the photometric redshifts of the galaxies. In
addition to the CFHT data, we also use {\it H} band, imaged with the
NIR wide-field camera OMEGA2000, operating at the prime focus of the
CAHA 3.5-m telescope. We further use {\it K$_s$} band observed with
KPNO 4-m telescope using the FLAMINGOS instrument and ISPI camera on
the CTIO 4-m telescope. Data from these instruments were combined to
obtain a single {\it K$_s$} band image, which was retrieved from the
COSMOS archive. The information on the CFHT data reduction can be
found in the Appendix \ref{app:data}. Further details on the {\it H}
band data can be found in \citet{gabasch08} and on the {\it K$_s$}
band in \citet{capak07}.

We first measure the seeing in all bands and convolve them with a
Gaussian kernel to match to the $K_s$ band, which had the worst seeing
($1.5^{\prime\prime}$). Thus, we proceed to the creation of
multi-colour catalogues. In order to assure that the centre of the
detected objects are the same in all observed bands, objects are
detected running
SExtractor\footnote{http://www.astromatic.net/software/sextractor}
\citep{sextractor} in dual-image mode configuration on the unconvolved
{\it i'} band image. The flux is measured in an aperture with diameter
size of $1.86^{\prime\prime}$. Table \ref{tab:photdata} provides a
summary of the data used to produce the photometric catalogues.

\begin{table}
 \centering
   \caption{Summary of the data used to compute the Photometric Redshifts}
   \begin{tabular}{ccc}
     \hline
     \hline
     Filter & {$\rm EXP\_TIME$} & Seeing \\ 
     \hline
     {\it u*} & $6.9$ h  & $0.95^{\prime\prime}$\\
     {\it g'} & $9.3$ h  & $0.85^{\prime\prime}$ \\
     {\it r'} & $25.5$ h & $0.74^{\prime\prime}$ \\
     {\it i'} & $53$ h   & $0.73^{\prime\prime}$ \\
     {\it z'} & $14.1$ h & $0.71^{\prime\prime}$ \\
     {\it H}  & $0.80$ h & $1.08^{\prime\prime}$\\
     $K_s$  & $0.84$ h & $1.50^{\prime\prime}$ \\
     \hline
   \end{tabular}
   \label{tab:photdata}
\end{table}

\subsection{Photometric Redshift Catalogues}\label{cats:photo-z}

The photometric redshifts are computed in the same way as presented in
\citet{fabrice08}, using the Bayesian PHOTO-Z code from
\citet{bender01}. In this section we provide a brief summary of the
method.

The templates of the spectral energy distribution (SED) used are
described in \citet{bender01} and \citet{gabasch04a,gabasch04b}. A
total of 31 templates are used: 18 default templates plus 13 from
\citet{ilbert06}. The SED templates can be seen in the right panel of
Fig. 12 of \citet{mikelbt10}. For each SED template, the code
computes the full redshift likelihood function. The step-size for the
redshift grid is 0.01.

\begin{figure}
\centering
\subfigure{\includegraphics[scale=0.40]{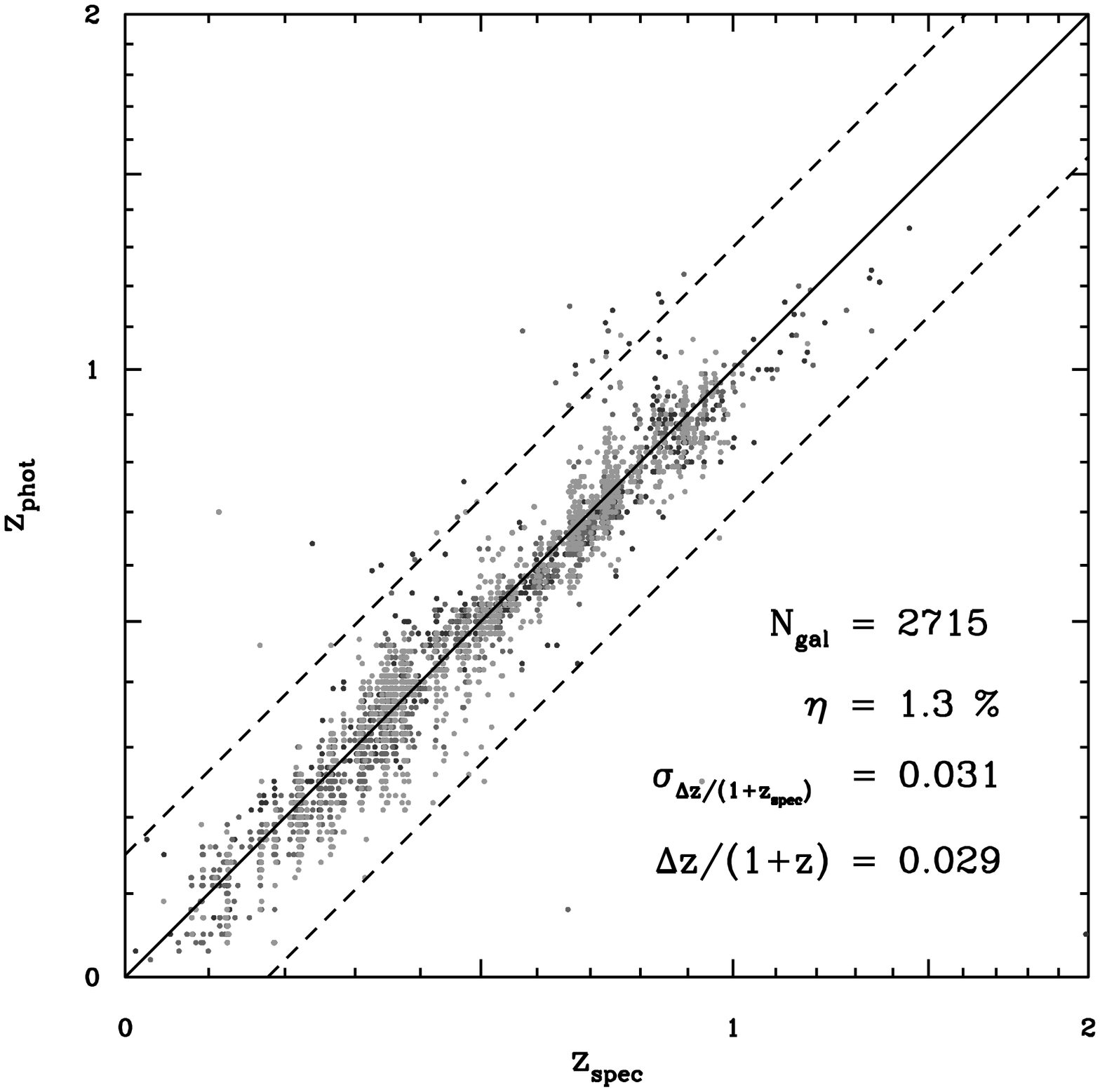}} \\
\subfigure{\includegraphics[scale=0.40]{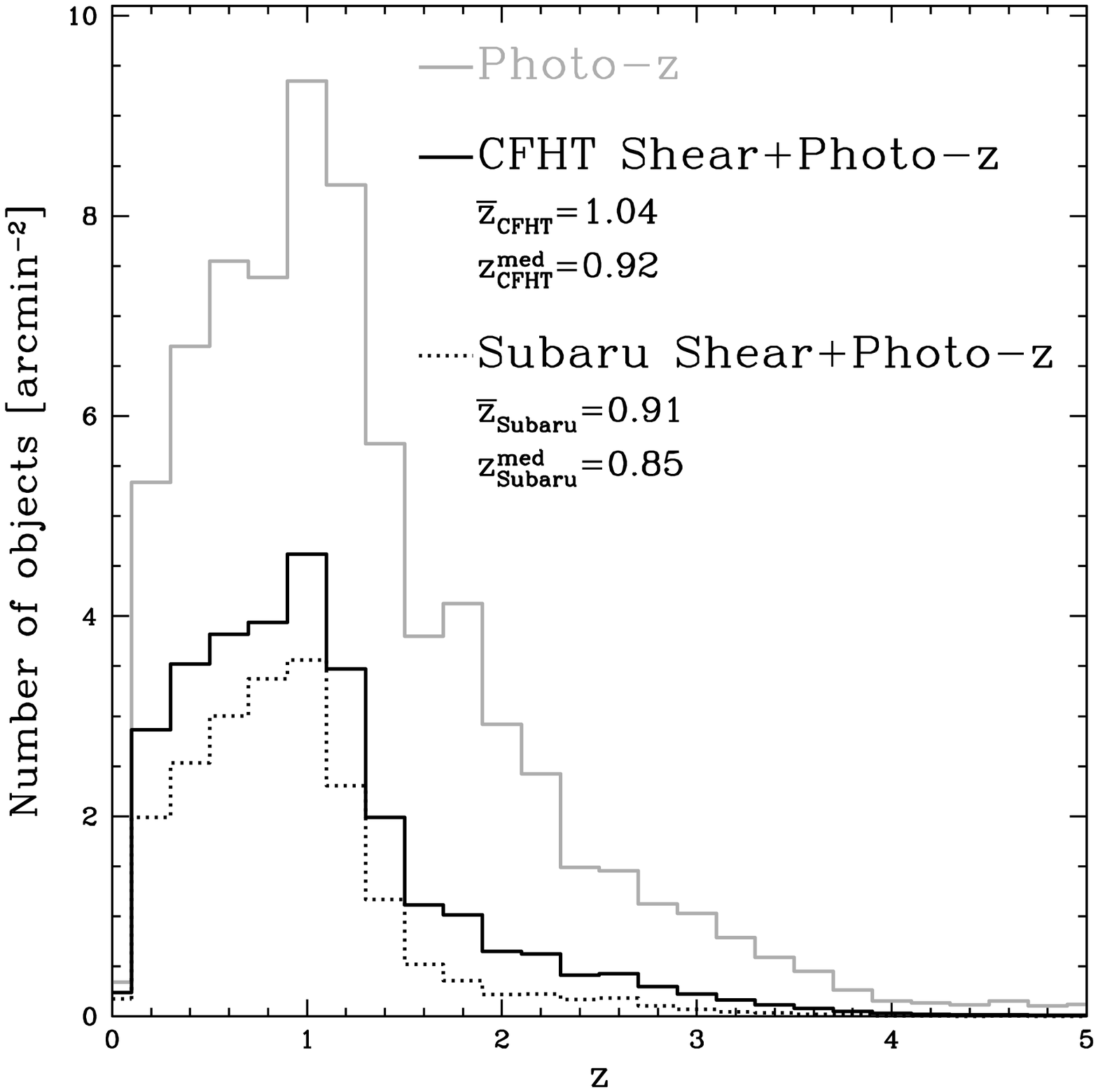}}
\caption{Top: photometric redshifts of non-stellar objects against
  spectroscopic redshifts from the zCOSMOS sample. Dotted lines are
  for $z_{\rm phot} = z_{\rm spec}\pm0.15 \times (1+z_{\rm
    spec})$. The fraction $\eta$ of catastrophic outliers is defined
  as a fraction of galaxies for which $ |z_{\rm spec}-z_{\rm
    phot}|/(1+z_{\rm spec})>0.15$ holds. Bottom: Distribution of the
  computed photometric redshifts. The grey solid histogram shows the
  distribution for all the objects with photometric redshifts. The
  black solid histogram shows the redshift distribution of the
  galaxies in the final CFHT catalogue (shear plus photo-z) and black
  dashed histogram shows the same but for the Subaru final
  catalogue.\label{photo-z}}
\end{figure}

We compare the measured photo-z with the zCOSMOS sample of
spectroscopic redshifts \citep{lilly07}. We retrieve redshifts for
2715 galaxies, spread over the area for which there was also {\it H}
band. The accuracy of the photo-z is $\sigma_{\Delta z/(1+z)} = 0.031$
and the redshift scatter is $\Delta z / (1+z)=0.029$, where $\Delta
z=|z_{\rm spec}-z_{\rm phot}|$. The fraction of catastrophic outliers
is $\eta=1.3\%$, where $\eta$ is defined as a fraction of galaxies for
which $ |z_{\rm spec}-z_{\rm phot}|/(1+z_{\rm spec})>0.15$ holds. In
the top panel of Fig. \ref{photo-z} we show a comparison of the
zCOSMOS spectroscopy redshifts to the ones computed in this work.

In order to get a catalogue free of insecure photo-z estimates all the
objects received a quality flag. Stars, saturated objects and objects
with high photo-z error have their flag value greater than 3. We kept
only objects with good photo-z quality flags (Flag$\leq3$). The
meaning of these flags can be found in Table A.1. of
\citet{fabrice08}. The bottom panel of Fig. \ref{photo-z} shows the
distribution of the redshifts in our final catalogue.

\subsection{Shear Catalogues}\label{cats:shear}

The weak lensing analysis is done with the KSB method
\citep{ksb,lk97,hoekstra98}. The KSB version used in this work (also
called KSB+) is described in detail in \citet{erben01} and
\citet{schrabback07}. In this section we summarise the method focusing
on the choices we made to create the shear catalogues used in this work.

The detection of sources is performed with SExtractor, using the {\it
  i'} and {\it i$^+$} bands. The weight and flag maps of the final
stacked images are used to detect the objects. This allows a more
precise evaluation of the signal-to-noise ratio of the detected
objects. For each extracted object, the weighted second moments of the
surface brightness distribution $Q_{\alpha\beta}$ are computed and the
observed complex ellipticity is derived as
\bea
   \label{eq:ell-mom}
     \evec^{\rm obs}&=&\frac{Q_{11}-Q_{22}+2\rm{i}\,Q_{12}}{Q_{11}+Q_{22}} \nonumber \\
     &=&e_1+{\rm i}\,e_2 
\eea
where $Q_{\alpha\beta}$ are measured using a Gaussian filter. The size
of the Gaussian window $r_g$ is equal to the half-light radius $r_h$
of the detected objects.

The observed ellipticity of galaxies is a sum of intrinsic
ellipticity, PSF shearing, PSF anisotropy and gravitational
lensing. The KSB approach assumes that the PSF can be described as an
anisotropic contribution convolved with an isotropic kernel that
mimics the seeing (assumed to be circularly symmetric), thus each
component of the observed ellipticity can be split into the components
as
\beq \label{eq:ell}
   e_{\alpha}^{\rm obs}= e^{\rm s}_{\alpha}+P^{\rm \,g}_{\alpha\beta}g_{\beta} + P^{\rm \,sm}_{\alpha\beta}q_{\beta}\,, \quad\quad  \alpha,\beta \in \{1,2\}\,,
\eeq
where $q$ is the term that accounts for the PSF anisotropy, $P^{\rm
  sm}$ is the smear polarisability calculated from the galaxy
brightness profile and weight function. $P^{\rm g}$ is the pre-seeing
shear polarisability which is calculated as
\beq
   P^{\rm g}= P^{\rm sh}-P^{\rm sm}\frac{P^{\rm sh *}}{P^{\rm sm *}}
\eeq
where $P^{\rm sh}$ is the shear polarisability tensor and the asterisk
* denotes quantities measured from stellar objects. $P^{\rm sh}$ can
be interpreted as the response of the galaxy ellipticity to
gravitation if there are no PSF effects.

Stars present in the catalogue provide a way to model the PSF anisotropy
across the field because they have zero intrinsic ellipticity,
i.e. $e^{\rm s*}_{\alpha}=0$, and their shapes are not gravitationally
distorted, i.e. $g_{\alpha}^*=0$. Thus, equation (\ref{eq:ell}) yields
\beq \label{eq:psfanisotropy}
   q_{\beta}=q^{*}_{\beta}=(P^{\rm sm *})_{\alpha\beta}^{-1} e^{\rm obs *}_{\alpha}\,.
\eeq

\begin{figure}
\centering
\subfigure{\includegraphics[scale=0.40]{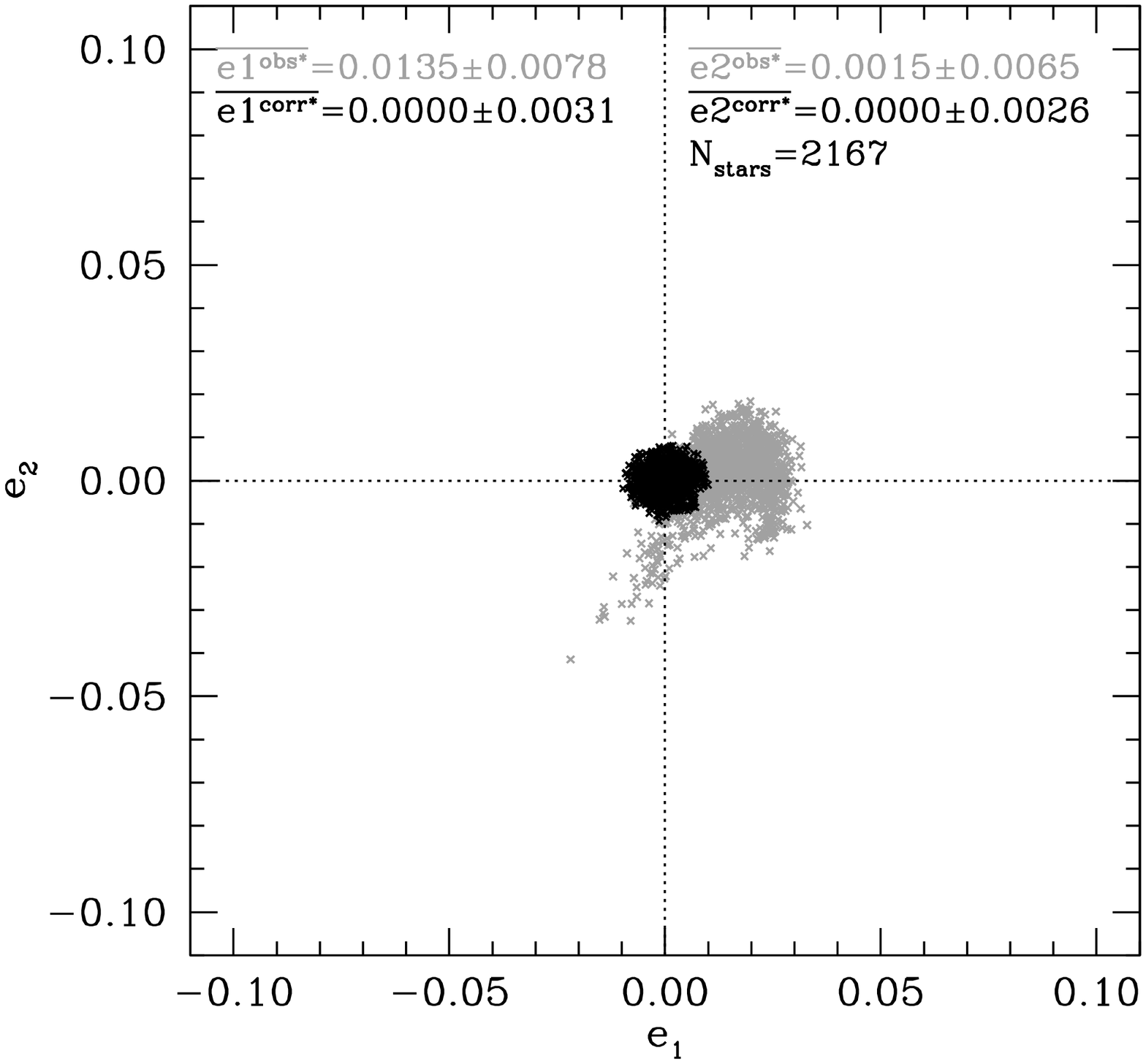}} \\
\subfigure{\includegraphics[scale=0.40]{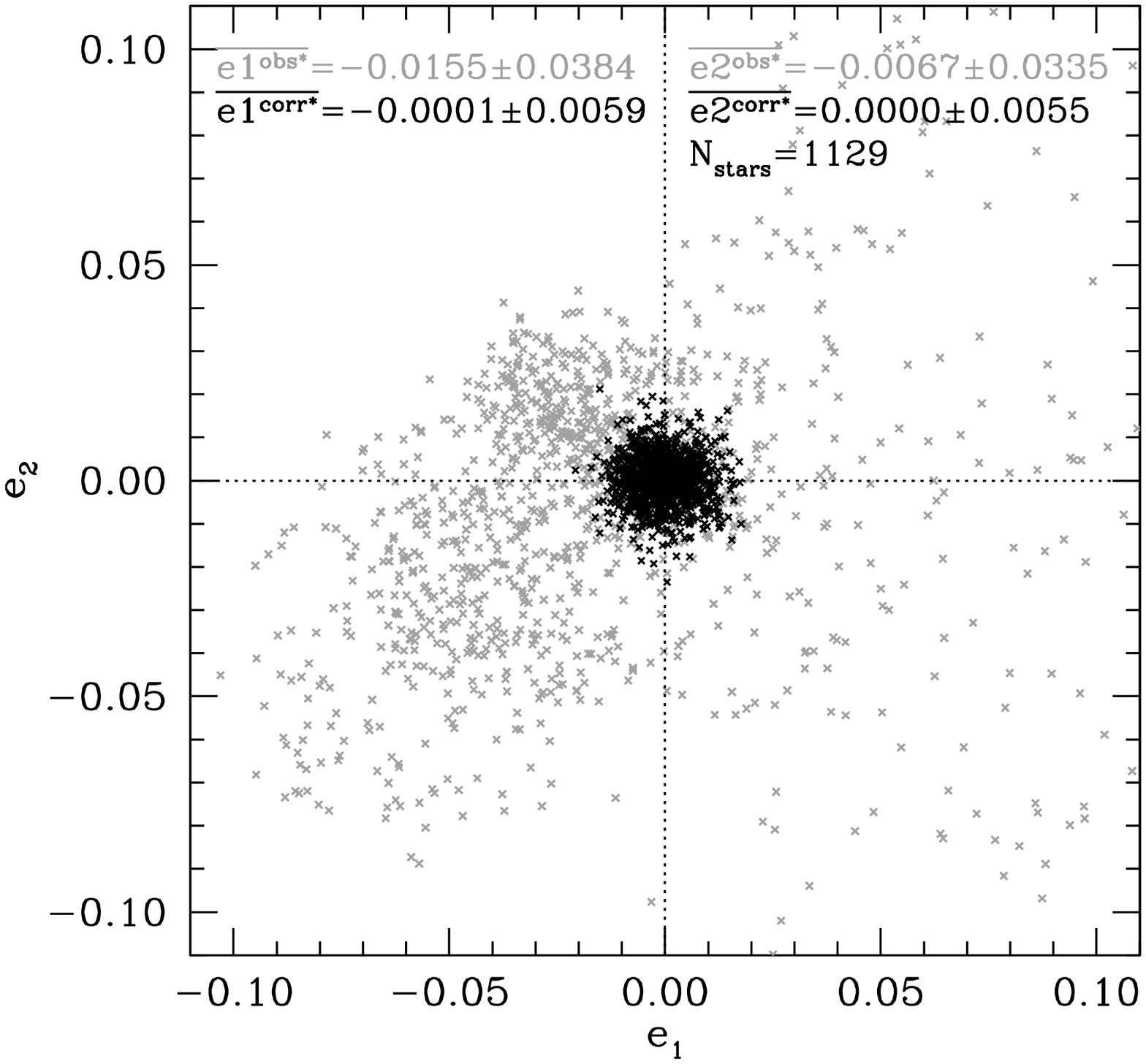}}
\caption{Observed ellipticity components (grey) and corrected (black)
  values after the PSF model subtraction. Top panel shows the CFHT
  data for which the entire FOV is modelled with a polynomial function
  of order 5. The bottom panel shows the same but for the Subaru
  data. For Subaru, the PSF is modelled in a chip-wise basis using a
  polynomial order of 4. The plotted quantities are measured with a
  Gaussian window of $r_g = 3\times r_h$.\label{psf}}
\end{figure}

The spatial variation of $q$ across the field-of-view is usually
described by a polynomial function. Stars are used to predict the PSF
anisotropy at the position of galaxies, and are selected by the visual
inspection of the $r_h$-mag diagram. For the CFHT catalogue, stars are
selected in the range of $0.35^{\prime\prime}<r_h<0.43^{\prime\prime}$
and $17.9<{\rm mag}<21.5$. We also exclude the stars close to CCD
borders yielding $\sim2100$ stars, which are used to fit a polynomial
function of order 5. For the Subaru data, the PSF modelling was more
complicated because the pattern varies discontinuously across the
field-of-view. We found that a polynomial function could not model
properly the PSF across the entire field. Thus, we perform the
correction on a chip basis, where only stars belonging to the region
imaged by one CCD are used. Subaru stars are selected in the range of
$0.28^{\prime\prime}\lesssim r_h \lesssim 0.36^{\prime\prime}$ and
$19.8\lesssim {\rm mag} \lesssim 21.7$, yielding on average $\sim72$
stars per chip with a minimum of 56 and a maximum of 87 stars. Yet,
the diagnostic plots (like the ones shown in Fig. \ref{psf}) were not
sufficient to justify which was the best polynomial order to fit
Subaru data. Thus, we make use of the diagnostic proposed by
\cite{rowe10} to help in the identification of the optimal polynomial
order: polynomial order 3 underfits the data, whereas order 5 overfits
(though there is a variation with the Gaussian window $r_g$ used to
measure stellar quantities). We conclude that a polynomial order of 4
yields the best performance.

Eventually, after the PSF modelling, ellipticities are
corrected by calculating
\beq 
e^{\rm corr}_{\alpha}=e_{\alpha}^{\rm
  obs}-P^{\rm sm *}_{\alpha\beta}q^{*}_{\beta}. 
\eeq 
Fig. \ref{psf} shows the ellipticity components of the stars before
and after the correction for PSF anisotropy. For the CFHT data, the
PSF model is excellent, with a residual stellar ellipticity dispersion
of $\sigma_{e_{i}^{\rm corr*}}\sim0.0028$ per each component. For the
Subaru data, the residual ellipticities have a higher but satisfactory
dispersion of $\sigma_{e_{i}^{\rm corr*}}\sim0.0057$.

Since $\langle \evec \rangle / 2=\gvec$, thus
\beq
   g_\alpha =\langle (P^{\rm g})_{\alpha\beta}^{-1}e^{\rm corr}_{\alpha} \rangle
\eeq
is an unbiased estimate for the reduced gravitational shear at the
galaxy positions. Hereafter we will also assume $ \gvec \simeq
\gammavec$ since we are in the weak lensing regime and $\kappa \ll 1$.

According to \citet{schrabback07} the KSB+ implementation requires, on
average, calibration factor of $c_{cal}=1/0.91$ so that $\langle
\gamma_{\alpha} \rangle = \langle e_{\alpha}\rangle /0.91$. This
calibration factor was derived from STEP1 simulations \citep{step1}.

We apply a final cut in the catalogue to select only galaxies with
relative high signal-to-noise. For the CFHT data we follow the
criteria: signal-to-noise of the detection $\nu>5$, $17.9<{\rm
  mag}<26$, $r_h>0.43^{\prime\prime}$ and $|\evec|<1$. The final
catalogue has a density of galaxies $n_{shear}=32.8$ galaxies
arcmin$^{-2}$. For the Subaru data the criteria is similar:
signal-to-noise of the detection $\nu>5$, $19.7<{\rm mag}<25$,
$r_h>0.37^{\prime\prime}$ and $|\evec|<1$ resulting in a catalogue with
density of $n_{shear}\sim 23.7$ galaxies arcmin$^{-2}$.

We match the CFHT shear and photo-z catalogues, producing a final
catalogue with a density of $n_{eff}=29.7$ galaxies arcmin$^{-2}$,
mean redshift of $\bar{z}=1.04$ and two-component ellipticity
dispersion of galaxies of $\sigma_{\evec^s}=0.47$. For the Subaru
data, the shear plus photo-z final catalogue has $n_{eff}=21.7$
galaxies arcmin$^{-2}$, mean redshift of $\bar{z}=0.91$ and
two-component ellipticity dispersion of galaxies of
$\sigma_{\evec^s}=0.42$. Hereafter we call these catalogues {\it
  shear-photo-z}. The values of the ellipticity dispersion for both
CFHT and Subaru data are in agreement with values previously found in
the literature \citep[e.g.][]{schirmer07,ume10,schrabback10}. Table
\ref{tab:lens} summarises the properties of the derived shear
catalogues and compares with previously published results on shear
measurements in the COSMOS field.

\begin{table*}
 \centering
\begin{minipage}{160mm}
 \caption{Summary of the Lensing Catalogues. Columns:(1) Telescope; (2)
    and (3): one-component dispersion of the stellar ellipticities
    after the PSF correction; (4) number of detected galaxies per arcmin$^{2}$; 
    (5) number of galaxies per arcmin$^{2}$ in the shear catalogue; 
    (6) number of galaxies per arcmin$^{2}$ in the shear-photo-z matched catalogue; 
    (7) and (8): one-component ellipticity dispersion of galaxies;
    (9): two-component ellipticity dispersion of galaxies, which is
    defined as $\sigma_{\evec^s}^2=\sigma_{e_1^s}^2+\sigma_{e_2^s}^2$;
    (10): mean redshift; (11): median redshift; (12): References.
    G07: \citet{gavazzi07}; K08: \citet{kasliwal08}; L07:
    \citet{leau07}; L10: \citet{leau10}; S10: \citet{schrabback10};
    M07: \citet{miyazaki07}; K08: \citet{kasliwal08}; B11:
    \citet{bellagamba11}.}
\begin{tabular}{cccccccccccc}
  \hline
  \hline
  Telescope & $\sigma_{e_{1}^{res*}}$ & $\sigma_{e_{2}^{res*}}$
  &$n_{det}$ & $n_{shear}$ &
  $n_{eff}$ & $\sigma_{e_1^s}$ & $\sigma_{e_2^s}$ & $\sigma_{\evec^s}$ &
  $\bar{z}$ & ${z^{\rm med}}$ & {Ref.} \\
  \hspace{5pt}
  (1) & (2) & (3) &(4) & (5) & (6) & (7) &
  (8) & (9) & (10) & (11) & (12) \\
  \hline
  CFHT   &  0.0031$^{\rm c}$ & 0.0026$^{\rm c}$ & 52.3 & 32.8 & 29.7 &
  0.33$^{\rm b}$ & 0.34$^{\rm b}$ & 0.47$^{\rm b}$& 1.04$^{\rm b}$ &
  0.92$^{\rm b}$ & This work \\
  Subaru &  0.0059$^{\rm c}$ & 0.0055$^{\rm c}$ & 40.3 & 23.7 & 21.7 &
  0.30$^{\rm b}$ & 0.30$^{\rm b}$ & 0.42$^{\rm b}$ & 0.91$^{\rm b}$ & 0.85$^{\rm b}$ & This work \\
  \hline
  CFHT   & 0.0040$^{\rm d}$ & 0.0040$^{\rm d}$ &  & 30.6 & & $\sim$0.23$^{\rm f}$ & $\sim$0.233$^{\rm f}$ & 0.333$^{\rm f}$ & 0.92 & & G07 \\
  HST & & & 71.0 & & & & & & & & K08 \\ %
  HST & & & & 66.0 & 34.0 & $\sim$0.27$^{\rm a}$&
  $\sim$0.27$^{\rm a}$ & 0.38$^{\rm a}$ &
  1.0$^{\rm b}$ & & L07, L10 \\ %
  HST    & & & & 80.0 & 76.0$^{\rm e}$ & $\sim$0.31$^{\rm b}$ & $\sim$0.31$^{\rm b}$ & 0.44$^{\rm b}$ & & & S10 \\
  Subaru & & & & 37.1 & & & & & & & M07 \\ %
  Subaru & & & 42.0 & & & & & & & & K08 \\ %
  Subaru & & & & 42.0 & & & & & & & B11, M07 \\
  \hline
\end{tabular}
\label{tab:lens}
\begin{flushleft}
  \scriptsize
   $^{\rm a}$ Quantity measured using the shear catalogue.\\
   $^{\rm b}$ Quantity measured using the shear-photo-z matched catalogue.\\
   $^{\rm c}$ Quantity measured using a Gaussian window of $r_g=3 \times r_h$.\\
   $^{\rm d}$ Unknown Gaussian window size used to measure the quantity.\\
   $^{\rm e}$ The used photo-z catalogue had a density of $\sim30$ galaxies
   arcmin$^{-2}$. Then, the photo-z distribution \\
   was used to estimate the redshifts of the remaining galaxies.\\
   $^{\rm f}$ Quantity measured using observed dispersion of ellipticities over the 40 nearest\\
   neighbours in the $r_h$-mag plane.
\end{flushleft}
\end{minipage}
\end{table*}

\subsection{The X-Ray Group Catalogue}\label{cats:x-ray}

We use the COSMOS X-ray catalogue of extended sources (Finoguenov et
al. in preparation) to trace the distribution of massive galaxy
associations in the field. The catalogue was obtained from a composite
mosaic of the XMM-Newton and Chandra X-ray data and it is an update
version of the X-ray catalogue presented in \citet{finoguenov07}. With
the usage of both data sets it has been possible to detect and measure
the flux of extended sources down to a limit\footnote{In fact, COSMOS
  X-ray data have an uniform depth of $10^{-15}$ erg cm$^{-2}$
  s$^{-1}$. However, in some areas, namely at the central parts, a
  depth of $4-6\times 10^{-16}$ erg cm$^{-2}$ s$^{-1}$ is
  measured. The expected performance of the future X-ray mission
  eROSITA on individual haloes is $2\times 10^{-14}$ erg cm$^{-2}$
  s$^{-1}$.}  of $10^{-15}$ erg cm$^{-2}$ s$^{-1}$.  The catalogue
contains a total of 231 systems, from $z=0.07-1.8$.

The mass estimates in this X-ray catalogue are based on the $L_{\rm
  X}$--$M_{200}$ scaling relation derived from a weak lensing analysis
of the galaxy groups and clusters in the COSMOS field
\citep{leau10}. The redshifts of the X-ray systems were assigned by
calculating the mean value of the photometric redshifts of the red
sequence galaxies within the X-ray extended region. The photometric
redshifts used were taken from \cite{ilbert09}.

All extended X-ray sources have a quality flag. We select objects with
X-ray quality Flag=1--3. Flag 1 is a zone free of projections, which
the significance of the X-ray detection is high. For these systems the
centre corresponds to the X-ray peak of the detection. Flag 2 systems
are subject to contamination (mainly due to unresolved AGNs), whose
centre correspond to a weighted optical centre of the system. The
X-ray detections for which Flag 1 and 2 were assigned are
spectroscopically confirmed systems. Flag 3 systems are similar to
Flag 1 and 2 but without the spectroscopic confirmation. Selecting
only Flags=1--3 and the X-ray detections that are in the CFHTLS-D2
field, our sample comprises 165 systems. The selected galaxy groups
have masses between $M_{200}=0.6-21.9 \times 10^{13}$\,M$_\odot$ with
median mass of $M_{200}=3.1\times 10^{13}$ M$_\odot$. The groups are
spread over the redshift range $z=0.07-1.84$ with a median value of
$z=0.68$.

\begin{figure}
\centering
\subfigure{\includegraphics[scale=0.33]{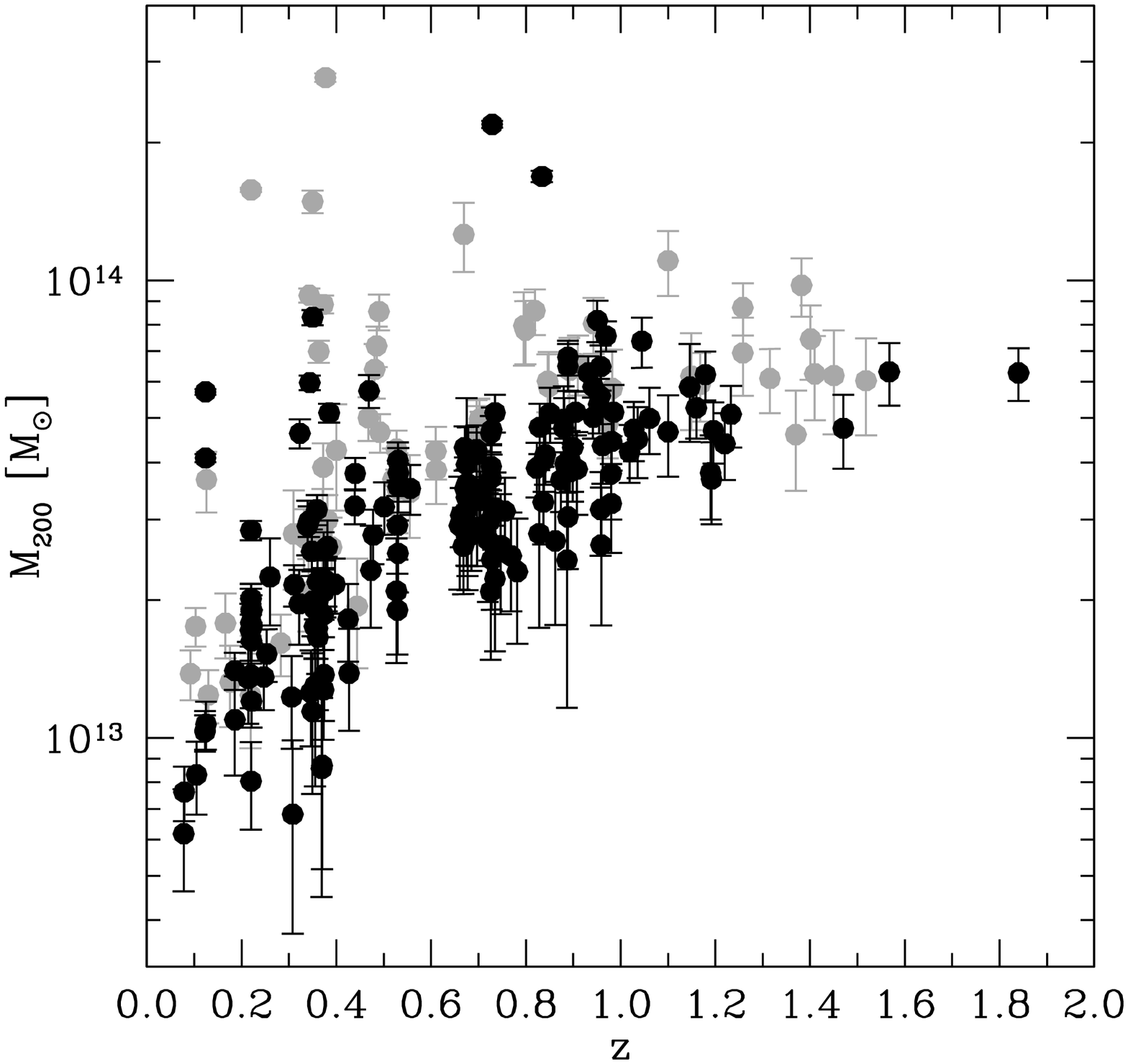}} \\
\subfigure{\includegraphics[scale=0.33]{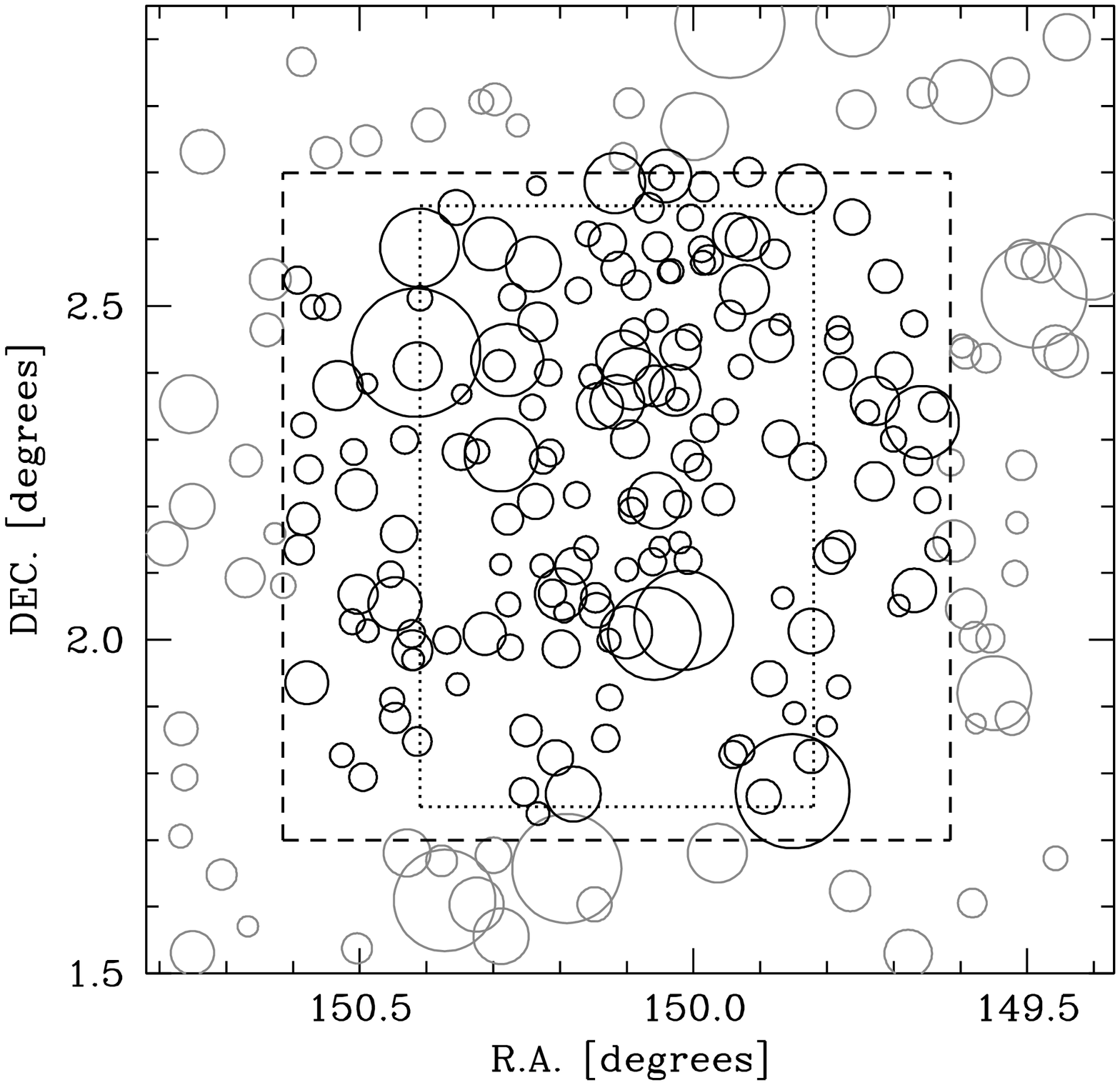}} \\
\subfigure{\includegraphics[scale=0.33]{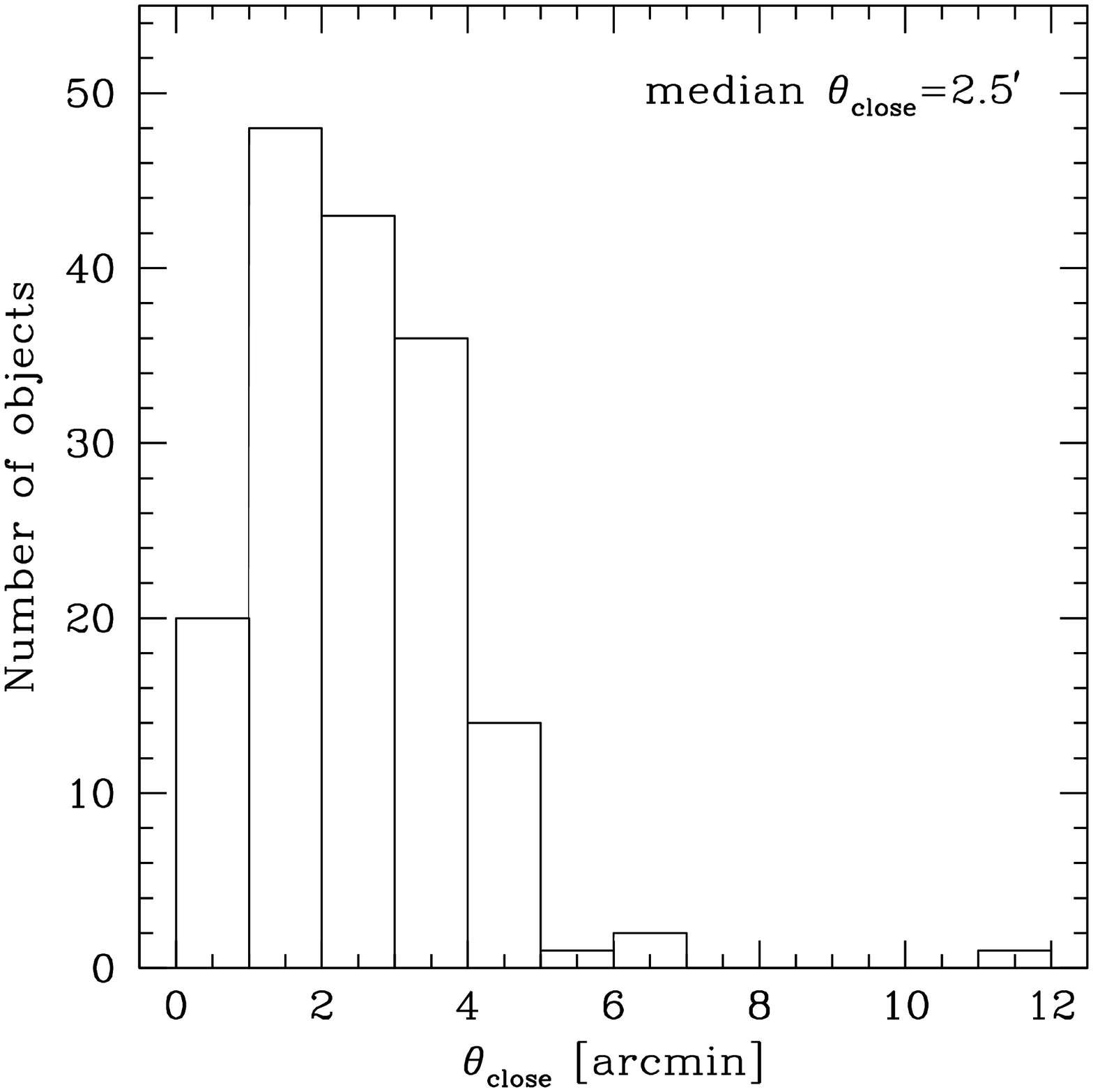}}
\caption{Top: Masses of the X-ray luminous haloes as a function of
  redshift. The black points show the selected systems whereas the
  grey points show the other systems that are out of our
  field-of-view. Middle: Positions of the X-ray detected haloes in the
  COSMOS field. The black circles show the systems used in this
  work. The grey circles show the remaining detected systems in the
  COSMOS field which were excluded from our sample. The size of the
  circles gives the $r_{200}$ value of the groups. The inner dotted
  line delimits the Subaru FOV and the external dashed line delimits
  the CFHTLS-D2 FOV. Bottom: Distribution of the projected distance to
  the closest neighbouring halo. Half of the sample has a secondary
  halo within $\theta_{\rm
    close}<2.5^{\prime}$. \label{xraydistribution}}
\end{figure}

The top panel of Fig. \ref{xraydistribution} shows the mass $M_{200}$
of selected systems as a function of the redshift $z$. The middle
panel shows the distribution of the groups in the field-of-view. Due
to the high depth of the X-ray catalogue, we can see that the field is
highly populated by haloes and, therefore, they have a small angular
separation between them. The bottom panel of
Fig. \ref{xraydistribution} shows the distribution of the projected
distance $\theta_{\rm close}$ between the galaxy groups and their
closest neighbour.

\section{Shear Simulations}\label{shear-sim}

In order to create shear mock catalogues, we use the position and the
number of galaxies from the shear-photo-z catalogues. This means that
galaxies are placed on the exact same positions as in the CFHT {\it
  i'} and Subaru {\it i$^+$} band data. Redshifts of the galaxies in
the mock catalogues are the same as the computed photo-z. The
distribution of the massive galaxy associations in the field and their
masses are taken from the X-ray catalogue. We notice that the
simulations could have been set without the need of obtaining a shear
and a photo-z catalogue. Simulations could have set, for instance, using a
known redshift distribution and by assigning random positions to the
field galaxies. However, we opt for obtaining the shear and photo-z
catalogues because current pipelines used to estimate these quantities
tend to suppress the observed density of galaxies. This is specially
true for shear pipelines, for which shape measurements fail for
certain types of galaxies, or for galaxies lying around bright
stars. Obtaining the shear and photo-z catalogues guarantees a more
realistic density distribution of galaxies.

We calculate the shear of each background galaxy, assuming that an
isolated halo is described by a Navarro-Frenk-White (NFW) profile
\citep{nfw}
\beq \label{eq:nfwprofile} 
  \rho(r) = \frac{\delta_{c}
  \rho_{cr}}{\frac{r}{r_{s}} \left( 1 + \frac{r}{r_{s}} \right)^{2}}
\eeq 
where $\rho_{cr}=3 H^2(z_d)/8 \pi G$ is the critical density of the
universe at the lens redshift $z_d$ and $H(z_d)$ is the Hubble's
parameter at the same redshift. The scale radius $r_s$ is related to
the virial radius $r_{vir}$ by the concentration parameter
$c_{vir}=r_{vir}/r_s$. Instead of the virial radius, it is common to
use the radius inside which the mass density is equal
$\rho=200\rho_{cr}$. We shall also use this convention from this point
on, therefore $c_{200}=r_{200}/r_s$, and $M_{200} \equiv
M(r_{200})=200 \rho_{cr} 4\pi/3r_{200}^{3}$. Thus, the density
contrast $\delta_{c}$ of equation (\ref{eq:nfwprofile}) is defined as
\beq
  \delta_{c_{200}} = \frac{200}{3} \frac{c_{200}^3}{\ln(1+c_{200})- \frac{c_{200}}{(1+c_{200})}} \,\,.
\eeq
The concentration value $c_{200}$ is calculated using a
$c_{200}(M_{200},z_d)$ relation presented in \citet{duffy08}, where
\beq
  c_{200} = \frac{5.71}{(1+z_d)^{0.47}} \times \left( \frac{M_{200}}{2.0\times10^{12}\,h^{-1}\rm{M_{\odot}}} \right)^{-0.084}
\eeq
with $h=0.72$. This relation was measured from N-body simulations
assuming WMAP5 cosmology and the same $M_{200}$ mass definition as we
use in this work. This $c_{200}(M_{200},z_d)$ relation was found as
the best fit for all haloes in the simulation (relaxed and unrelaxed)
between $z_d=0-2$. The galaxy groups of our sample have concentration
values between $2.7<c_{200}<5.1$.

Lensing measures the projected mass inside of a disk of radius
$R^2=r^2-z^2$, which in angular units is defined as $R=\theta \times
D_d$. The analytic expression for the shear of a lens characterised by
an NFW density profile, as a function of a dimensionless radius
$x=R/r_{s}$, is given by \citet{bartelmann96} and \citet{wb00}:
\begin{eqnarray}\label{eq:gammanfw}
   \gamma(x)=
   \left\{
     \begin{array}{l}
       \frac{r_{s}\delta_{c}\rho_{cr}}{\Sigma_{cr}} \left[
         \frac{8\,{\rm atanh}  \sqrt{(1-x)/(1+x)}}{x^2\sqrt{1-x^2}}+
         \frac{4}{x^2}\ln\left(\frac{x}{2}\right) \right.   \\
       \left. -\frac{2}{(x^2-1)} +\frac{4\,{\rm atanh}
           \sqrt{(1-x)/(1+x)}}{(x^2-1)(1-x^2)^{1/2}} \right], \,\, {\rm
         if}\quad x<1  \\
         \\
       \frac{r_{s}\delta_{c}\rho_{cr}}{\Sigma_{cr}}
       \left[\frac{10}{3}+4\ln\left(\frac{1}{2}\right)\right], \,\, {\rm
         if}\quad x=1  \\
         \\
       \frac{r_{s}\delta_{c}\rho_{cr}}{\Sigma_{cr}} \left[\frac{8
           \arctan
           \sqrt{(x-1)/(1+x)}}{x^2\sqrt{x^2-1}}+\frac{4}{x^2}\ln\left(\frac{x}{2}\right)
       \right.   \\
       \left. -\frac{2}{(x^2-1)}
         +\frac{4 \arctan \sqrt{(x-1)/(1+x)}}{(x^2-1)^{3/2}}\right], \,\,
       {\rm if}\quad x>1 
 \end{array}\right.
\end{eqnarray}
where $\Sigma_{cr}$ is the critical surface mass density depending
on the distances of lens and source,
\beq
   \Sigma_{cr}=\frac{c^{2}}{4\pi G}\frac{D_s}{D_d\,D_{ds}}
\eeq
with $c$ being the speed of light and $G$ the gravitational constant. 

With the above set of equations it is possible to calculate the shear
imposed by each lens on each background galaxy. Thus, for all galaxy
groups in our sample, a shear mock catalogue is generated, assuming
that the groups are isolated in the sky. Field galaxies with redshift
smaller than the redshift of the lens had their shear value set to
zero. Hereafter, we call these catalogues {\it isolated-pure-shear},
which are in total 165, each one of them corresponding to one galaxy
group of the sample. The ``pure-shear'' refers to the fact that galaxy
shape noise is not included at this point.

In reality, groups are not isolated but immersed in the field and what
is measured is the shear caused by all lenses.  The total shear of the
$j$-th galaxy is thus calculated by summing the shear over all the
lenses/groups
\beq
   \label{eq:total-shear}
   \gammavec_{{\rm total}_j}(\theta)=\sum_{j=1}^{N_{\rm Lens}=165}(\gamma_{1_j} + {\rm i}\,\gamma_{2_j})\,.
\eeq
From this point on, we call this catalogue {\it
  multiple-lens-pure-shear}. For each data set (CFHT and Subaru),
isolated-pure-shear catalogues and a multiple-lens-pure-shear
catalogue are generated.

Finally, the intrinsic ellipticity of galaxies has to be taken into
account. The observed shape of the $j$-th galaxy is given by the sum
of the intrinsic ellipticity $\evec_j^s$ and the induced shear
$\gammavec_j$, so that
\beq
  \label{eq:totale}
  \evec_j=e_{1_j}+{\rm i}\,e_{2_j}=(\gamma_{1_j} + e_{1_j}^s) + {\rm i}\,(\gamma_{2_j}+e_{2_j}^s)\,. 
\eeq 
We assign an intrinsic ellipticity to each galaxy drawn at random with
a Gaussian probability distribution. The width of the Gaussian
distribution is equal to the observed ellipticity dispersion obtained
from the shear-photo-z catalogues. The values are $\sigma_{e_1^{\rm
    s}}\sim\sigma_{e_2^{\rm s}}=0.34$ for CFHT data and
$\sigma_{e_1^{\rm s}}\sim\sigma_{e_2^{\rm s}}=0.30$ for Subaru data.
We generate 100 sets of random ellipticities and add to the shear
according to equation (\ref{eq:totale}) to both isolated-pure-shear
and multiple-lens-shear catalogues. Hereafter, we call these
catalogues {\it isolated-shape-noise} and {\it
  multiple-lens-shape-noise}.

It should be noted that the weak lensing study presented in this paper
is idealised by the fact that the analysed objects follow precise NFW
profiles with spherical symmetry. However, it has been shown in
previous studies that halo profiles can strongly deviate from simple
spherical models with canonical NFW slopes
\citep[e.g.][]{shaw06,ck07}. Furthermore, haloes can exhibit
substructures which are correlated to the main object
\citep[e.g.][]{abate09}. It has recently been shown by
\cite{marian10,bk10} that these deviations from simple NFW profiles
introduce additional scatter in the shear measurements and,
consequently, in the physical parameters derived from the shear. This
means that in practise, for real data, these deviations would be hard
to distinguish from LSS projections.

\section{Results}\label{resul}

The shape distortion is sensitive to all the matter along the
line-of-sight. From the bottom panel of Fig. \ref{xraydistribution}
we notice that more than half of the the groups in the field have at
least one neighbouring halo in a distance $\theta_{\rm
  close}<2.5^\prime$. The proximity of the haloes will likely perturb
the shear field of each single halo. In this section, we investigate
how the proximity of haloes with masses in the group regime modifies
the shear field. By using the mock catalogues, we can disentangle the
shear contribution of individual lenses and check whether the
detection of haloes and density profile estimates are affected by multiple
lensing.

\subsection{Halo detection via weak lensing}\label{resul:det}

The detectability of haloes by their weak lensing signal depends on
how much the coherent distortion is significant in comparison to the
local shape and shot noise. The aperture mass statistics ($M_{ap}$),
firstly introduced by \cite{map96}, has been broadly used to search
for haloes. In this method, the tangential shear contribution of all
sources that fall inside a circular aperture of a radius $\theta_0$ is
summed up with a weight function $Q(\theta)$. The $M_{ap}$ values are
usually calculated by placing the aperture on a grid that covers the
data region. The $M_{ap}$ signal is defined as
\beq \label{eq:map}
   M_{ap}=\frac{1}{N_{\theta_0}}\sum_{i=1}^{N_{\theta_0}}e_{\rm{t}_{i}}(\theta_i)Q_i(\theta_i)
\eeq 
with $N_{\theta_0}$ being the number of source galaxies within the
aperture. The distance $\theta_i$ is the projected angular distance
between the aperture centre $\theta_c$ and the $i$-th source galaxy
and $e_{\rm t}$ is the tangential shear defined as
$e_{\rm{t}}(\theta)=-\Re{[\evec(\theta) \times \rm{exp}(-2i\phi_c)]}$,
with $\phi_c$ being the polar angle between the horizontal axis and
the position $\theta_i$ of the object. Since the $M_{ap}$ value
estimated for the cross component of the shear, defined as
$e_{\rm{x}}(\theta)=-\Im{[\evec(\theta) \times \rm{exp}(-2i\phi_c)]}$,
has expectation value equals zero, it will be used as tool to search
for systematics errors. The cross component of the mass aperture
statistics ($M_{{ap}_{\rm x}}$) is calculated by the substitution of
$e_{\rm t}$ by $e_{\rm x}$ in equation (\ref{eq:map}). The
decomposition of the shear signal in tangential and cross components
are also referred as E-modes and B-modes\footnote{This naming
  convention has its origins with the CMB polarisation, which the
  pattern in the sky can be split into two components: the E-mode,
  which is an electric-field like decomposition or gradient-mode and
  the B-mode, which is a magnetic-field like decomposition or
  curl-mode. The gravity has zero curl because it is a conservative
  force, therefore B-modes are expected to be zero.}.  The $Q(\theta)$
of equation (\ref{eq:map}) is the filter function, used to enhance the
signal-to-noise of the detection since the shear is a very noisy
quantity. The noise within the aperture is given by
\beq 
   \label{eq:map-noise}
   \sigma_{M_{ap}}^2=\frac{1}{N_{\theta_0}^2}\sum_{i=1}^{N_{\theta_0}}
   \langle e_{\rm{t}_{i}}^2 \rangle Q_i^2(\theta_i) 
\eeq 
where $\langle e_{\rm{t}}^2 \rangle={\sigma_{\evec^{\rm s}}}^2$/2, with
$\sigma_{\evec^{\rm s}}$ being the two-component ellipticity dispersion
of galaxies. 

The significance of the detection is computed via the
signal-to-noise ratio
\beq 
   \label{eq:snr}
   {\rm S/N}=\frac{M_{ap}}{\sigma_{M_{ap}}}=\frac{\sqrt{2}}{\sigma_{\evec^s}}\,
   \frac{\sum_{i=1}^{N_{\theta_0}}e_{\rm{t}_{i}}(\theta_i)Q_i(\theta_i)}{\sqrt{\sum_{i=1}^{N_{\theta_0}}Q_i^2(\theta_i)}}.
\eeq
It should be noted that, for several times in previous works, when an
arbitrary choice of $\sigma_{\evec^{\rm s}}$ is made, this quantity is
set to 0.30. However, for most of the cases, this value is consistent
with the dispersion of only one component. If such a wrong value is
used, then S/N can be incorrectly improved by a a factor
$\sim30\%$. We refer to \citet{bs01} for further information on the
$M_{ap}$ statistic technique.

\begin{figure}
\center
\includegraphics[scale=.40]{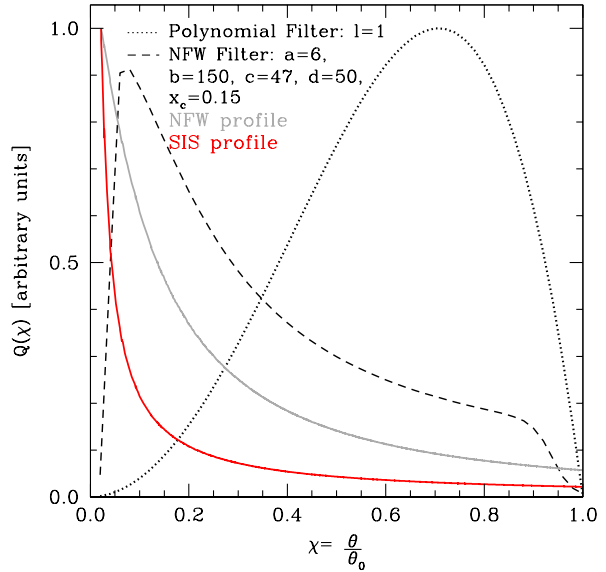}
\caption{The filter $Q$ as a function of the dimensionless radius
  $\chi$. Black lines show the shape of the $Q$ function for the set
  of parameters described in the text. The dotted line shows the
  polynomial and the dashed line the NFW filter function. The solid
  grey line shows the tangential shear profile of a group with
  $M_{200}=2.5\times10^{13}$ M$_{\odot}$ at $z_d=0.30$ assuming that
  the halo is described by an NFW density profile. The solid red line
  shows the tangential shear profile of a group with $\sigma_v=310$ km
  s$^{-1}$, also at $z_d=0.30$, but which is described by singular
  isothermal sphere (SIS) profile. The tangential shear profile is
  scaled to match the peak of the filter function. \label{filter}}
\end{figure}

Several types of filter functions $Q(\theta)$ have been used in the
literature. In \citet{schneider98} a family of polynomial filters was
proposed 
\beq 
   \label{eq:qpoly}
   Q(\chi)=\frac{(1+l)(2+l)}{\pi\theta_0^2} \chi^2 (1-\chi^2)^l 
\eeq 
where $\chi=\theta/\theta_0$ and $l$ gives the polynomial
order. Although this filter was extensively used in the past, it is
not optimal because haloes do not have a density profile that follows a
polynomial function. Therefore, a filter that has the shape similar
to an NFW profile\footnote{Although this filter is not strictly an NFW
  filter, we called it NFW hereafter in agreement with previous works
  like \cite{schirmer04}.} and enhances the signal-to-noise was
proposed by \citet{schirmer04}
\beq
   \label{eq:qnfw}
   Q(\chi)=\frac{1}{1+\rm{exp}(a-b\chi)+\rm{exp}(-c+d\chi)} \left[
   \frac{\tanh \left(\frac{\chi}{x_c}\right)} {\pi\theta_0^{2}\left( \frac{\chi}{x_c}\right) } \right]. 
\eeq 
We take these two filter functions to study the signal-to-noise ratio
of the selected galaxy groups. We use $l=1$ for the polynomial
filter. For the NFW filter function, we use $a=6$, $b=150$, $c=47$,
$d=50$ and $x_c=0.15$.  The motivation for choosing these values is
found in \cite{hetterscheidt05}: $a$ and $b$ set with these values
make an exponential drop of $Q$ in $\chi=0$; $c$ and $d$ make an
exponential cut-off in $\chi=1$ and $x_c\sim0.15$ maximises the S/N
for several aperture sizes; polynomial order $l=1$ also makes the $Q$
function drop sharply so that $Q(\chi=1)=0$. Fig. \ref{filter} shows
the behaviour of these two filters $Q$ as function of the
dimensionless radius $\chi$.

It has been pointed out by \cite{maturi05,maturi10} that weak lensing
cluster detection can be significantly improved if NFW-like filters
are adapted in such a way that the LSS signal contribution is
suppressed. \cite{gruen11} performed a study showing that their simple
strategy are not always successful, in particular when the background
density of galaxies is high (as it is for the COSMOS space data). This
is because pure LSS filters place higher weights in the innermost
regions of the haloes, where variations of the shear profile due to
correlated structures are more important, and thus, the virial mass
estimates can become more insecure if a LSS filter is used. We
therefore restrict ourselves to pure NFW and polynomial filters in
this work.

\begin{figure}
\center
\includegraphics[scale=.40]{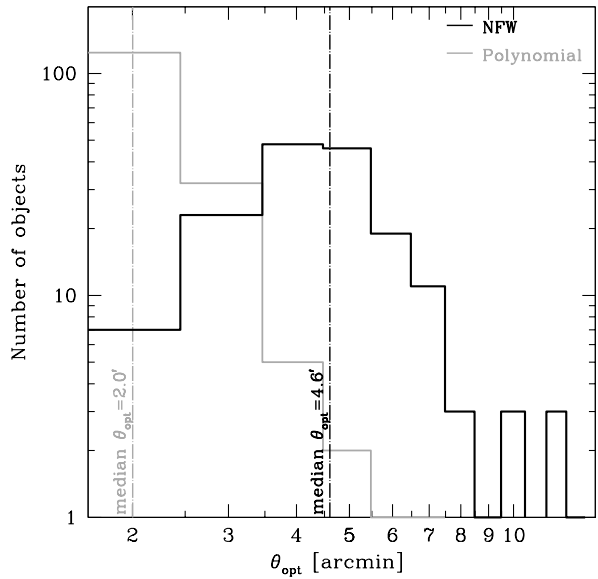}
\caption{Distribution of aperture
    sizes that maximise the $\langle {\rm S/N} \rangle$ of each
    isolated galaxy group. The grey histogram shows the optimal sizes
    for a polynomial filter function and the black histogram shows the
    same but for an NFW filter function. The median value of the
    distribution is $\theta_{\rm opt}=2.0^{\prime}$ for the polynomial
    filter and $\theta_{\rm opt}=4.6^{\prime}$ for the NFW
    filter. \label{optimal_aper}}
\end{figure}

The signal-to-noise ratio of a weak lensing detection depends on the
aperture size $\theta_0$. In order to calculate the aperture size
$\theta_{\rm opt}$ that maximises the signal-to-noise S/N for each
lens and filter, we place several aperture sizes on the position of
the haloes, checking the aperture value for which the signal-to-noise
S/N is maximised. Only galaxies with $z_s>z_d+0.05$ and $z_s\geq0.40$
are taken into account. The signal-to-noise S/N of each galaxy group
is calculated using the isolated-pure-shear catalogue with a CFHT-like
configuration. The input value of the ellipticity dispersion in
equation (\ref{eq:snr}) is $\sigma_{\evec^s}=0.47$. We call the
attention that, calculating the signal-to-noise S/N using the
isolated-pure-shear catalogues results in the mean expected
signal-to-noise $\langle {\rm S/N} \rangle$, i.e., the mean value of
the S/N distribution obtained from different realisations of random
samples of intrinsic ellipticity. Fig. \ref{optimal_aper} shows the
distribution of the optimal apertures sizes for the galaxy groups in
our sample.  We find a median value of $\bar{{\theta}}_{\rm
  opt}=2.0^\prime$ for the polynomial filter and $\bar{{\theta}}_{\rm
  opt}=4.6^\prime$ for the NFW filter function. In general, the NFW
filter requires larger aperture sizes than the polynomial
filter. Using a Subaru-like configuration, the distribution of optimal
apertures changes a bit but the median values for the two filter
functions remain the same.

\begin{figure}
\centering
\subfigure{\includegraphics[scale=0.38]{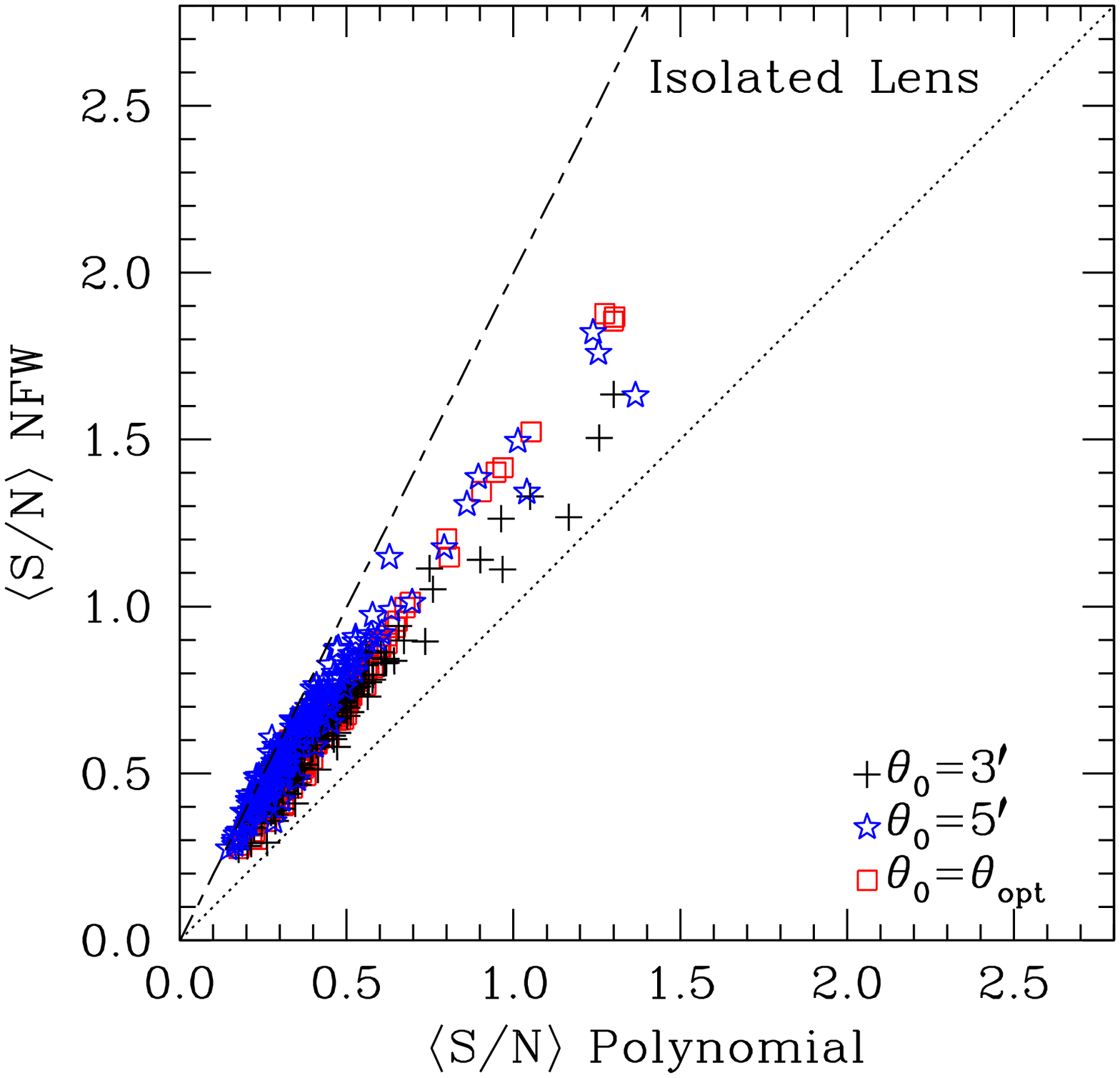}} \\
\subfigure{\includegraphics[scale=0.38]{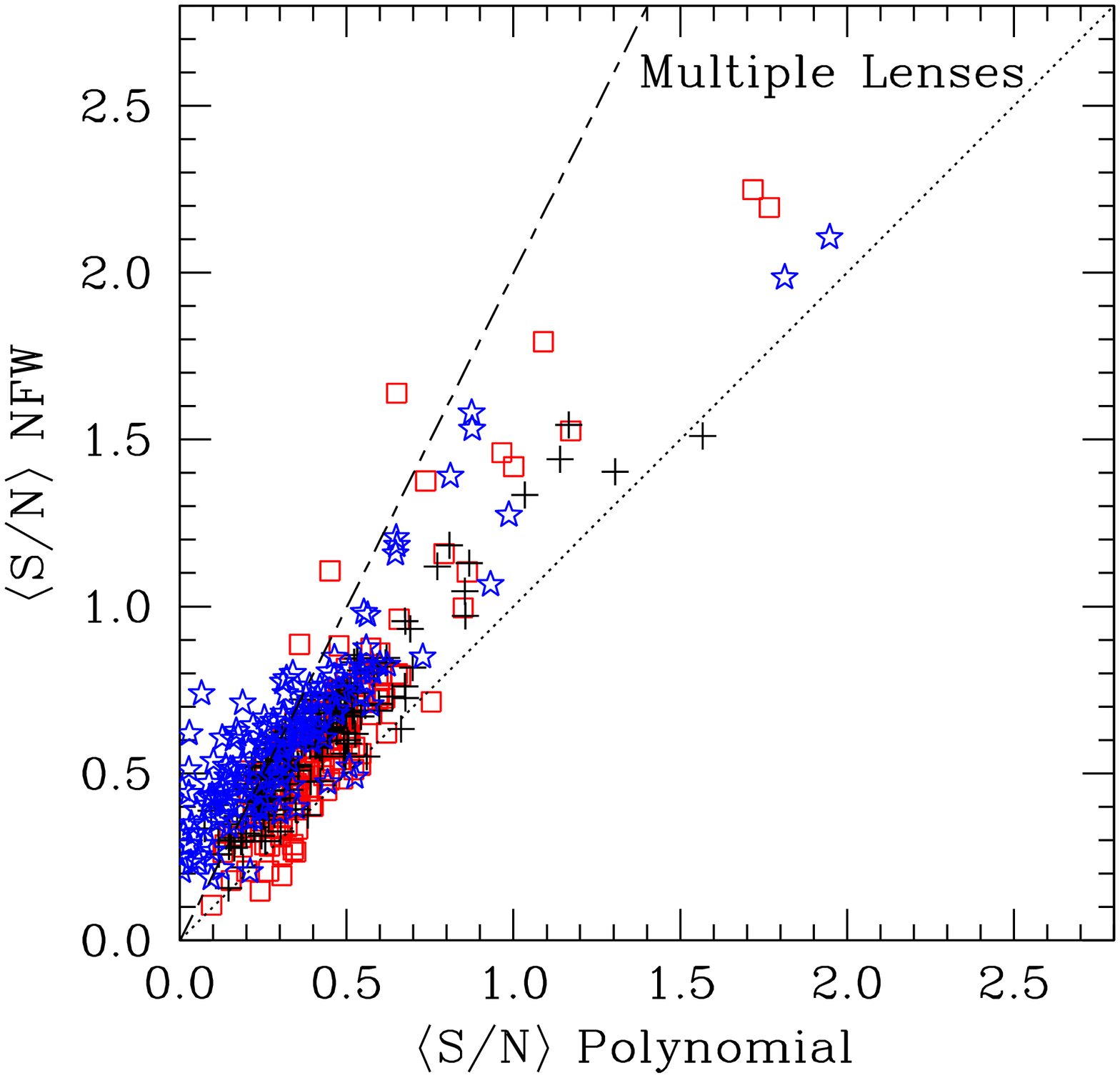}}
\caption{Mean expected signal-to-noise $\langle {\rm
    S/N} \rangle$ calculated using the $M_{ap}$ statistics for an NFW
  filter versus a polynomial filter function. The signal-to-noise is
  obtained from a CFHT-like galaxy distribution for 3 different
  aperture sizes: $3^\prime$ (black crosses), $5^\prime$ (blue stars),
  and the aperture value which maximises the signal-to-noise (red
  squares) for each halo and filter. The top panel shows the values
  computed when the lenses are considered isolated in the sky. The
  bottom panel shows the values calculated considering the
  contribution of all lens in the FOV. The dotted line has unitary
  slope and the short-dashed line twice the unity. Both lines are
  shown for guiding purpose.\label{map_snr}}
\end{figure}

A comparison of the mean expected signal-to-noise $\langle {\rm S/N}
\rangle$ between the two filter functions evaluated in the
isolated-lens context is shown on the top panel of
Fig. \ref{map_snr}. In this plot, we use three apertures sizes to
calculate the signal-to-noise ratio: the value which corresponds to
the optimal aperture ${\theta}_{\rm opt}$ of each halo and filter,
$3^{\prime}$ and $5^{\prime}$. The comparison between the two filter
functions evaluated in the multiple-lens context is shown on the
bottom panel of the Fig. \ref{map_snr}. Once again, we calculate the
signal-to-noise using the CFHT-like configuration with the described
criteria to select background galaxies. Fig. \ref{map_snr} shows
that if lenses are treated as isolated, the mean expected
signal-to-noise using an NFW filter is always higher, even when the
optimal aperture $\theta_{\rm opt}$ of each halo and filter is used.

\begin{figure}
\centering
\subfigure{\includegraphics[scale=0.38]{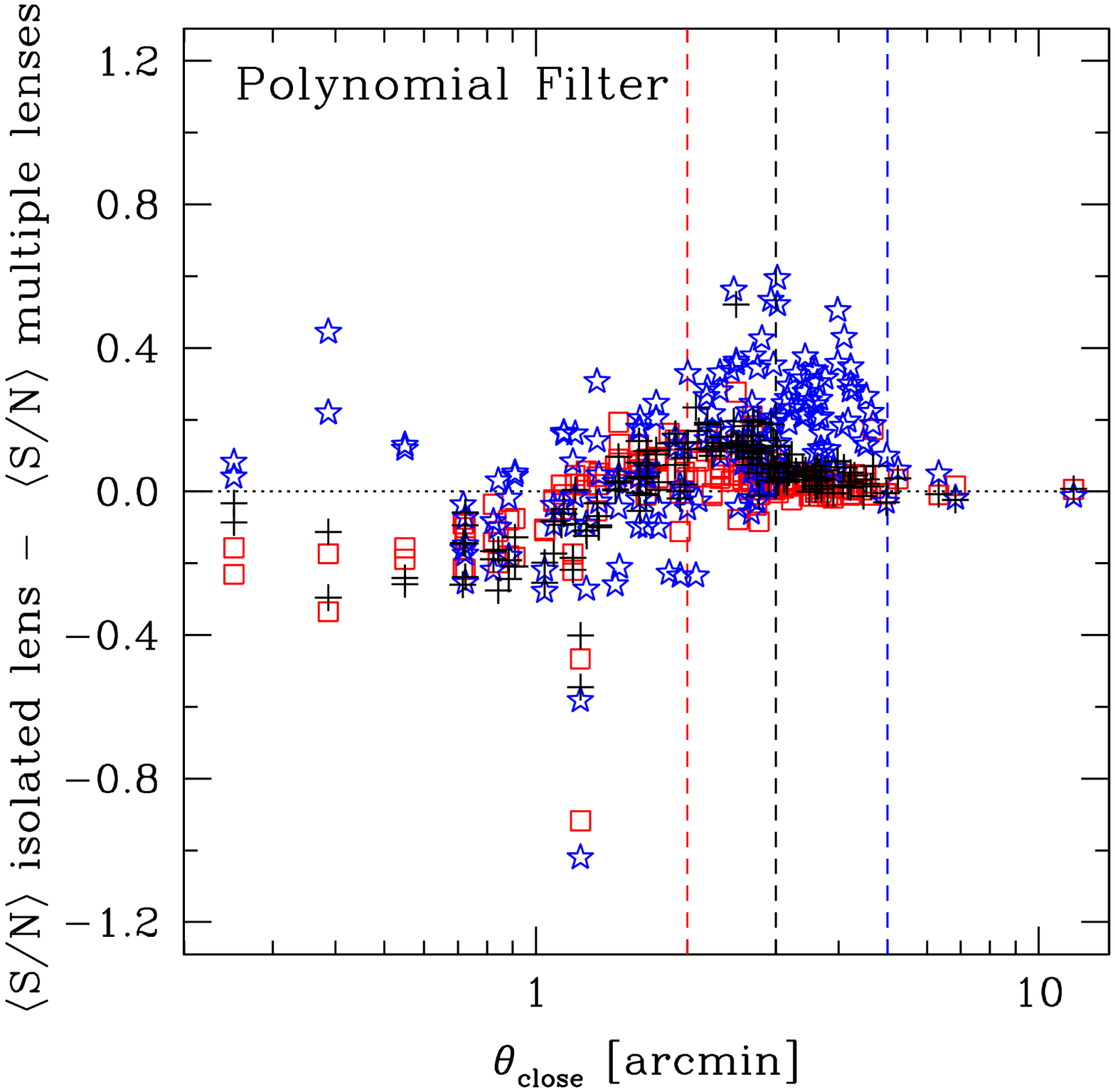}} \\
\subfigure{\includegraphics[scale=0.38]{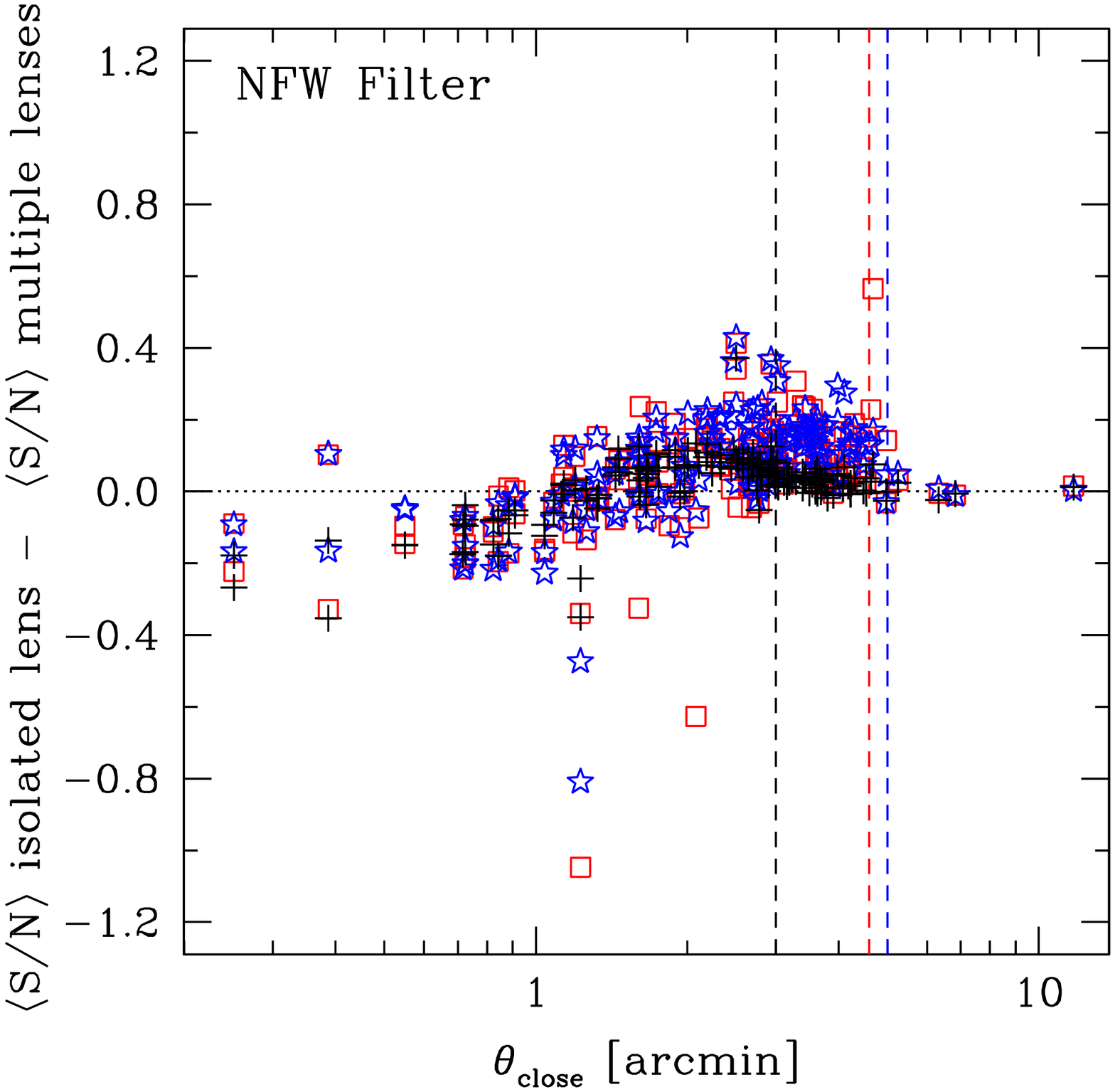}}
\caption{Difference of the mean expected signal-to-noise $\langle {\rm
    S/N} \rangle$ obtained for the isolated and multiple lens
  calculations as a function of the distance of the closest
  neighbouring halo $\theta_{\rm close}$. On the top panel we show the
  difference of the $\langle {\rm S/N} \rangle$ calculated using the
  polynomial filter and on the bottom panel we show the same, but for
  the NFW filter function. Symbols follow the same convention as in
  Fig. \ref{map_snr}. Vertical lines show the size of the apertures
  used to calculate the signal-to-noise, except for the red line,
  which shows the median value of the optimal aperture, being
  $\bar{\theta}_{\rm opt}=2.0^{\prime}$ for the polynomial and
  $\bar{\theta}_{\rm opt}=4.6^{\prime}$ for the NFW filter. The rms
  values of the difference in the signal-to-noise $\langle {\rm S/N}
  \rangle$ are: 0.13, 0.23 and 0.12 for the polynomial filter
  ($3^{\prime}$, $5^{\prime}$ and $\theta_{\rm opt}$ respectively) and
  0.09, 0.16 and 0.17 for the NFW filter function (also $3^{\prime}$,
  $5^{\prime}$ and $\theta_{\rm opt}$
  respectively). \label{snr_iso_vs_web}}
\end{figure}

Fig. \ref{snr_iso_vs_web} shows the difference of the mean expected
signal-to-noise $\langle {\rm S/N} \rangle$ obtained for the isolated
and multiple lens calculations as a function of the projected distance
$\theta_{\rm close}$ between the galaxy groups and their closest
neighbour. Multiple haloes along the line-of-sight can both give rise
to a larger shear signal or suppress it. The root mean square (rms)
values of the difference in the signal-to-noise $\langle {\rm S/N}
\rangle$ are: 0.13, 0.23 and 0.12 for the polynomial filter
($3^{\prime}$, $5^{\prime}$ and $\theta_{\rm opt}$ respectively) and
0.09, 0.16 and 0.17 for the NFW filter (also $3^{\prime}$,
$5^{\prime}$ and $\theta_{\rm opt}$ respectively). It is worth noting
that for one galaxy group the difference in the signal-to-noise
$\langle {\rm S/N} \rangle$ is up to a factor of 0.8-1, depending on
the aperture size used. We can generalise these results to an
arbitrary background galaxy density $n_{gal}$, obtaining
\beq
  \Delta{\rm S/N} \thickapprox 15\% \times \sqrt{\frac{n_{gal}}{30}} 
\eeq
and
\beq
   \Delta{\rm S/N}_{\rm max} \thickapprox 90\% \times \sqrt{\frac{n_{gal}}{30}}\,.
\eeq
The polynomial filter shows more scatter in the difference of $\langle
{\rm S/N} \rangle$ than the NFW filter, which can be explained by the
steepness of the NFW filter function. Fig. \ref{snr_iso_vs_web} also
shows that the difference in the signal-to-noise is larger when the
closest halo in projection falls within the aperture or, in other
words, the difference is larger when the distance to closest halo is
smaller than the aperture size used. A Subaru-like configuration does
not change the results shown in Figs. \ref{map_snr} and
\ref{snr_iso_vs_web}, it only yields in smaller values of the
signal-to-noise due to lower density of background galaxies.

Table \ref{max-snr} shows the maximum value of the signal-to-noise
$\langle {\rm S/N} \rangle$ which can be obtained for the selected
galaxy groups using optimal aperture sizes in the measurements. As
expected, the maximum signal-to-noise is very low due to the low mass
range of the haloes studied. It is unlikely that the galaxy groups
investigated in this work can be detected by their weak lensing signal
with the Subaru- and the CFHT-like configurations.

\begin{table}
\begin{center}
\caption{Maximum of the S/N expectation values for the COSMOS haloes\label{max-snr}}
\begin{tabular}{ccccc}
\hline
\hline
Telescope & \multicolumn{2}{c}{Isolated Lens} & \multicolumn{2}{c}{Multiple lenses} \\
\cline{2-5}
          & Polynomial  & NFW  & Polynomial  & NFW  \\
\hline
CFHT      & 1.30        & 1.87 & 1.77        & 2.25 \\ 
Subaru    & 1.08        & 1.59 & 1.07        & 1.54 \\ 
\hline
\end{tabular}
\end{center}
\end{table}

\begin{figure}
\center
\includegraphics[scale=.58]{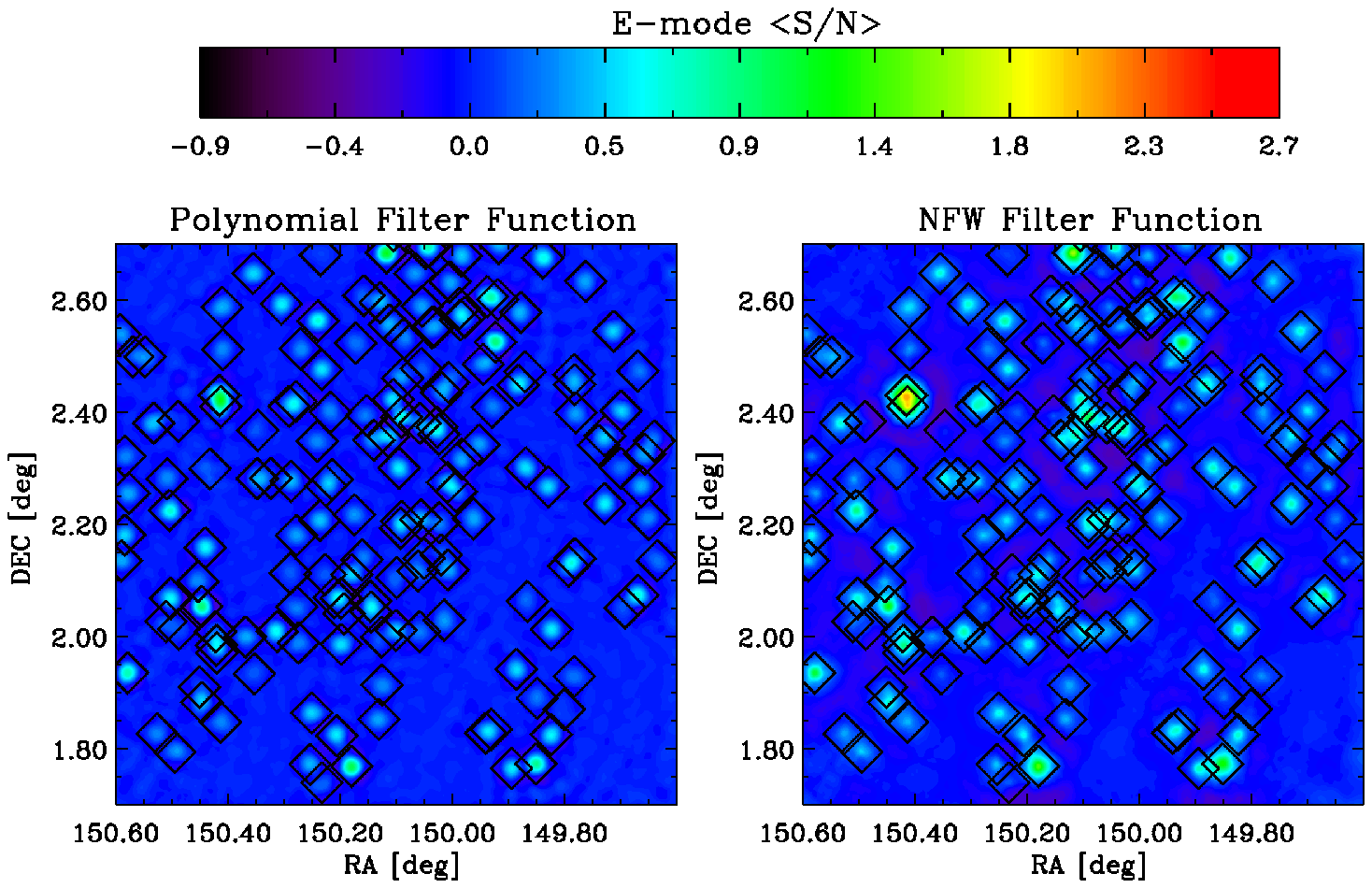}
\caption{Mean expected signal-to-noise $\langle {\rm S/N} \rangle$
  calculated using a COSMOS simulated catalogue which have a CFHT-like
  distribution of galaxies.  On the left panel is shown the $\langle
  {\rm S/N} \rangle$ map computed with an aperture of $\theta_{0_{\rm
      poly}}=2.0^\prime$ and a polynomial filter. On the right panel
  is shown the $\langle {\rm S/N} \rangle$ map computed with an
  aperture of $\theta_{0_{\rm NFW}}=4.6^\prime$ and an NFW
  filter. Diamonds mark the positions of the selected galaxy
  groups. If the grid points are shifted by $6^{\prime\prime}$, the
  maximum change in $\langle {\rm S/N} \rangle$ is 0.24 for the
  polynomial and 0.27 for the NFW filter, with $\Delta \langle {\rm
    S/N} \rangle_{\rm rms}=0.02$ for both filters. \label{map_d2_noellip}}
\end{figure}

We conduct the $M_{ap}$ statistics of the whole field area by
splitting it into a grid with $12^{\prime\prime}$ of resolution and
evaluate the mean signal-to-noise $\langle {\rm S/N} \rangle$ at each
grid point. An array of $ 300\times 300$ grid points is necessary to
cover the CFHTLS-D2 field and $185 \times 270$ to cover the Subaru
imaged area. An aperture of $\theta_{0_{\rm poly}}\equiv
\bar{\theta}_{\rm opt}=2.0^\prime$ is used to evaluate the
signal-to-noise with the polynomial filter and $\theta_{0_{\rm
    NFW}}\equiv \bar{\theta}_{\rm opt}=4.6^\prime$ with the NFW
filter. We make a cut in the catalogues to select only source galaxies
with $z_s\geq0.40$. Any other information on the redshift of source
galaxies is not taken into account so that all galaxies lying within
the aperture are used to evaluate the signal. This is done because
when blind searches are conducted to detect haloes, the redshift of the
haloes $z_d$ are not known a priori, making background galaxy selection
not possible. When galaxy redshifts are available, it is possible to
carry out the analysis using redshifts slices, but this goes beyond
the aim of this work. We calculate the signal-to-noise at each grid
point using the 100 multiple-lens-shape-noise catalogues and evaluate
the mean. Fig. \ref{map_d2_noellip} shows the mean expected
signal-to-noise $\langle {\rm S/N} \rangle$ map using the CFHT-like
configuration. Subaru configuration results in a similar map with
smaller area but with smaller values of S/N. We check the influence of
the grid position to the signal-to-noise by displacing the grid points
by $6^{\prime\prime}$, i.e., half of the grid size. The maximum change
in $\langle {\rm S/N} \rangle$ is 0.24 and 0.27 for the polynomial and
NFW filters respectively, with an rms of the difference equals 0.02.

\begin{figure}
\centering
\subfigure{\includegraphics[scale=0.40]{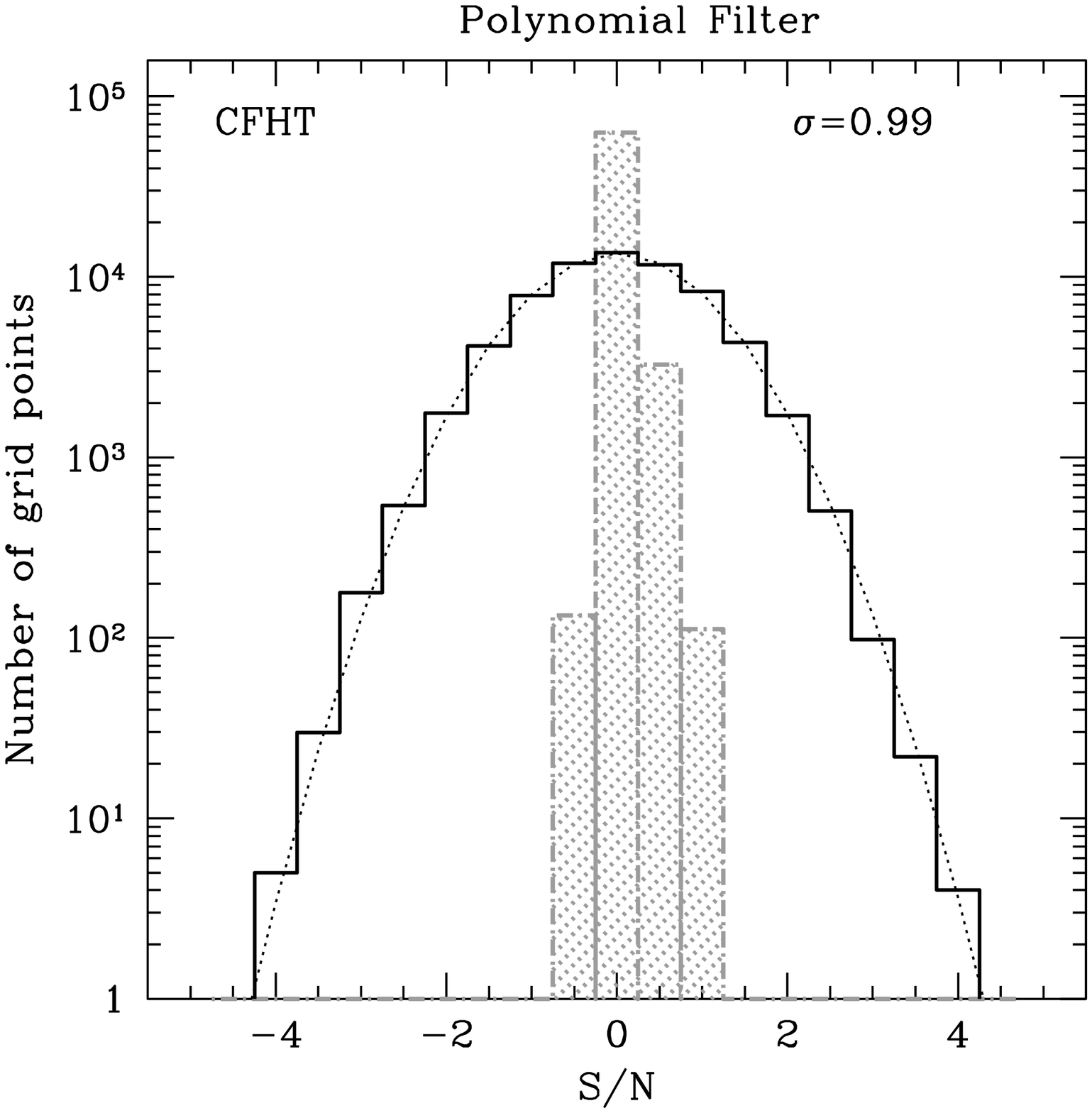}} \\
\subfigure{\includegraphics[scale=0.40]{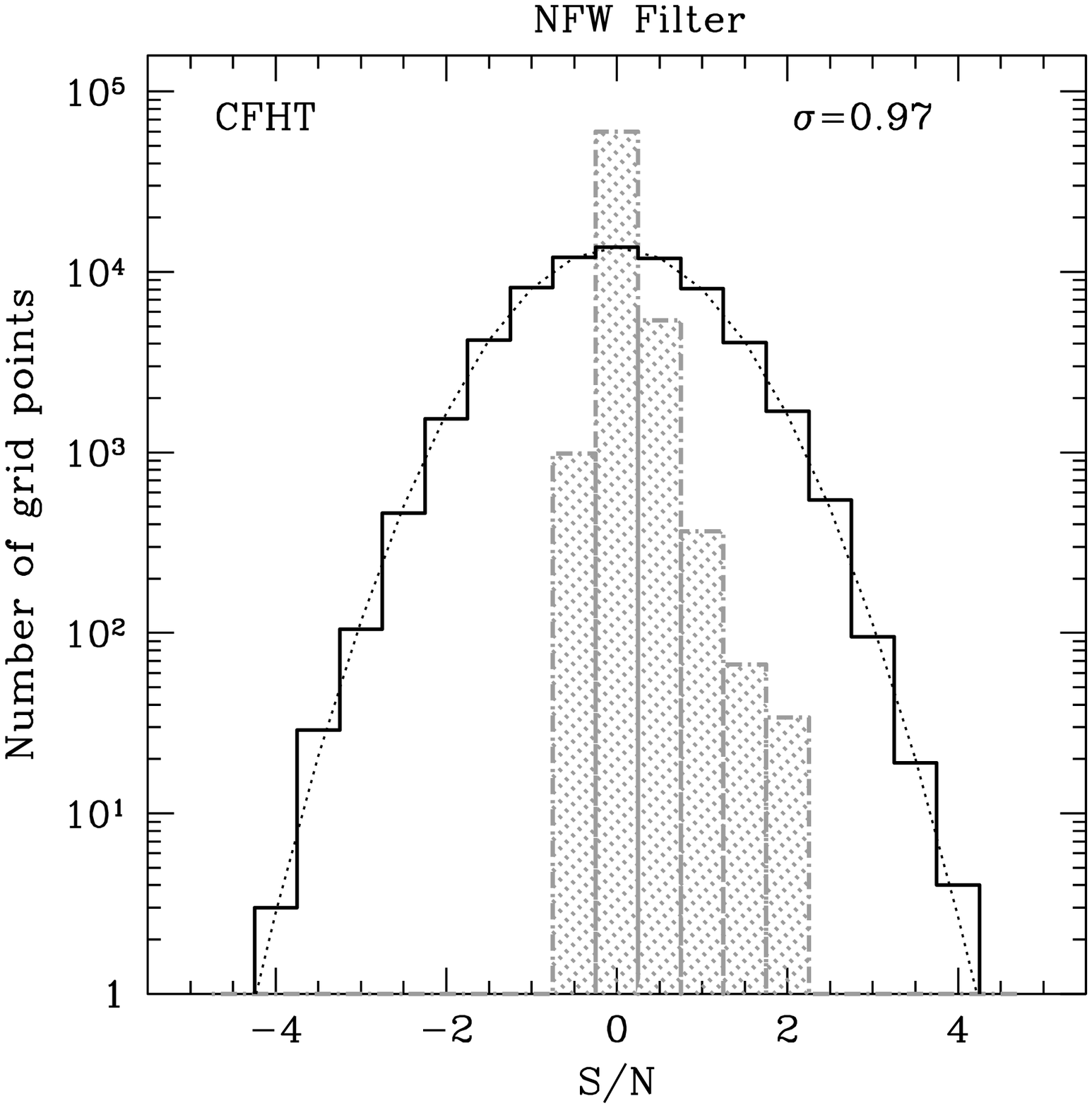}}
\caption{Signal-to-noise S/N distribution of the grid points. The grid
  covers the CFHTLS-D2 field in an array of $300\times300$ grid points
  with a step-size of $12^{\prime\prime}$. Only grid points within an
  aperture lying totally inside of the field are plotted, yielding
  $260\times257$ grid points. The hatched histogram shows the
  distribution of the mean expected signal-to-noise $\langle {\rm S/N}
  \rangle$ obtained from the CFHT multiple-lens-pure-shear
  catalogue. The thick black histogram shows the distribution of S/N
  obtained from shape noise only, assuming an intrinsic ellipticity
  dispersion equal to the CFHT configuration, i.e.,
  $\sigma_{\evec}^s=0.47$.  The thin dotted line shows the best-fit of
  a Gaussian function to the intrinsic ellipticity distribution. The
  fitted sigma $\sigma$ is written on the plot and is consistent with
  1.\label{map_d2_pure-ellip}}
\end{figure}

\begin{figure}
\centering
\subfigure{\includegraphics[scale=0.40]{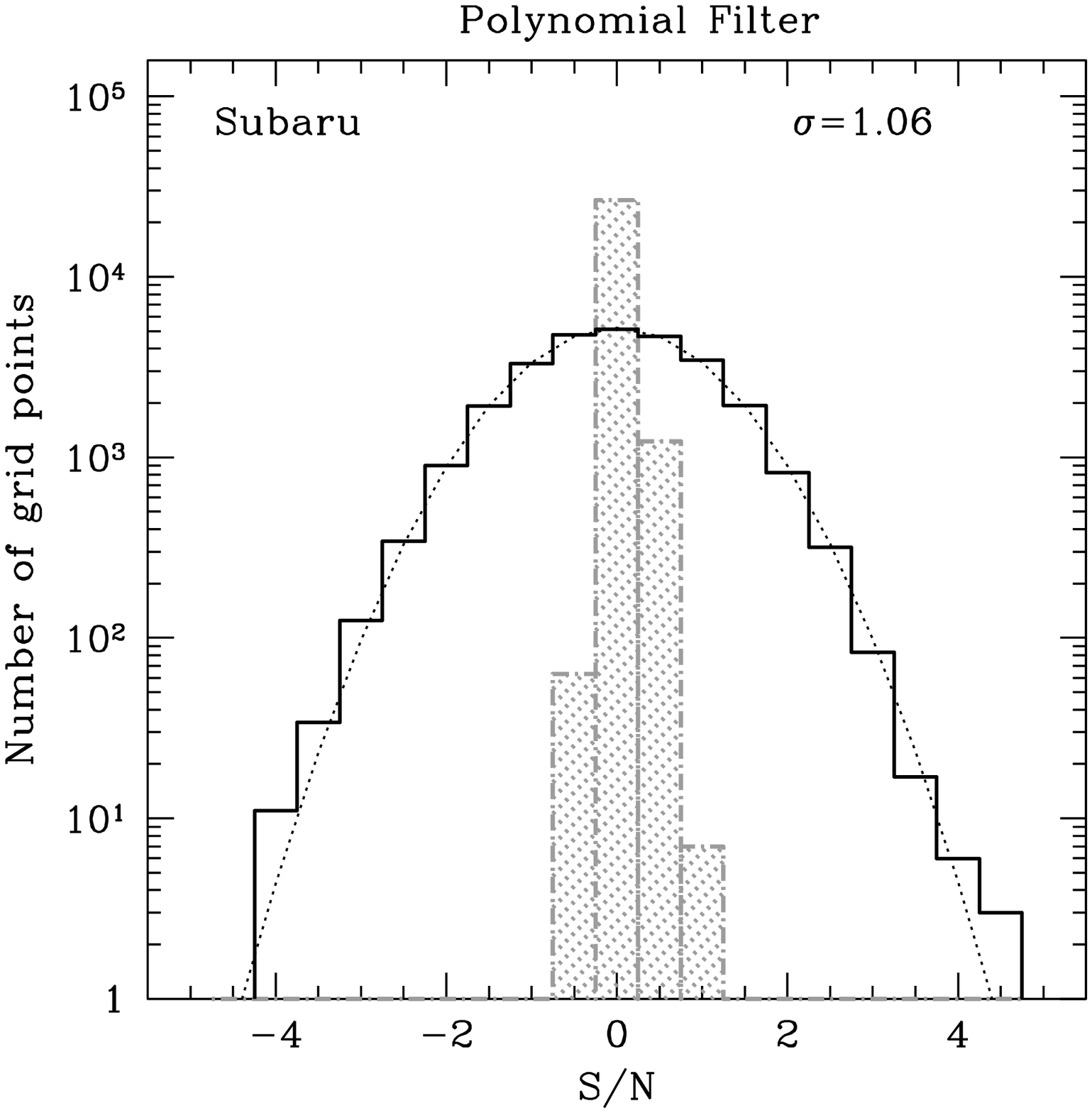}} \\
\subfigure{\includegraphics[scale=0.40]{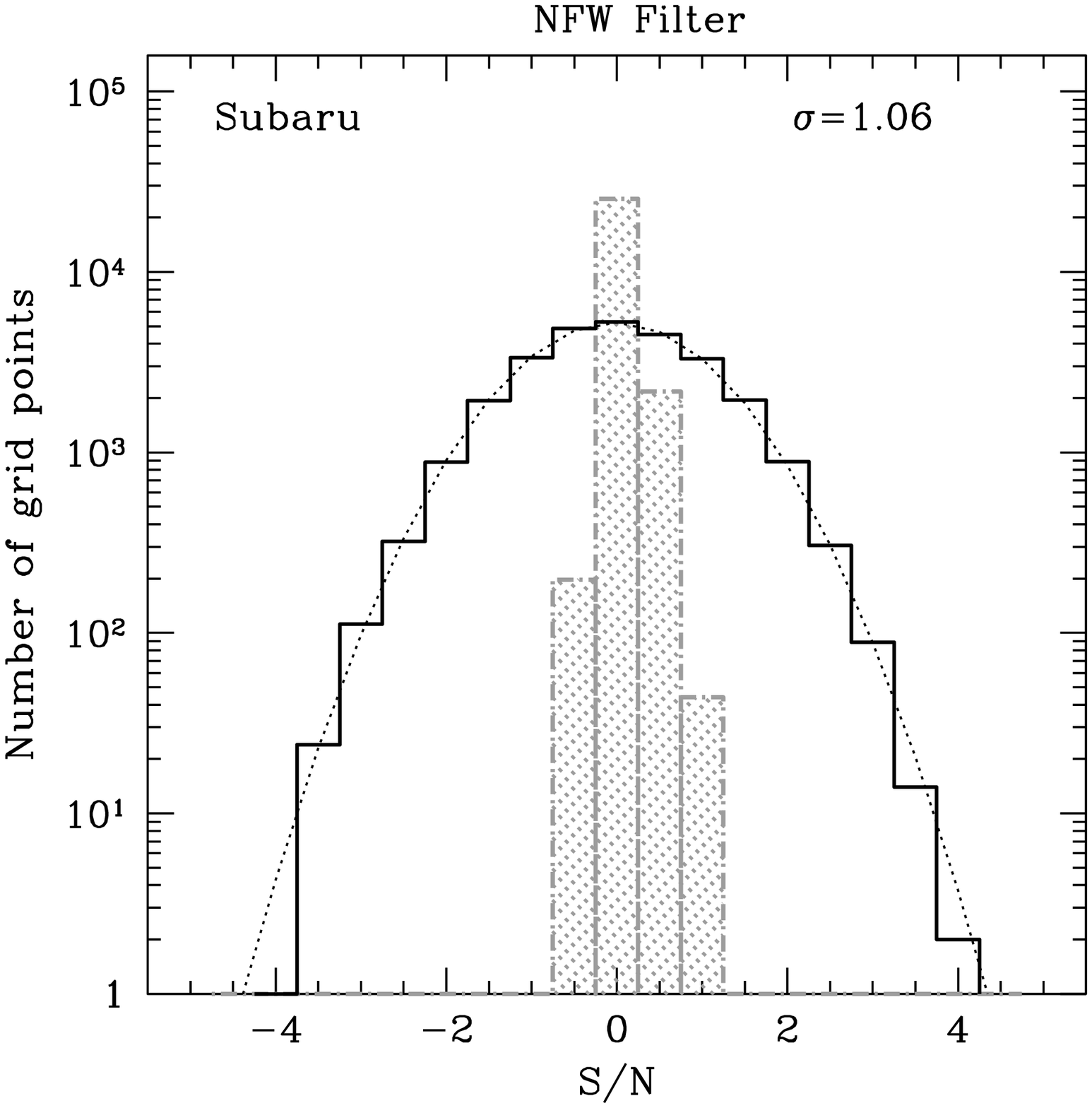}}
\caption{The same as Fig. \ref{map_d2_pure-ellip} but for a
  Subaru-like configuration. An array of $185\times270$ grid points
  was necessary to cover the data, but plotted here are just the grid
  points that fully lie on the data field, yielding $125\times223$
  grid points. The hatched histogram shows the distribution of the
  mean expected signal-to-noise $\langle {\rm S/N} \rangle$ obtained
  from the Subaru multiple-lens-pure-shear catalogue. Thick black
  histogram shows the distribution of S/N obtained from shape noise
  only, assuming an intrinsic ellipticity dispersion equal to the
  Subaru configuration, i.e.,
  $\sigma_{\evec}^s=0.42$. \label{map_subaru_pure-ellip}}
\end{figure}

Likewise, we perform the $M_{ap}$ statistics of the field using a pure
intrinsic ellipticity realisation and check how the S/N distribution
of this {\it pure-shape-noise} catalogue compares to the one obtained
from the multiple-lens-pure-shear catalogue. Once again, we use an
aperture of $\theta_{0_{\rm poly}}=2.0^\prime$ for the polynomial
filter and $\theta_{0_{\rm NFW}}=4.6^\prime$ for the NFW filter
function. The S/N distributions of the multiple-lens-pure-shear and
pure-shape-noise catalogues with the CFHT-like configuration are shown
in Fig. \ref{map_d2_pure-ellip}.  The same is shown in Fig.
\ref{map_subaru_pure-ellip} but for the Subaru-like
configuration. Only grid points falling inside an aperture that fully
lies inside of the data fields are considered, yielding $260\times257$
grid points for the CFHT and $125\times223$ grid points for the Subaru
configuration. As we can see in Figs. \ref{map_d2_pure-ellip} and
\ref{map_subaru_pure-ellip}, the pure intrinsic ellipticity follows a
Gaussian probability distribution, centred at zero and width
$\sigma\sim1$, i.e., consistent to the S/N units. Therefore,
independently of intrinsic ellipticity dispersion of the data, it is
possible to have positive and negative E-modes (and also B-modes)
originating from the intrinsic ellipticity in a various range of S/N:
$|{\rm S/N}|\leq1$ accounts for about 68\% of the set of the grid
points, $|{\rm S/N}|\leq2$ accounts for 95\%, $|{\rm S/N}|\leq3$ for
99.7\%, $|{\rm S/N}|\leq4$ for 99.99\%, and so on. Thus, for the CFHT
described grid configuration, this means that $\sim 200$ grid points
are expected to have $|{\rm S/N}|\geq3$ originating from intrinsic
alignments. On the other hand, the gravitational shear originating
from the galaxy groups in our sample result in a signal-to-noise
smaller than 3, meaning that neither a CFHT nor a Subaru-like
configuration are sufficient to detect COSMOS-like haloes without
contamination generated by false peaks.

\begin{figure}
\center
\includegraphics[scale=.43]{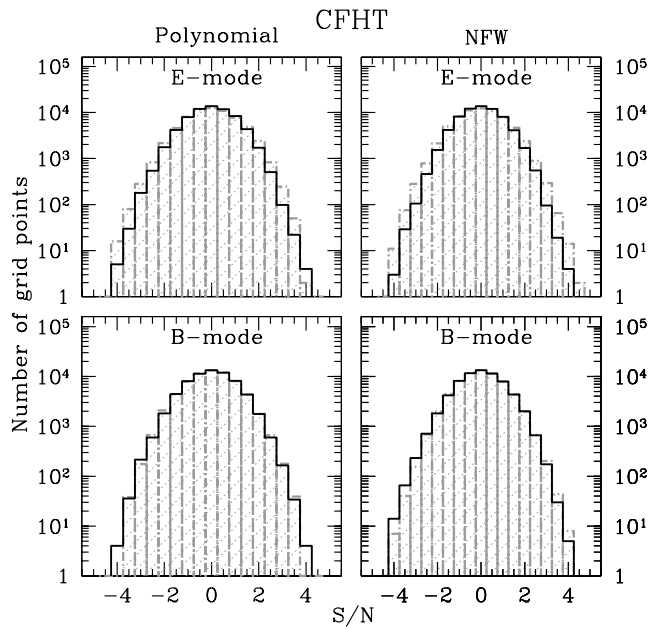}
\caption{Same as Fig. \ref{map_d2_noellip} but for the observed CFHT shear-photo-z
  catalogue. Upper panels show the distribution of E-modes and lower
  panels the distribution of B-modes. The hatched histogram shows the
  distribution of the signal-to-noise S/N obtained from the CFHT
  shear-photo-z catalogue. Thick black histogram shows the distribution
  of S/N obtained by shape noise only, assuming an intrinsic
  ellipticity dispersion equal to the CFHT configuration, i.e.,
  $\sigma_{\evec}^s=0.47$. \label{map_d2_real}}
\end{figure}

\begin{figure}
\center
\includegraphics[scale=.43]{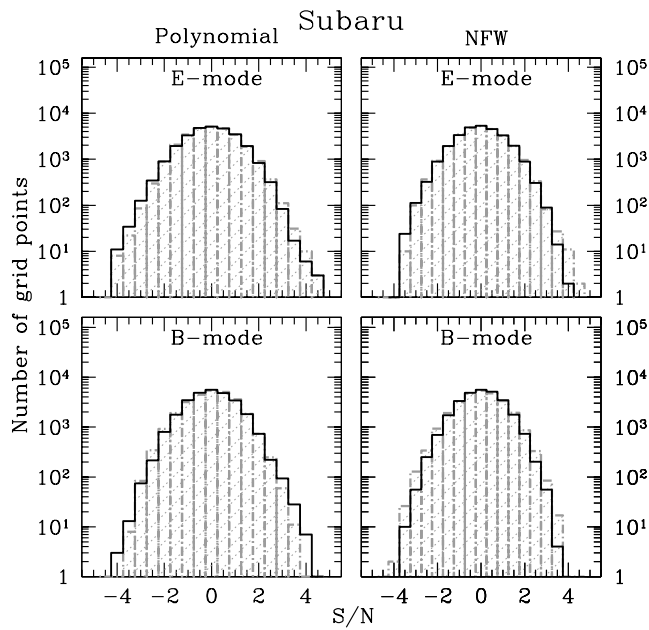}
\caption{Same as Fig. \ref{map_d2_noellip} but for the observed Subaru
  shear-photo-z catalogue. The hatched histogram shows the distribution
  of the signal-to-noise S/N obtained from the Subaru shear-photo-z
  catalogue. Thick black histogram shows the distribution of S/N
  obtained by shape noise only, assuming an intrinsic ellipticity
  dispersion equal to the Subaru configuration, i.e.,
  $\sigma_{\evec}^s=0.42$. \label{map_subaru_real}}
\end{figure}

Finally, we perform the same analysis but using the observed CFHT and
Subaru shear-photo-z catalogues and plot the S/N distribution of the
grid points. We evaluate the S/N of E-modes and B-modes and show in
Figs. \ref{map_d2_real} and \ref{map_subaru_real}. These two figures
demonstrate that the observed shear-photo-z catalogues yield in similar
S/N distributions to the ones obtained from the pure-shape-noise
catalogues. Furthermore, the S/N distribution of E-modes and B-mode are
almost the same.  Once more, this shows that the galaxy groups in our
sample can not be detected by their weak lensing signal without being
contaminated by the false peaks generated by intrinsic ellipticity.

\subsection{Previous halo detections in the COSMOS field}\label{resul:cosmos-lit}

In this section we present previously published results on shear
measurements in the COSMOS field. Table \ref{tab:lens} summarises the
results that are discussed in this section.

Our conclusion that COSMOS haloes can not be detected (at a
significant level ensuring low false detections) using CFHT and
Subaru-like configurations also holds for an HST-like configuration,
as it was shown by \citet{leau10}. They used the approach introduced
by \citet{hamana04} to predict the signal-to-noise ratio of the same
haloes studied in this work but for an HST-like galaxy distribution,
with $n_{shear}=66$ galaxies arcmin$^{-2}$. Their ellipticity
dispersion includes shape noise ($\sigma_{e_{i}^{\rm s}}\sim0.27$) and
shape measurement errors yielding $\sigma_{e_{i}^{\rm
    total}}=\sigma_{e_{i}^{\rm s}}+\sigma_{e_{i}^{\rm err}}=0.31$ per
component. Following \citet{hamana04}, they computed the convergence
map convolved with a Gaussian kernel. They found that even with an
HST-like galaxy distribution, COSMOS systems can not be detected
individually, except for nine haloes that have ${\rm S/N}>4$ (see
Fig. 1 from Leauthaud et al. 2010). From these nine systems, only two
of them are in our field-of-view. These two haloes have $4<{\rm
  S/N<5}$ with an HST-like configuration. We check the S/N of these
systems with Subaru and CFHT-like configurations using the Hamana et
al. method (H04) and the same parameters choices as presented in
Leauthaud et al. (L10) which used a Gaussian kernel
$\theta_{G}=1^{\prime}$ and background galaxies fixed at $z_s=1$. The
signal-to-noises are related through
\beq 
{\rm S/N} = {\rm S/N_{H04}} \left( \frac{0.40}{\sigma_{\evec^{\rm s}}} \right) \sqrt{ \frac
  {n_{shear}} {30}} \left( \frac{D_{ds}}{D_s} \right) \left[ {\left(
      \frac{D_{ds}}{D_{s}} \right)}\Big |_{z_s=1} \right]^{-1} 
\eeq
which yields ${\rm S/N_{L10}}={\rm S/N_{H04}}\times 1.35$, ${\rm
  S/N_{CFHT-this\,work}}={\rm S/N_{H04}}\times0.89$ and ${\rm
  S/N_{Subaru-this\,work}}={\rm S/N_{H04}}\times0.84$, using the
galaxy density of the CFHT and Subaru shear catalogues ($n_{shear}=32.8$
galaxies arcmin$^{-2}$ and $n_{shear}=23.7$ galaxies arcmin$^{-2}$
respectively). We visually inspect the Fig. 3 of \citet{hamana04}
paper and conclude that the galaxy groups in our sample would have a
maximum S/N$\sim3$ using their approach. Therefore, for these very
same haloes the S/N would not be greater than 2.7 for the CFHT-like
configuration and 2.5 for the Subaru-like configuration. Since the
HST-like configuration yields a S/N $\sim1.52$ higher than the
CFHT-like, it is possible to detect some of these haloes with deep
space-based observations as shown in \citet{leau10}.

Nevertheless there are weak lensing detections in the COSMOS field
claimed in the literature. \citet{kasliwal08} detected 3 systems using
both HST and Subaru data. The E-mode peaks were measured using the
convergence map with a kernel $\theta_{G}=1^{\prime}$, and only
detections with ${\rm S/N}>4$ were considered as safe. The detection
named as A in their paper is a real cluster with an X-ray counterpart
at $z_d=0.73$. The detection named as B matches an X-ray peak at
$z_d=0.83$ but the signal is claimed to be originating from a group at
$z_d\sim0.3$. Within a region of $\sim4^{\prime}$ around the detection
B there are two X-rays peaks at $z_d\sim0.3$ and three X-rays peaks at
$z_d\sim0.85$. Hence, this could be a case where five structures along
the line-of-sight act together to produce a signal that is interpreted
to be originating from one of these five structures alone. The
detection named as C is also real with an X-ray counterpart at
$z_d=0.22$ but it lies outside of both CFHTLS-D2 and Subaru
field. This cluster is one of those that could be detected with high
S/N in Fig.  1 of \citet{leau10}. The number of the detected objects
in Kasliwal et al. work is $n_{det}$=42 galaxies arcmin$^{-2}$ for
Subaru and $n_{det}$=71 galaxies arcmin$^{-2}$ for HST
data.\footnote{The reason why previous works that used Subaru
  observations of COSMOS field show much higher density of galaxies
  than we have found in this work is due to stacking strategy used to
  co-add individual exposures. While we have chosen only exposures
  taken with similar dither pattern, previous works have stacked all
  the exposures, regardless of the shift between them and the camera
  orientation. When combining all the exposures into a final mosaic
  using the data reduction procedure described in the Appendix
  \ref{app:data:subaru} we are also able to get $n_{shear}\sim40$
  galaxies arcmin$^{-2}$.}

\citet{gavazzi07} performed a study of the four CFHTLS Deep fields.
Using also the convergence map to detect E-mode peaks, with kernel
$\theta_G=1^{\prime}$, $n_{shear}$=30.6 galaxies arcmin$^{-2}$,
$z_d\sim1$ and $\sigma_{\evec^{\rm s}}\sim0.33$, they found 3 peaks
with S/N$\sim3.6$ in the CFHTLS-D2 field. Safe detections were
classified as the ones with ${\rm S/N}>3.5$. The peak called Cl-08 matches
the detection A from Kasliwal et al., although the redshift computed
using shear tomography $z_d=0.44$ does not match the actual redshift
of the cluster. Peaks called Cl-09 and Cl-13 have no X-ray association
in a distance of $\sim2.5^{\prime}$. Also, the redshifts found with
the shear tomography do not match the redshift of the nearest groups
at this distance.  

\begin{figure}
\center
\includegraphics[scale=.40]{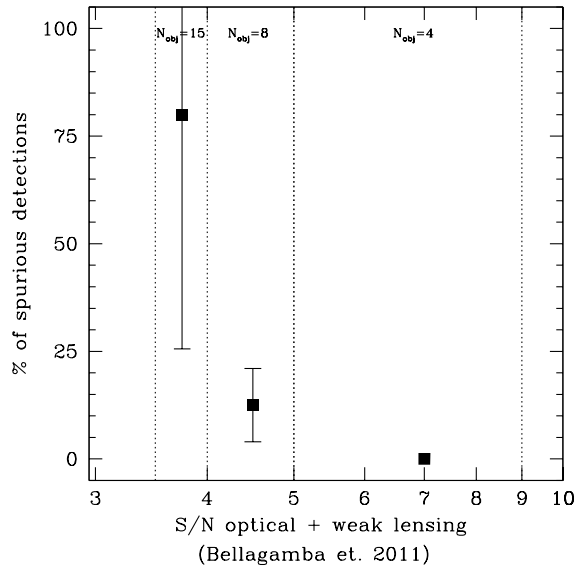}
\caption{Percentage of spurious detections as a function of
  signal-to-noise of the detected systems in \citet{bellagamba11}. The
  signal-to-noise is divided into three bins, which in the figure is
  delimited by the dashed vertical lines. For each bin, the percentage
  of spurious detections is calculated from the total number of
  objects ${\rm N_{obj}}$ within the bin. The percentage is shown as
  points with 1$\sigma$ error bars. \label{bellagamba}}
\end{figure}

In a recent paper \citet{bellagamba11} presented an optimal linear
filtering technique for optical and weak lensing data. The weak
lensing detection was performed in a similar way to the $M_{ap}$
statistics using the filter function proposed by
\citet{maturi05,maturi10}, which was designed specifically to suppress
the contribution from the large-scale structure. The input shear
catalogue was taken from \citet{miyazaki07}, which used Subaru data with
a density of $n_{shear}=42$ galaxies arcmin$^{-2}$ and assumed mean
redshift of background galaxies of $\bar{z_s}=0.8$. Using weak
lensing solely, they detected 82 peaks with S/N$>3$ but 40\% of the
detections are expected to be spurious. The matched optical and weak
lensing catalogue reduces the number of detections to 27 systems, where
only detections with S/N$\geq3.5$ were considered as safe. We check
for the X-ray counterparts of these 27 systems and calculate the
percentage of spurious detections as a function of signal-to-noise
provided in Table 1 of \citet{bellagamba11}. We make use of the full
COSMOS field, since both Bellagamba et al. and the COSMOS X-ray
catalogues cover more or less the same area (slightly larger than the
CFHTLS-D2 field). In order to calculate the percentage of spurious
detections, we split the signal-to-noise into three bins: $3.5<{\rm
  S/N}\leq4$, $4<{\rm S/N}\leq5$ and $5<{\rm S/N}\leq9$, covering the
S/N range of the 27 systems. Then, we check for the X-ray counterpart,
matching the spatial position and redshift of the systems to the X-ray
haloes found in the COSMOS catalogue. When it is not possible to find
the match, the system is classified as spurious. Fig. \ref{bellagamba}
shows the result: the percentage of spurious detections drops with the
increase of signal-to-noise. For systems with ${\rm S/N}>5$ the
percentage is zero. For the 12 systems within the bin $4<{\rm
  S/N}\leq5$, 11 have an X-ray peak associated. The positions of the
X-ray COSMOS catalogue and the positions found using the optical plus
weak lensing filtering technique of \citet{bellagamba11} are in very
good agreement, apart from one system, where the offset is
$\sim2^{\prime}$. This result makes this technique very promising for
searches of groups and clusters of galaxies with a low rate of spurious
detections if a threshold in S/N of 4 is used.

\subsection{Tangential Shear Dispersion}\label{resul:tsd}

In this section we investigate the tangential shear
dispersion that the haloes in the field introduce in the tangential
shear profile of individual groups. The aim of this analysis is to
understand the relevance of this ``large-scale structure'' noise to
the total error budget of the shear measurements.

\cite{hoekstra01b,hoekstra03} investigated the effect of the large-scale
structure on mass measurements and how it perturbs the tangential
shear profile. By splitting the observed shear into the components
\beq
   \gamma_{\rm{t}}(\theta)=\gamma_{\rm{t}}^{\rm{halo}}(\theta)+\gamma_{\rm{t}}^{\rm{LSS}}(\theta)
\eeq 
one conclusion obtained was that the distant large-scale structure
does not affect the mass estimates of clusters of galaxies but does
contribute to the uncertainty of the measurement. The work proposed by
\cite{hoekstra01b,hoekstra03} considered a massive cluster
($M_{200}\geq5\times10^{14}$ h$^{-1}$ M$_\odot$) at $z_d=0.3$ plus a
power spectrum of the density fluctuations. Our sample can not be
treated in the same way, because the field is populated by several
lenses.  Following \cite{hoekstra01b,hoekstra03}, we split the
observed shear in components but also subdivide the shear due to LSS
into two components, in a way that
\begin{eqnarray}
  \label{eq:g-split}
 \gamma_{\rm{t}}(\theta)&=&\gamma_{\rm t}^{\rm halo}(\theta)+\gamma_{\rm  t}^{\rm  LSS}(\theta) \nonumber \\
  &=&\gamma_{\rm{t}}^{\rm{halo}}(\theta)+\gamma_{\rm{t}}^{\rm{close-haloes}}(\theta)+\gamma_{\rm{t}}^{\rm{distant-haloes}}(\theta)
\end{eqnarray}
where the component $\gamma_{\rm{t}}^{\rm{close-haloes}}$ includes the
shear introduced by all haloes with a maximum distance of $5^\prime$
from the centre of the main galaxy group. The shear signal introduced
by the other haloes in the field is taken into account by the
$\gamma_{\rm{t}}^{\rm{distant-haloes}}$ term.  The motivation for
choosing $5^\prime$ as the dividing line between {\it close-haloes}
and {\it distant-haloes} is: (1) the optimal aperture value for
detections of the individual haloes using an NFW filter is
$\theta\sim5^\prime$ (see Fig. \ref{optimal_aper} ) and; (2) the
nearest halo separation peaks at $\theta\sim2.5^\prime$ dropping
almost to zero at $\theta\sim5^\prime$ (see
Fig. \ref{xraydistribution}). Using a dividing line of $5^\prime$
implies that all constellations have at least one extra halo in
addition to the main one. Consequently the {\it close-haloes} term can
be interpreted as a second-halo term seen in projection.

A good approximation for the dispersion in the averaged
tangential shear within a measured radius $\theta$ is
\beq
 \label{eq:g-split-err}
    {\sigma_{\gamma_{\rm t}}^{\rm obs}}^2(\theta)\sim \frac{{\sigma_{\evec^s}}^2}{2N}(\theta) + 
{\sigma_{\gamma_{\rm t}}^{\rm{close-haloes}}}^2(\theta) + {\sigma_{\gamma_{\rm t}}^{\rm{distant-haloes}}}^2(\theta) 
\eeq
since the correlation between $\gamma_{\rm{t}}^{\rm{close-haloes}}$
and $\gamma_{\rm{t}}^{\rm{distant-haloes}}$ is small, and the
tangential shear dispersion introduced by the term $\gamma_{\rm
  t}^{\rm halo}(\theta)$ \, is due to intrinsic ellipticity only.  In
this equation, $N$ is the number of galaxies for which the tangential
shear is measured.

Next, we investigate how the tangential shear of the main halo is
affected by the presence of the other galaxy groups in the field. For
this analysis we use the CFHT shear simulations due to larger sky
coverage than the Subaru simulations. Using the isolated-pure-shear
catalogues of each group, we compute the tangential shear within an
aperture for the three terms of equation (\ref{eq:g-split}). In order
to quantify $\gamma_{\rm{t}}^{\rm{close-haloes}}$ and
$\gamma_{\rm{t}}^{\rm{distant-haloes}}$ for each main halo, we first
identify the galaxy groups matching the {\it close-haloes} and {\it
  distant-haloes} criteria. Then, the total shear of the $j$-th galaxy
is calculated by summing the shear over all the groups classified as
close and distant separately. This procedure is similar to what we did
to generate the multiple-lens-pure-shear catalogue using equation
(\ref{eq:total-shear}), but now the number of {\it close-haloes} and
{\it distant-haloes} is different for each galaxy group.

\begin{figure}
\centering
\subfigure{\label{fig:edge-a}\includegraphics[scale=0.40]{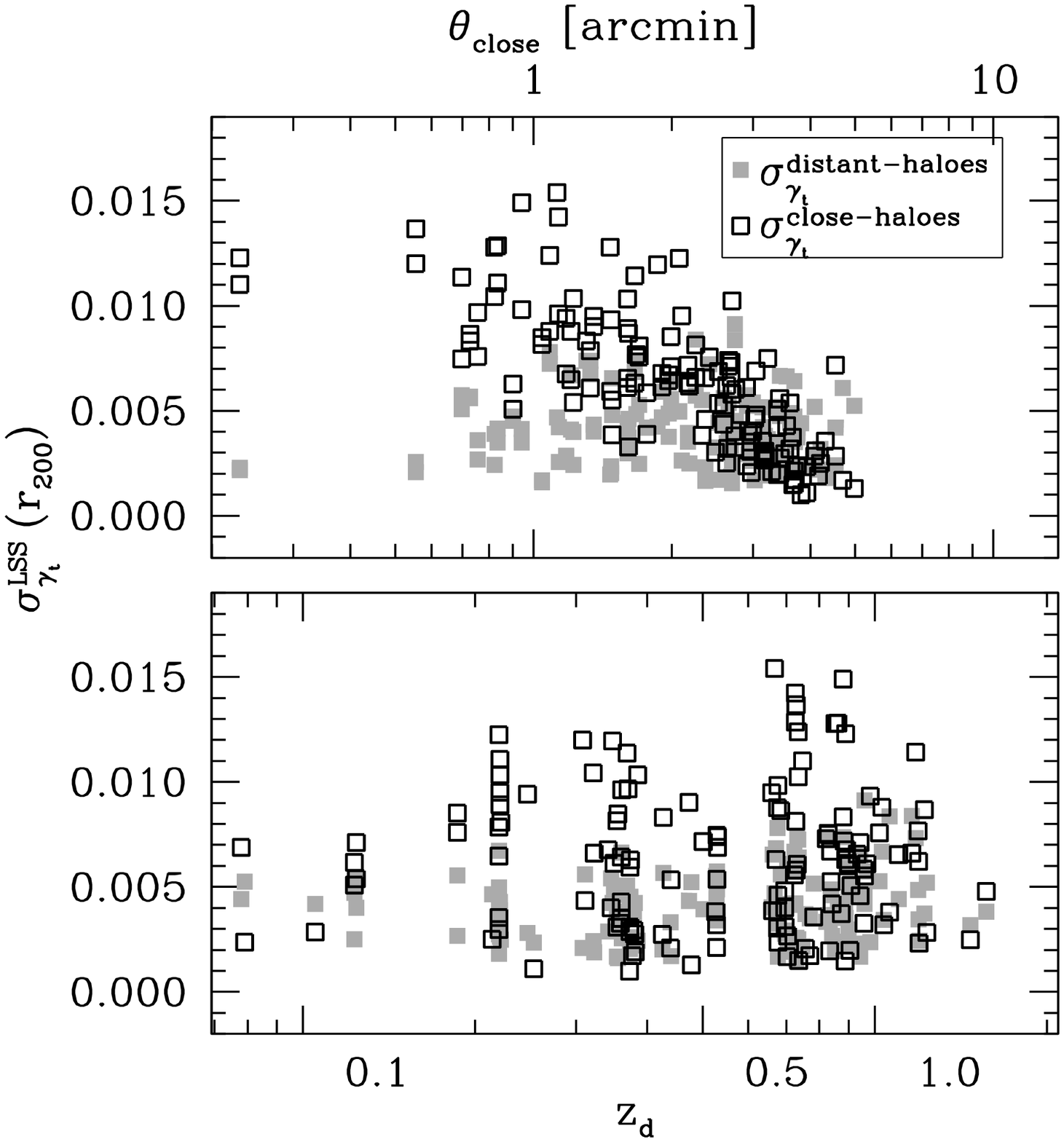}}
\subfigure{\label{fig:edge-b}\includegraphics[scale=0.40]{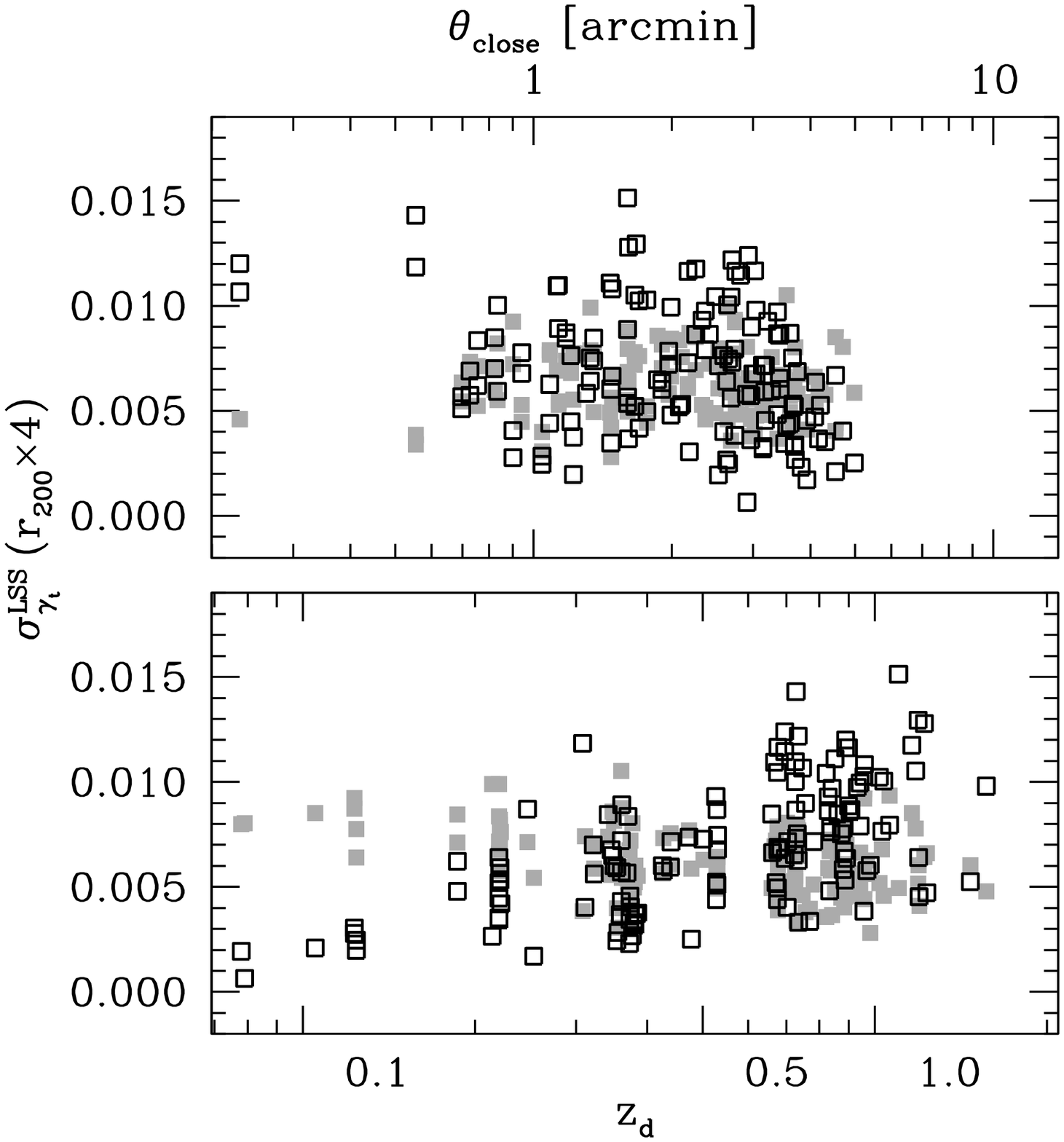}} \\
\caption{Dispersion in the averaged tangential shear measured within
  two radii: $R=r_{200}$ (upper) and $r_{200}\times4$ (bottom). The
  dispersion is shown as a function of redshift the of main lens $z_d$
  and as a function of distance to the nearest lens $\theta_{\rm
    close}$. The tangential shear dispersion introduced by haloes that
  are close in projection to the main galaxy group is shown by the
  black-open squares, whereas the tangential shear dispersion
  introduced by distant haloes is shown by the grey-filled
  squares. See the text for details. \label{gamma_std}}
\end{figure}

The tangential shear dispersion is measured for two aperture sizes:
$R=r_{200}$ of the main galaxy group and $R=r_{200}\times4$, which is
equivalent $\sim5^{\prime}$ or $\sim2$ Mpc for COSMOS galaxy
groups. We have only used the groups for which the measured radii are
fully inside the data field, totalling 137 groups. Fig. \ref{gamma_std}
shows the terms ${\sigma_{\gamma_{\rm t}}^{\rm close-haloes}}(R)$ and
${\sigma_{\gamma_{\rm t}}^{\rm distant-haloes}}(R)$ as a function of
the redshift of the main galaxy group $z_d$ and as a function of the
projected distance to closest neighbour $\theta_{\rm close}$. On the
top panel, the measurements are performed for $R=r_{200}$ and on the
bottom panel for $R=r_{200}\times4$.

From Fig. \ref{gamma_std}, for the measurements within
$r_{200}$ we conclude that: (1) the shear dispersion of the {\it
  close-haloes} term is a steep function of the closest halo proximity;
(2) the shear dispersion of the {\it distant-haloes} term is smaller
than the {\it close-haloes} term. The mean values of the tangential shear
dispersion are: $\sigma_{\gamma_{\rm t}}^{\rm close-haloes}(r_{200})
\sim 0.006$ and $ \sigma_{\gamma_{\rm t}}^{\rm distant-haloes}(r_{200})
\sim0.004$. For measurements within $r_{200}\times4$ we conclude that:
(1) the contribution of {\it close-haloes} and {\it distant-haloes} are
of the same order of magnitude, meaning that they can be treated
together as a single source of external noise. The mean values of the
tangential shear dispersion are: $\sigma_{\gamma_{\rm t}}^{\rm
  close-haloes}(r_{200}\times4)\sim0.007$ and $\sigma_{\gamma_{\rm
    t}}^{\rm distant-haloes}(r_{200}\times4)\sim0.006$.

On average $\sigma_{\gamma_{\rm t}}^{\rm LSS}\sim0.006$ per
component. This value corresponds to $\sim1.8\%$ of the intrinsic
ellipticity value of one component, and is consistent with the values
found in \citet{hoekstra10}: ${\sigma_{\gamma_{\rm t}}^{\rm
    LSS}}=0.0060-0.0045$ for $\theta<5^{\prime}$. We briefly
investigate how $\sigma_{\gamma_{\rm t}}^{\rm LSS}$ varies with the
aperture size. We measure the azimuthally averaged tangential shear as
a function of the distance from 100 random positions spread over the
field. The dispersion of the azimuthally averaged tangential shear is
measured within several apertures and annuli, with a step-size equals
$0.5^{\prime}$.  We find that our $\sigma_{\gamma_{\rm t}}^{\rm LSS}$
estimate is a factor of two higher for large aperture sizes
($\theta=5-15^{\prime}$) than it is in comparison to
\citet{hoekstra10} results and to \cite{gruen11} results. This can be
explained by the overdense region that the COSMOS field lies, which
causes a higher cosmic shear signal \citep[e.g.][which found higher
clustering amplitudes in the COSMOS field than for other sky
patches]{mccracken07, meneux09, kovac10}.

\begin{figure}
\center
\includegraphics[scale=.40]{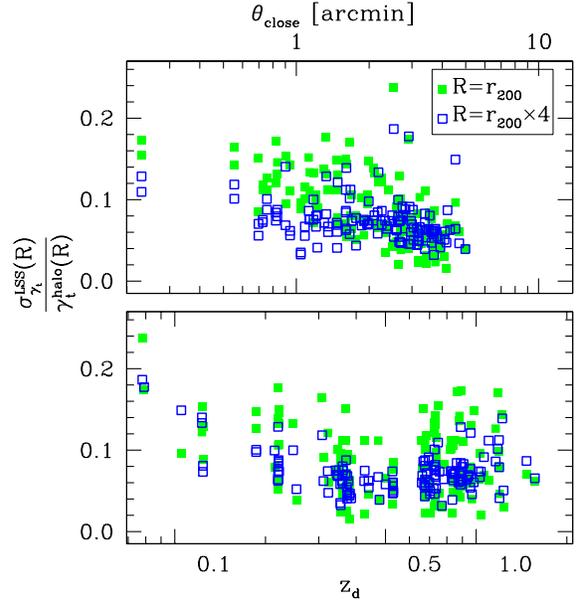}
\caption{The ratio of $\sigma_{\gamma_{\rm t}}^{\rm LSS}$ and
  $\gamma_{\rm t}^{\rm halo}$ measured for two aperture sizes:
  $R=r_{200}$ (green-filled squares) and $R=r_{200}\times4$ (blue-open
  squares). The ratio is shown as a function of the redshift of the
  main lens $z_d$ and as a function of the distance to the nearest
  lens $\theta_{\rm close}$.\label{noise_frac}}
\end{figure}

Fig. \ref{noise_frac} shows the ratio $\sigma_{\gamma_{\rm t}}^{\rm
  LSS}/\gamma_{\rm t}^{\rm halo}$ as a function of the redshift of the
main lens $z_d$ and as a function of the distance to the nearest lens
$\theta_{\rm close}$. The ratio is also measured using two aperture
sizes: $R=r_{200}$ and $R=r_{200}\times4$. On average, the LSS
contamination represents $8.8\% \pm 4.2\%$ and $7.3\% \pm 2.6\% $ of
the shear signal of the selected galaxy groups when they are
considered isolated in the sky, for aperture values equal to $r_{200}$
and $r_{200}\times4$, respectively. However, the percentage of the LSS
contamination depends on the redshift of the main lens, the aperture
size used to measure the shear signal and whether there are close
neighbours.

Since $\sigma_{\gamma_{\rm t}}^{\rm LSS}\sim0.006$ and
$\sigma_{e_i^s}\sim0.33$ for our data, it is possible to calculate the
number of galaxies N for which the LSS and intrinsic ellipticity
noises reach the same order of importance. From equation
(\ref{eq:g-split-err}) we can infer
\beq
   \frac{{\sigma_{e_i^s}}^2}{N}={\sigma_{\gamma_{\rm t}}^{\rm LSS}}^2
\eeq
yielding $N\sim3000$. This corresponds to a density of $\sim26$
galaxies arcmin$^{-2}$ if an aperture of $6^{\prime}$ is
considered. Therefore, for a deep survey like COSMOS, it is already
possible to achieve the density of galaxies for which the LSS noise
error becomes equal to the intrinsic ellipticity noise.

\subsection{Density profiles}\label{resul:mdp}

In this section we present an analysis of the density
contrast profiles of the galaxy groups of our sample.

As discussed in the Section \ref{resul:det}, the detection of low mass
systems via weak lensing is limited to shape noise contamination. One
way to overcome this problem is by averaging the shear signal of
several galaxy groups with similar properties. The density contrast
$\Delta\Sigma(R)$ \citep{me91} is an estimator often used to stack the
shear profile of haloes. It is defined as
\beq \label{eq:dens_cont}
   \Delta\Sigma(R) \equiv \bar\Sigma(<R)-\langle \Sigma(R) \rangle =\gamma_{\rm{t}}(R)\times\Sigma_{cr}\,,
\eeq
where $\bar\Sigma(<R)$ is the mean surface density interior a radius
$R$ and $\langle \Sigma(R) \rangle$ is the azimuthal average of
$\Sigma(R)$ at radius $R$. Since the tangential shear is multiplied by
$\Sigma_{cr}$, the density contrast $\Delta\Sigma$ is a redshift
independent quantity. The density contrast is related to the mass $M$
of the halo via
\beq
   M(R)=\int_0^{R}\Sigma(r)dr^2 \sim \bar\Sigma(<R) \times \pi R^2\,,
\eeq
therefore the $M_{200}$ mass is given by
\beq
   M_{200} = \pi r_{200}^2 (\Delta\Sigma(r_{200}) + \langle \Sigma(r_{200})\rangle)\,.
\eeq

The stacking technique has been adopted in the literature few times:
\citet{hoekstra01a} used CNOC2 data and made use of the shear signal
of an ensemble of 50 groups at $z_d=0.12-0.55$ and velocity dispersion
ranging from $\sigma_v=50-400$ km s$^{-1}$. The averaged velocity
dispersion obtained from the stacked profiles was
$\sigma_v=274^{+48}_{-59}$ km s$^{-1}$. \citet{parker05} adopted the
same technique as \citet{hoekstra01a}, but using a sample of 116 CNOC2
groups with median redshift of $z_d=0.33$. Both works have stacked the
tangential shear profile of groups, which is not a redshift
independent quantity. A remarkable achievement was presented by
\citet{johnston07} who did use the density contrast profile of 130,000
SLOAN systems between $z_d=0.1-0.3$. The systems were divided in 12
bins of optical richness and 16 bins of {\it i} band luminosity. Then,
an averaged density contrast was obtained for each bin. For the first
time, the stacking technique of an ensemble of systems at higher
redshifts was presented by \citet{leau10}.  This study consisted of
the analysis of 127 galaxy groups with $z_d\le1$, which were also
selected from the COSMOS X-ray catalogue. The galaxy groups were split
in nine bins of redshift and X-ray luminosity. The obtained density
contrast of each bin was used to estimate $M_{200}$, which was
eventually used to derive a \lx--$M_{200}$ relation.

In this section, we also make use of the stacking technique to analyse
the density contrast profiles of the galaxy groups in our sample. One
disadvantage of this method is that, in order to constrain physical
parameters of the systems investigated, it is necessary to average the
density contrast of galaxy groups with similar properties. Hence, such
properties should be known a priory (e.g. mass, richness,
luminosity). We stack the lensing signal of the galaxy groups in our
sample using the same binning system as proposed by
\cite{leau10}. Table \ref{tab:binning} shows the properties of the
seven bins of redshift and X-ray luminosity used.  Moreover, as it was
done in Section \ref{resul:tsd}, we also split the density contrast
into the contribution originating from the main galaxy group to the
contribution originating from the LSS ({\it close-} and {\it
  distant-haloes} terms). If the latter is not zero, then the mass
estimates from the density contrast profiles are not
reliable. Therefore, we study the density contrast profiles of the
individual groups as well as the averaged density contrast profiles
obtained from the ensembles of groups. We check how the contribution
originating from the LSS affects the density contrast of the groups if
they were isolated in the sky and the averaged density contrast
obtained from the ensembles.

\begin{table*}
  \centering
  \begin{minipage}{140mm}
  \caption{Average properties of binning system}
  \begin{tabular}{cccccccc}
    \hline
    \hline
    Bin$^{\rm a}$ & $N_{\rm haloes}$ & z & $M_{200}$ &
    $r_{200}^{\rm b}$ & $L_{{\rm X} (0.1-2.4 {\rm keV})} E(z)^{-1}$ &
    $\theta_{\rm close}^{\rm c}$ & Scale$^{\rm d}$ \\
    & & & [$10^{13}$ M$_{\odot}$] & [arcmin] & [$10^{42}$ erg s$^{-1}$]
    &[arcmin] & [$10^{-3}$ kpc arcmin$^{-1}$] \\
    \hline
    A2 & 2  & 0.35 & 4.9 & 2.3 & 5.5 & 1.9 & 3.433\\
    A3 & 8  & 0.36 & 2.5 & 1.8 & 2.0 & 2.7 & 3.370\\
    A4 & 12 & 0.22 & 1.6 & 2.3 & 0.9 & 2.2 & 4.792\\
    A5 & 15 & 0.36 & 1.7 & 1.6 & 1.1 & 2.2 & 3.370\\
    A6 & 9  & 0.50 & 3.3 & 1.5 & 3.4 & 2.9 & 2.767\\
    A7 & 24 & 0.70 & 3.4 & 1.2 & 4.1 & 2.2 & 2.352\\
    A8 & 20 & 0.86 & 4.4 & 1.2 & 7.8 & 2.4 & 2.179\\
    \hline
  \end{tabular}
  \begin{flushleft}
    \scriptsize
    $^{\rm a}$ Naming convention as used in \cite{leau10}.  
    Bins named as A0 and A1 had no elements and were excluded
    from the table. \\
    $^{\rm b}$ Calculated using the averaged mass $M_{200}$ and
    the adopted cosmology at the averaged redshift z. \\
    $^{\rm c}$ Calculated by averaging out the value $\theta_{\rm
      close}$ of each group in a bin. \\
    $^{\rm d}$ Calculated using the averaged redshift and the
    adopted cosmology.
  \end{flushleft}
  \label{tab:binning}
\end{minipage}
\end{table*}

From equations (\ref{eq:g-split}) and (\ref{eq:dens_cont}), we can
infer that the density contrast written in terms of the LSS components
is given by
\bea
  \label{eq:dc-split}
  \Delta\Sigma^{\rm obs}(R) & = &
  \Delta\Sigma^{\rm{halo}}(R)+\Delta\Sigma^{\rm{close-haloes}}(R)+ \\ \nonumber
  & & \Delta\Sigma^{\rm{distant-ha
    loes}}(R)
\eea 
where the dividing line between {\it close-haloes} and {\it
  distant-haloes} is kept the same as in Section \ref{resul:tsd}.

For this analysis we have also used the CFHT shear simulations due to
larger sky coverage than the Subaru simulations. Using the
isolated-pure-shear catalogues of each group, we compute the density
contrast within an aperture for the three terms of the equation
(\ref{eq:dc-split}). In order to calculate the density contrast terms
due to external haloes, we proceed in a similar way to what we did in
Section \ref{resul:tsd}: (1) we identify the galaxy groups matching
the {\it close-haloes} and {\it distant-haloes} criteria; (2) we
compute the total shear of the $j$-th galaxy by summing (the shear)
over all the groups classified as close and distant separately; (3) we
calculate the tangential shear of each galaxy for the {\it
  close-haloes} and {\it distant-haloes} terms and; (4) we calculate
the density contrast of each term using equation (\ref{eq:dens_cont}).

The stacked density contrast profiles of each bin is calculated by
averaging the density contrast profiles of all galaxy groups belonging
to the bin. This is done for each term of equation (\ref{eq:dc-split})
separately.

\begin{figure}
\centering
\subfigure{\includegraphics[scale=0.40]{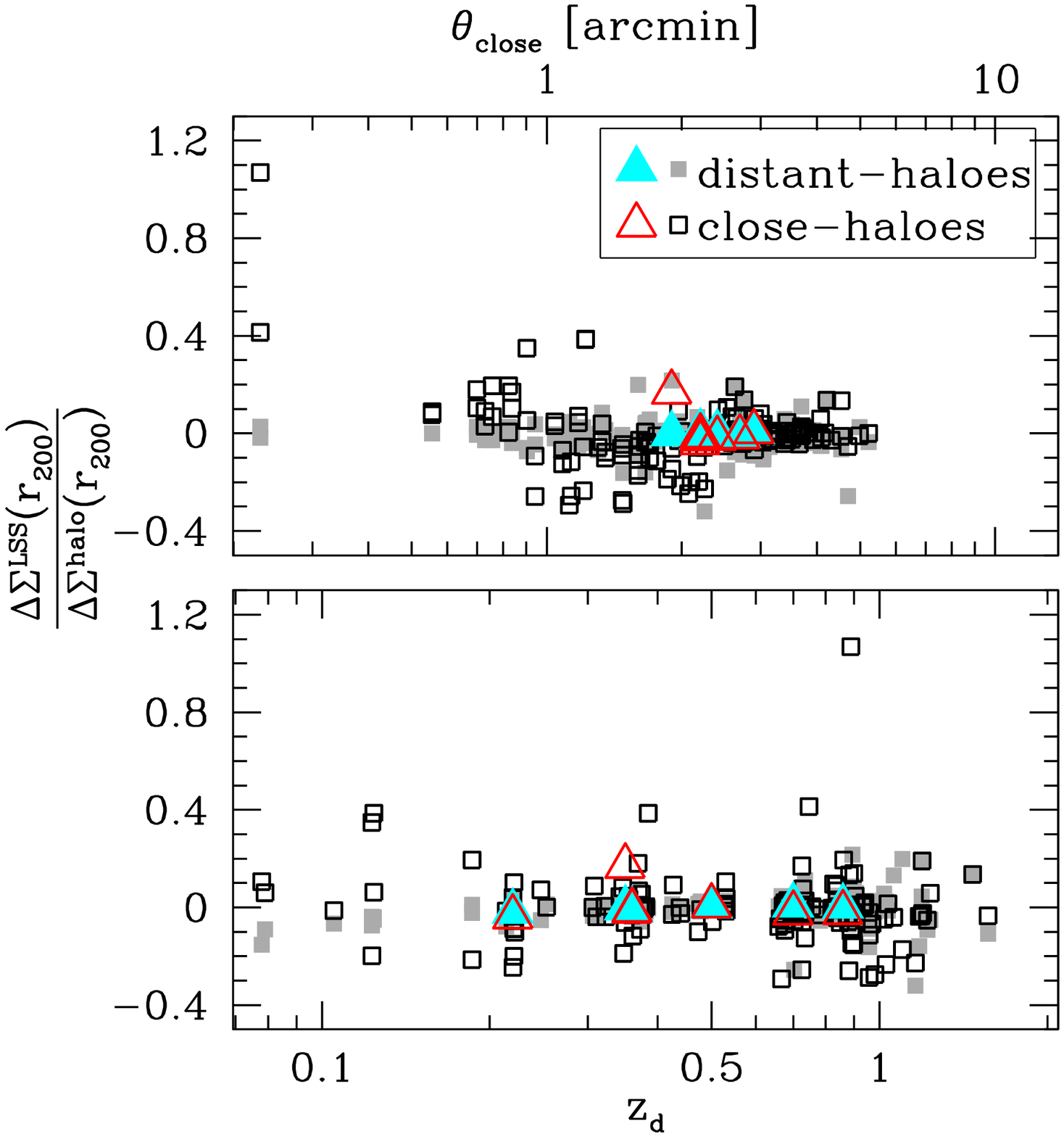}} \\
\subfigure{\includegraphics[scale=0.40]{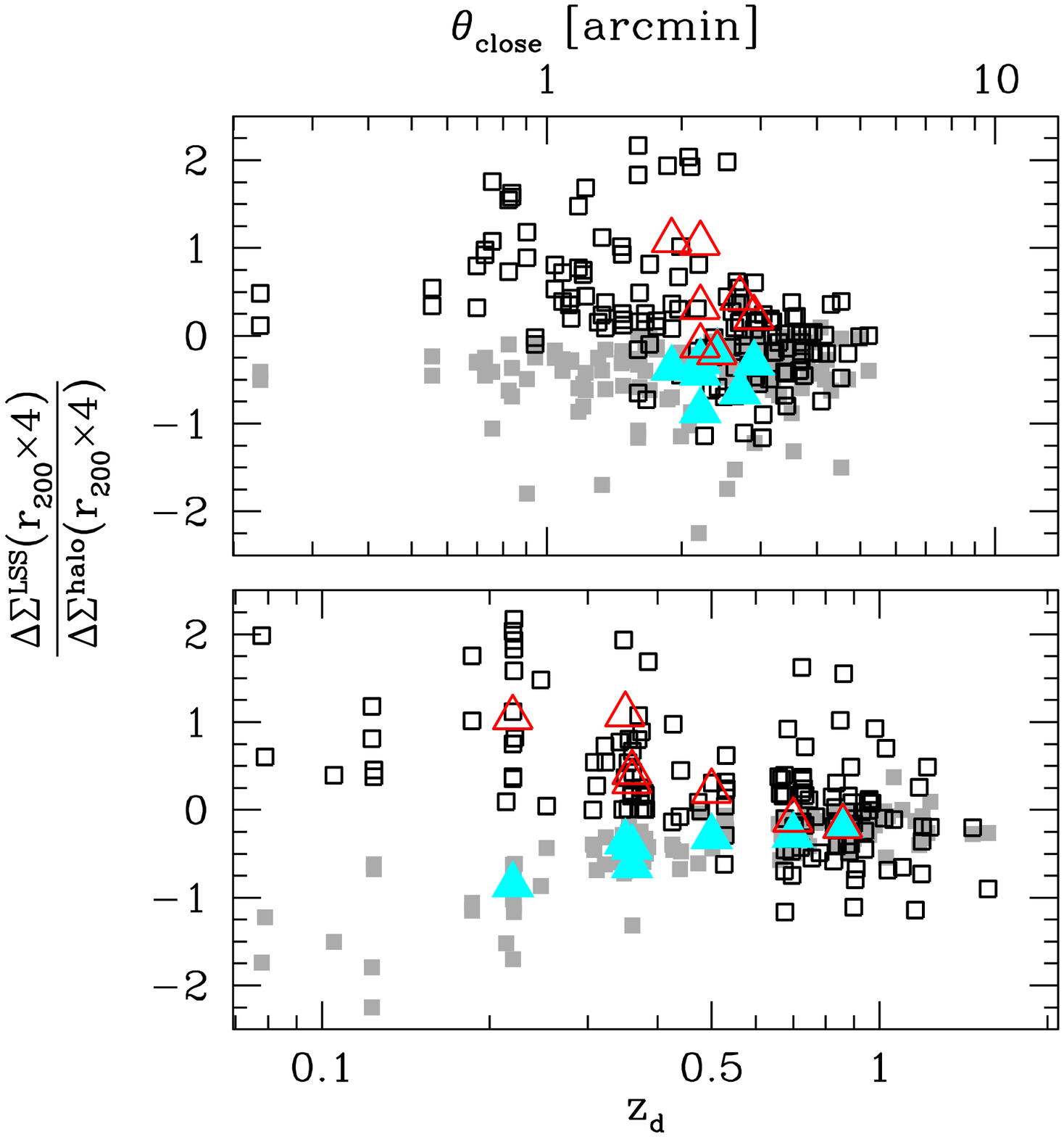}}
\caption{Contribution of the surrounding COSMOS haloes to the
  estimated integrated density contrast of individual groups. The LSS
  terms of the density contrast is measured within $R=r_{200}$ (top)
  and $R=r_{200}\times4$ (bottom) and divided by the integrated
  density contrast of the main group within same radii.  In each plot
  the lowest panel show the ratio as a function of the redshift of the
  main galaxy group and the uppermost panel as a function of the
  projected distance to the closest neighbour. The ratio
  $\Delta\Sigma^{\rm{close-haloes}}(R)/\Delta\Sigma^{\rm{halo}}(R)$ is
  shown by the black-open squares, whereas the ratio
  $\Delta\Sigma^{\rm{distant-haloes}}(R)/\Delta\Sigma^{\rm{halo}}(R)$
  is shown by the grey-filled squares.  Triangles show the average
  values obtained by stacking the density contrast of several groups
  that are binned according to Table \ref{tab:binning}. Red-open
  triangles show the ratio for {\it close-haloes} and cyan-filled
  triangles show the ratio for {\it distant-haloes} of the averaged
  profiles. \label{deltasigma_frac}}
\end{figure}

Next, the density contrast is measured for two aperture sizes:
$R=r_{200}$ and $R=r_{200}\times4$. Again, we have only used the
groups for which the measured radii are fully inside the data field,
totalling 137 groups.  Fig. \ref{deltasigma_frac} shows the ratio of
$\Delta\Sigma^{\rm{close-haloes}}(R)$ and
$\Delta\Sigma^{\rm{distant-haloes}}(R)$ over
$\Delta\Sigma^{\rm{halo}}(R)$ as a function of the redshift of the
main halo $z_d$ and as a function of the projected distance to closest
neighbour $\theta_{\rm close}$. The ratio can be either positive or
negative. This happens because the shear field is perturbed by the
extra lenses along the line-of-sight and, depending on the
configuration of the lenses, the additional tangential shear can
become negative or positive. The consequence of a negative value for
the tangential shear is an underestimation of the parameters obtained
from this quantity.

Fig. \ref{deltasigma_frac} shows that, when measured within
$r_{200}$, the contamination of the {\it close-haloes} term scatters
around zero, with an rms equals $0.15$. However, there are several
cases for which the contamination is of the order of 40\% and for one
case it reaches 100\%. For the latter, the main galaxy group is
located at high redshift and has a close galaxy group in projection
($\theta_{\rm close}<1^{\prime}$). On the other hand, the {\it
  distant-haloes} term does not affect the density contrast estimate
of the main galaxy group, with a mean value equal to zero and an rms
value of $0.06$. This latter result is in agreement with
\cite{hoekstra01b,hoekstra03} findings. When the density contrast
profile of several groups is stacked, the contamination of the {\it
  close-haloes} term cancels out and the rms of the ratio drops to
$0.07$. This value is expected to drop even more if the stacking was
performed over a larger number of galaxies groups within each bin. The
rms of the {\it distant-haloes} term is again consistent with zero
when the stacking approach is considered. The measurements using a
larger radius ($r_{200}\times4$) show more scatter than within
$r_{200}$. For both {\it close-haloes} and {\it distant-haloes} terms
the mean values of the density contrast ratio are not zero and the rms
value increases in comparison to measurements within $r_{200}$. The
stacking technique does not help to decrease the rms value either. As
it happened to the tangential shear dispersion measurements evaluated
within $r_{200}\times4$, the contribution of the terms {\it
  close-haloes} and {\it distant-haloes} are of the same order of
importance. If the two terms are considered together, the mean value
of the ratio $\Delta\Sigma(R)^{\rm LSS}/\Delta\Sigma(R)^{\rm halo}$
drops to zero, but the scatter remains high, around $\sim55\%$.

Tables \ref{meanDeltaSigma} and \ref{rmsDeltaSigma} summarise
the mean and rms values of the ratios $\Delta\Sigma(R)^{\rm
  LSS}/\Delta\Sigma(R)^{\rm halo}$ considering the individual and
stacking measurement scenario.

\begin{table*}
  \centering
 \begin{minipage}{100mm}
  \caption{Mean value of the ratio $\Delta\Sigma(R)^{\rm LSS}/\Delta\Sigma(R)^{\rm halo}$}
  \begin{tabular}{ccccc}
    \hline
    R          & \multicolumn{2}{c}{Individual halo measurement} & \multicolumn{2}{c}{Stacked halo measurement} \\
    \cline{2-5}
    & close-haloes & distant-haloes & close-haloes  & distant-haloes\\
    \hline         
    $r_{200}$         & 0.00  &  0.01   & 0.01   &  0.00   \\
    $r_{200}\times4$  & 0.25  & -0.43   & 0.40   & -0.44   \\
    \hline
  \end{tabular}
  \label{meanDeltaSigma}
\end{minipage}
\end{table*}

\begin{table*}
  \centering
  \begin{minipage}{100mm}
  \caption{RMS of the ratio $\Delta\Sigma(R)^{\rm LSS}/\Delta\Sigma(R)^{\rm halo}$}
  \begin{tabular}{ccccc}
    \hline
    R          & \multicolumn{2}{c}{Individual halo measurement} & \multicolumn{2}{c}{Stacked halo measurement} \\
    \cline{2-5}
    & close-haloes & distant-haloes & close-haloes  & distant-haloes\\
    \hline
    $r_{200}$         & 0.15  & 0.06    & 0.07   & 0.01  \\
    $r_{200}\times4$  & 0.72  & 0.58    & 0.62   & 0.49  \\
    \hline
  \end{tabular}
  \label{rmsDeltaSigma}
\end{minipage}
\end{table*}

\subsection{High redshift groups}\label{resul:highz}

Fig. \ref{deltasigma_frac} demonstrates that the density contrast
estimate can be biased by $\sim100\%$ if the main lens is located at
high redshift (high-z) and has another halo along the line-of-sight
very close in projected distance ($\theta_{\rm close}<1^{\prime}$). In
this section we briefly investigate the probability of finding such a
configuration, considering that the COSMOS survey provides a
representative distribution of haloes in the sky.

\begin{figure}
\center
\includegraphics[scale=.40]{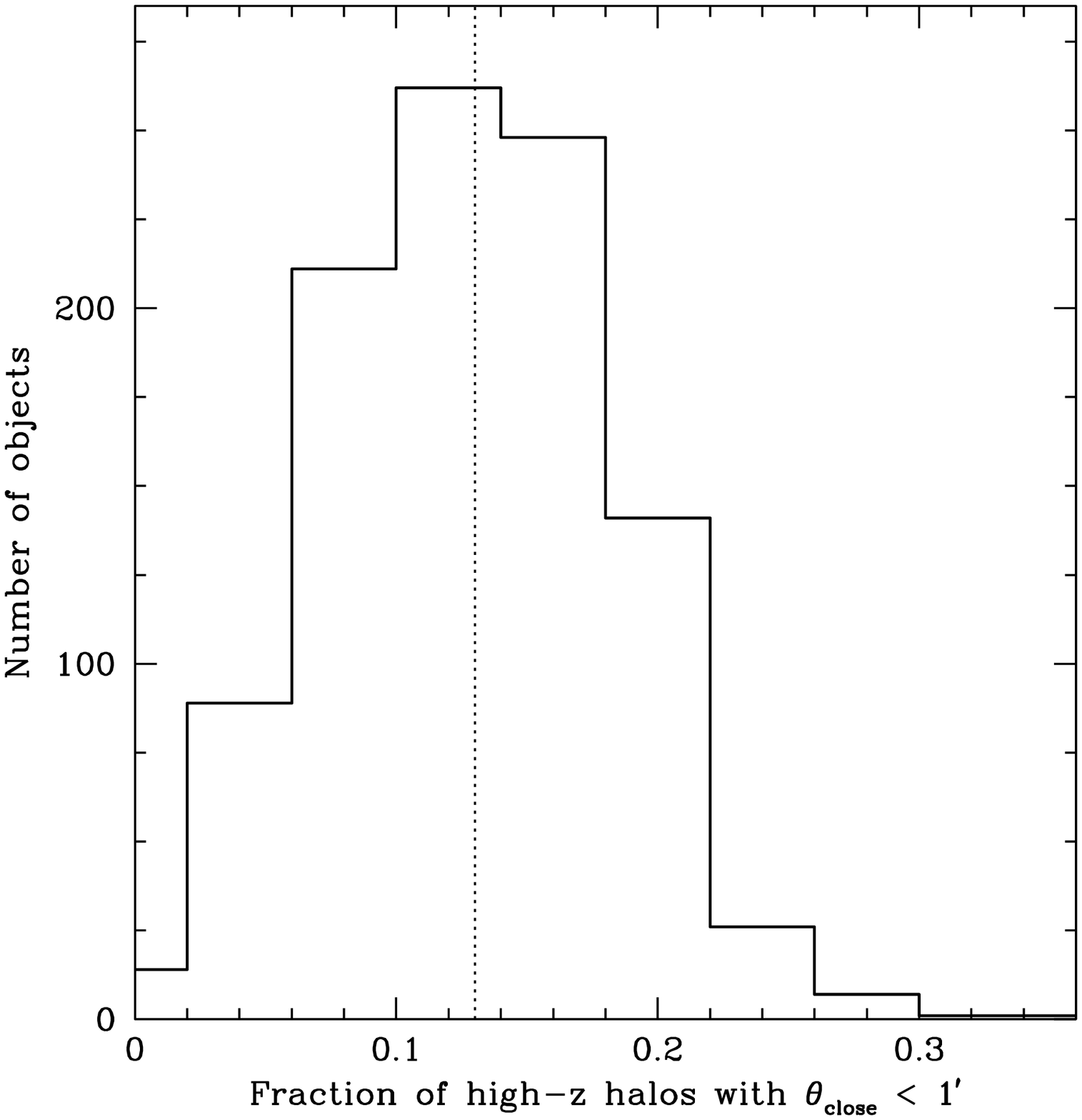}
\caption{Distribution of the fraction of groups at $z_{d}\geq0.8$ that
  have a neighbour within a distance $\theta_{\rm close}<1^{\prime}$
  over the total number of high-z groups. The distribution is drawn
  from 1000 realisations of random positions spread over the COSMOS
  field. From a total of 54 groups at $z_{d}\geq0.8$ there is mean
  probability that $13\%\pm5\%$ of the groups have a halo along the
  line-of-sight within $1^{\prime}$.\label{higz_dist}}
\end{figure}

We define high-z groups as the ones with $z_d\geq 0.8$, totalling 54
groups. In order to investigate the frequency of the high-z groups
with close companions, we generate 1000 realisations of random
positions for the groups in our sample. The groups are distributed
within the same area as they are observed. For each realisation and
galaxy group, we calculate the projected distance of the nearest
neighbour. Next, we evaluate the percentage of high-z groups with a
companion within $1^{\prime}$. We note that, the total number of
high-z groups is kept fixed to all realisations, since the redshift
distribution of the groups is not changed. Depending on the
realisation, the percentage of high-z groups with neighbours within
$\theta_{\rm close}<1^{\prime}$ varies from 0 to $30\%$. On average,
$13\%$ of high-z groups have a another halo along the line-of-sight
that is closer than 1$^{\prime}$. Fig. \ref{higz_dist} shows the
distribution of this fraction for the 1000 random realisations of
positions.

\begin{figure}
\center
\includegraphics[scale=.40]{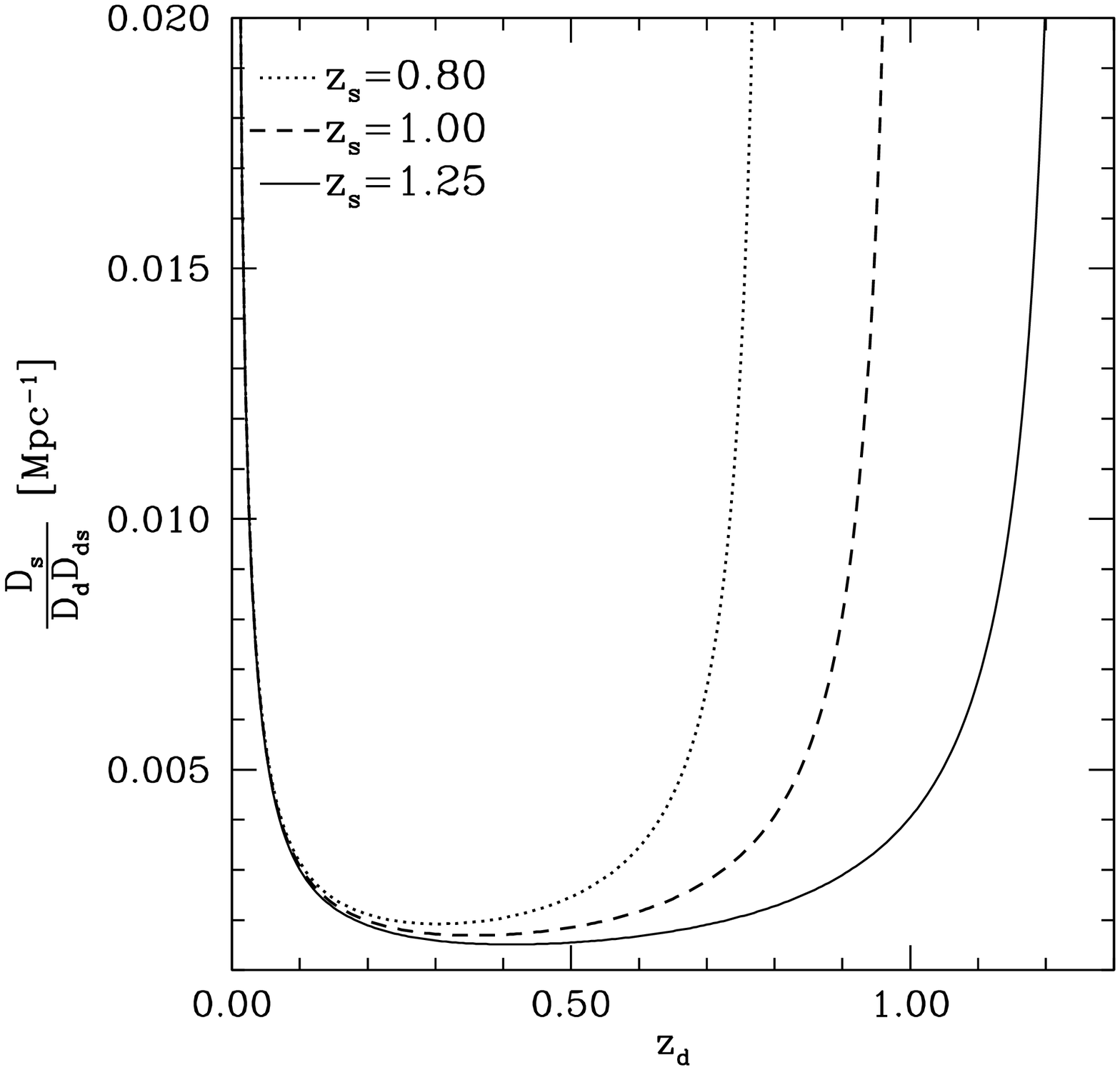}
\caption{The factor $D_s/D_dD_{ds}$ as a function of the redshift of
  the lens for three different background populations, fixed at
  redshifts: $z_s=0.8$ (dotted line), $z_s=1.0$ (dashed line) and
  $z_s=1.25$ (solid line). The ratio $\Sigma_{cr}^{\rm high-z}/
  \Sigma_{cr}^{\rm fg}$ is equal to 2.3, 2.4, and 2.6, for a
  foreground lens at $z_d^{\rm fg}=0.35$ and the respective
  configurations: (1) source population at $z_s=0.8$ and high-z lens
  at $z_d^{\rm high-z}=0.65$; (2) source population at $z_s=1.0$ and
  high-z lens at $z_d^{\rm high-z}=0.80$; and (3) source population at
  $z_s=1.25$ and high-z lens at $z_d^{\rm
    high-z}=1.00$. \label{higz_dd_dsdds}}
\end{figure}

In order to understand why high-z groups have their shear signal
contaminated by foreground masses more significantly than haloes at
intermediate redshifts, we recall the definition of the density
contrast. Considering that the total shear observed is the sum of the
shear introduced by the high-z halo $\gamma_{\rm t}^{\rm high-z}$ plus
the shear introduced by a foreground halo $\gamma_{\rm t}^{\rm fg}$,
we find the excess density estimate equals

\begin{eqnarray}
\label{eq:dc-hz}
   \Delta \Sigma &=& (\gamma_{\rm t}^{\rm high-z} + \gamma_{\rm t}^{\rm fg})\times \Sigma_{cr}^{\rm high-z} \nonumber \\
   &=& \Delta \Sigma^{\rm high-z} + \Delta \Sigma^{\rm fg} \left( \frac{\Sigma_{cr}^{\rm high-z}}{\Sigma_{cr}^{\rm fg}} \right)\,.
\end{eqnarray}
The quantity $\Delta \Sigma^{\rm fg}$ in the right-hand side of
equation (\ref{eq:dc-hz}) is multiplied by the ratio between the
critical density of the high-z halo and the critical density of
foreground halo. For most of the cases this ratio is greater than one
and hence the foreground halo contributes in a boosted way to the
total $\Delta \Sigma$ budget. Fig. \ref{higz_dd_dsdds} helps to
understand this: since the critical density is $\propto D_s/D_dD_{ds}$
we can analyse this factor as a function of the redshift of the halo
$z_d$ for fixed source population at redshift $z_s$. In
Fig. \ref{higz_dd_dsdds} we use three different redshifts values for
the background sources: $z_s=0.8$ which represents the median redshift
of a shallow survey, $z_s=1.0$ which is the median redshift of the
galaxies found in this work and $z_s=1.25$ which mimics the median
redshift of background sources of high-z groups. The figure
demonstrates that, for the three different source populations, haloes
at higher-z always have the factor $D_s/D_dD_{ds}$ higher than the
haloes at intermediate redshifts, meaning that the ratio
$\Sigma_{cr}^{\rm high-z}/ \Sigma_{cr}^{\rm fg}>1$. The same happens
to low-z haloes ($z_d<0.20$) as already previously noticed by
\citet{hoekstra01b}.

\section{Summary and Conclusions}\label{conc}

We have derived shear and photo-z catalogues using CFHT and Subaru
observations of the COSMOS field. The combined shear-photo-z
catalogues result in a density of $29.7$ and $21.7$ galaxies
arcmin$^{-2}$ respectively. The two-component intrinsic ellipticity
dispersion found is $\sigma_{\evec^{\rm s}}=0.47$ and
$\sigma_{\evec^{\rm s}}=0.42$ for CFHT and Subaru. The final
shear-photo-z catalogues plus the information on the X-ray luminous
groups of the COSMOS field (Finoguenov et al. in preparation) served
as input to compute the shear field assuming that haloes are described
by an NFW density profile. Based on this, the distortion on the shape
of the source galaxies due to each lens was calculated. Calculations
taking into account the contribution of all lenses in the field were
also computed. An intrinsic ellipticity distribution was randomly
generated according to the observations and attributed to the source
galaxies. Thus, a comparison between the shear signal of individual
isolated groups and the observed shear signal which is originating
from all galaxy groups embedded in the field was established.

The two different data sets provide information that can be used to
forecast results for future surveys, with a deeper or shallower
strategy. The main conclusions of this work are:

\begin{itemize}

\item With both a CFHT and Subaru-like configuration, COSMOS-like
  groups can not be detected using the $M_{ap}$ statistic approach,
  unless the intrinsic ellipticity acts cooperatively or a high-false
  detection rate is accepted.

\item Positive and negative E and B-modes with $|{\rm S/N}|\geq3$ are
  likely to happen by accident for about $\sim200$ positions out of
  66820 investigated. Hence only ${\rm S/N}>4$ peaks, which happen
  with a probability $<0.01\%$, can be considered as safe.
  
\item The filtering technique using optical plus weak lensing methods
  proposed in a recent paper by \cite{bellagamba11} is able to detect
  $\sim 7\%$ of total haloes with almost no spurious detection if the
  threshold for an optical plus weak lensing detection is
  ${\rm S/N}\geq4$. For this technique, lower values of S/N increase the
  number of spurious detections as to $\sim75\%$.

\item If the COSMOS field provides a representative picture of the
  full sky, half of the X-ray detected groups have a neighbour (also
  detected in X-rays and with the mass characteristics as shown in
  Fig. \ref{xraydistribution}) within a distance of $\theta_{\rm
    close}<2.5^{\prime}$.

\item In spite of the low masses of COSMOS groups, their presence in
  the field can perturb the signal-to-noise ratio of another halo. The
  rms of the difference in signal-to-noise is $\thickapprox 15\%
  \times \sqrt{n_{gal}/30}$ when an aperture with optimal size for the
  group detection is used. One noticeable case shows a difference of
  $\sim90\%$.

\item The observed density contrast profile, often used as mass
  estimator, can also be affected by the presence of extra objects
  along the line-of-sight. When measured for individual groups within
  $r_{200}$, the average bias introduced by close haloes is zero with
  an rms value of $\sim14\%$. Distant haloes also introduce an average
  bias equals zero but the rms is $\sim5\%$. When the density contrast is
  measured inside a radius four times larger than $r_{200}$, the average
  bias originating from all extra groups is still zero but the scatter
  increases to $55\%$. Stacking the density contrast profile of
  several groups cancels out the biases introduced by close and
  distant haloes, as expected.

\item The shear signal originating from other haloes than the main
  galaxy group introduces an uncertainty in the shear measurements
  that has to be added to the uncertainty from intrinsic
  ellipticity. The average value of the LSS uncertainty obtained from
  COSMOS haloes is $\sigma_{\gamma_{\rm t}}^{\rm LSS}\sim0.006$ per
  component, which corresponds to $\sim1.8\%$ of the one-component
  intrinsic ellipticity value.

\item The LSS and intrinsic ellipticity noise have the same order of
  magnitude if there are shape measurements of $N\sim3000$ galaxies
  within the aperture considered. Deep observations using current
  instruments can already achieve this density of galaxies and,
  therefore, the LSS error should be included in the total error
  budget.

\item The tangential shear dispersion within randomly placed apertures
  of $\theta=5-15^{\prime}$ is about a factor of two higher than the
  value predicted in the works of \cite{hoekstra10} and
  \cite{gruen11}. This can be explained by the overdense line-of-sight
  of the COSMOS field (cosmic variance). On the other hand, we show
  that the structures causing line-of-sight contamination up to
  $z_d=1$ can be detected with deep X-ray observations and modelled
  quantitatively.

\item High-z groups can have their shear signal more contaminated by
  foreground objects than groups at intermediate redshift. The crucial
  configuration is when there is a line-of-sight object within
  $1^{\prime}$ from the centre of the high-z galaxy group. Therefore,
  weak lensing study of low mass systems at high-z requires special
  attention regarding of the biases introduced by the LSS. From
  simulations, we concluded that on average 13\% of groups at
  $z_d\geq0.8$ have this configuration.

\end{itemize}

Our results are based on COSMOS ground-based observations but can be
extended to other fields. The weak lensing study of galaxy groups can
be favoured by the wide-sky coverage of future surveys such as the Dark
Energy Survey (DES), the Large Synoptic Survey Telescope (LSST) and
the Kilo-Degree Survey (KIDS), which will image more than 1,000 square
degrees of the southern sky.

If deep observations and wide-sky coverage are available, then the
study of individual groups is possible, though the contamination by
near haloes in projection has to be taken into account and
modelled. With wide-sky coverage alone, we can extract the mean
properties of ensembles of galaxy groups using the stacking technique
of density contrast profiles, so that the contribution introduced by
the large-scale structure is cancelled out. Nevertheless, the
uncertainty in the shear measurements introduced by the large-scale
structure can not be eliminated and has to be taken into account in
the total error budget.

\section*{Acknowledgements}

The authors want to thank Barnaby Rowe and Martin Kilbinger for the
useful discussions on PSF modelling and shear systematics respectively
and Daniel Gr\"{u}n for the useful comments. P.F.S. also wants to
thank Nuno Gomes for carefully reading this manuscript. M.L. thanks
the European Community for the Marie Curie research training network
``DUEL'' doctoral fellowship MRTN-CT-2006-036133. This research was
supported by SFB-Transregio 33 ``The Dark Universe'' by the Deutsche
Forschungsgemeinschaft (DFG).

Based on observations obtained with \MegaPrime, a joint project
of CFHT and CEA/DAPNIA, at the Canada-France-Hawaii Telescope (CFHT)
which is operated by the National Research Council (NRC) of Canada,
the Institut National des Science de l'Univers of the Centre National
de la Recherche Scientifique (CNRS) of France, and the University of
Hawaii. This work is based in part on data products produced at
TERAPIX and the Canadian Astronomy Data Centre as part of the
Canada-France-Hawaii Telescope Legacy Survey, a collaborative project
of NRC and CNRS.

Based on data collected at Subaru Telescope, which is operated by the
National Astronomical Observatory of Japan.


\appendix

\section{Data Reduction}\label{app:data}
\subsection{CFHT}\label{app:data:cfht}

The data obtained from the Canada-France-Hawaii Telescope (CFHT) used
in this work were collected in the framework of the
Canada-French-Hawaii-Telescope Legacy Survey (CFHTLS), observed with
the \MegaPrime instrument. The Deep Survey is constituted of four
independent patches in the sky. The patch D2 is centred in the COSMOS
field, covering 1 degrees$^2$ and it is used in this work. We perform
the data reduction with the GaBoDS/THELI pipeline described in details
in \citet{erben05,erben09} and \citet{hildebrandt07}. We refer the
reader to these publications for further details. In this section, we
summarise the most important steps of the data reduction.

We retrieve {\it u$^*$, g', r', i'} and {\it z'} bands data from the
CFHT public
archive\footnote{http://cadcwww.dao.nrc.ca/cadcbin/cfht/wdbi.cgi/cfht/quick/form}
and process in a colour basis. The archival data are already
preprocessed, being corrected for bias and flat field. Preprocessing
also includes the removal of instrumental signatures from the raw data
(such as bad and hot pixels) and removal of fringes in the case of
{\it i'} and {\it z'} bands. For each CCD chip, a weight map
containing information on the noise properties is created. The weight
maps are used in the co-addition process, but they are also helpful to
filter out blended and corrupted detections of the source catalogues,
which are used in the astrometric calibration.

The astrometric solution is obtained with SCAMP pipeline
\citep{scamp} using the sixth data release of the Sloan Digital Sky
Survey (SDSS-R6) as a reference catalogue \citep{SDSS-R6}. The
positional accuracy of the {\it i'} band data has an rms of
$0.14^{\prime\prime}$ with respect to the SDSS-R6 catalogue.

Photometric zero-points are derived for each colour, bringing all
individual images to the same flux scale. The images observed under
photometric conditions had the zero-point corrected by the airmass,
instrumental zero point and the colour dependency on extinction
coefficients. 

After the astrometric and photometric calibration, the sky background
is subtracted and the individual exposures are stacked using a
weighted mean combination. The original image pixels are remapped
using SWarp \citep{swarp} adopting a LANCZOS3 kernel. The final
stacked images have the same pixel size as the original images
($0.186^{\prime\prime}$). A weight map image containing information on
the noise properties of the final co-added image and a flag image
carrying the information on the saturated pixels are also
created. These final co-added images as well as their weight and flag
maps are used to generate the photometric catalogues.

We analyse the impact on the PSF homogeneity by co-adding exposures
taken during the three different CFHT \MegaPrime configuration
phases\footnote{See:
  http://www.cfht.hawaii.edu/Science/CFHTLS-DATA/cfhtlsgeneralnews.html\#0007.}
separately. These epochs concern to the phases of investigations on
the \MegaPrime image quality. The first and second phase consist of
data taken before and after November, 24th, 2004 when the lens L3 was
accidentally mounted back upside-down. The mirror flipping brought a
surprising improvement of the image quality. The third phase consists
of the data taken after August, 12th, 2005, when a change in the
height of the \MegaPrime corrector was made. This final adjustment has
improved the image quality in terms of homogeneity over the entire
field-of-view. We found that the stacked image produced using only
exposures taken during the third phase of the instrument indeed yields
in a more homogeneous PSF pattern, making the correction of stellar
ellipticities easier (see Section \ref{cats:shear}). Thus, to carry
out our lensing analysis, an extra stacked image of the {\it i'} band
data was produced using only the exposures taken during CFHT
\MegaPrime phase three.

Image areas that could potentially infer error on the shape
measurements (e.g. bright stars haloes and diffraction spikes,
under-density haloes around large galaxies, asteroids tracks, etc)
received a flag. This masking procedure is done semi-automatically as
described in \citet{erben09}. After that, masks are visually
inspected. When applying all masks there is a loss of $\sim 19\%$ of
the total area. 

\subsection{Subaru}\label{app:data:subaru}

Subaru data are reduced in a similar way as to the CFHT data so that
we could establish a more robust comparison between the different data
sets. For that, we use the standard Suprime-Cam data reduction package
(SDFRED) \citep{yagi02, ouchi04} as well as the AstrOmatic
softwares\footnote{http://www.astromatic.net/}: SExtractor, SCAMP,
SWarp and Weight Watcher \citep{wwatcher}.

The data were retrieved from the SUBARU
archive\footnote{http://smoka.nao.ac.jp/search.jsp}. For each CCD
frame, we estimate and subtract the bias and correct by
flat-field. Master flats used in the flat-field process are
constructed using sky-flats observed at the same night as the science
images. Since the observations were taken during two different nights,
for each night a master flat is created using 14 single exposures
normalised to the unity, using a 3 sigma-clipping algorithm to reject
offset pixels. The master flats are created using the {\it imcombine}
task of the Image Reduction and Analysis Facility
(IRAF)\footnote{http://iraf.noao.edu/}. After the flat-field
correction a residual scattered light is still visible on the
images. This is corrected with a super-flat, which is created out of
the already flat-fielded data. As a last step, areas shaded off by the
AG probe are masked out.

For each CCD frame we create a weight and a flag image using the
Weight Watcher pipeline. The weight maps took into account the
pixel-to-pixel variation in sensitivity, cosmic rays hits and bad
regions (bad and hot pixels) assigning a zero weight for the affected
pixels. The information about saturated pixels is carried by the flag
images.

Source catalogues are created using SExtractor and are used as inputs
to compute a global astrometric solution with SCAMP, taking the
SDSS-R6 as reference catalogue. This leads to an rms value of the
position difference of $\sim0.22^{\prime\prime}$ with respect to
SDSS-R6 catalogues. Subaru {\it i$^+$} band image has an rms of the
position deviation of $0.05^{\prime\prime}$ with respect to the CFHT
{\it i'} image.

\begin{figure}
\centering
\subfigure{\includegraphics[scale=0.40]{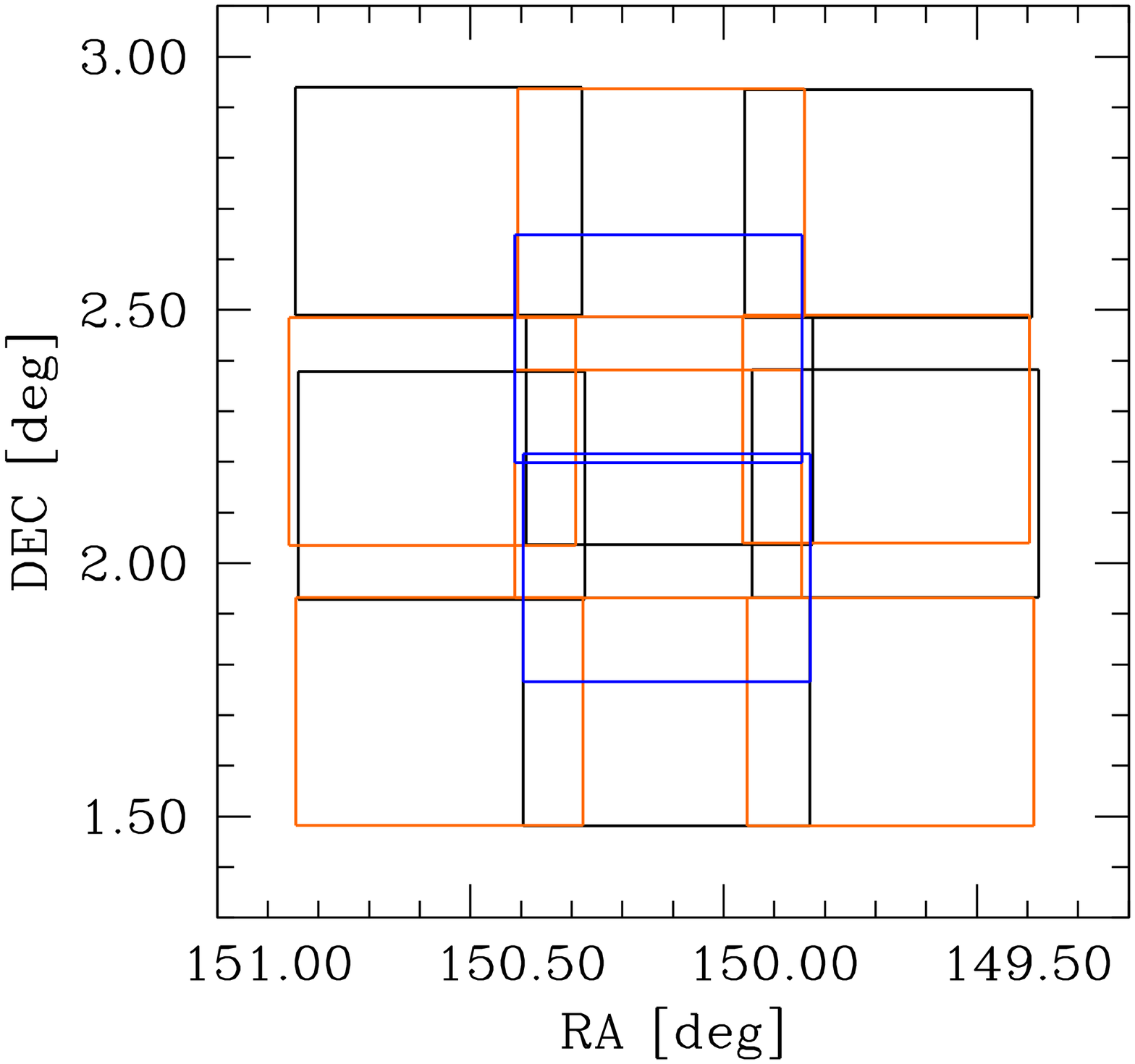}} \\
\subfigure{\includegraphics[scale=0.40]{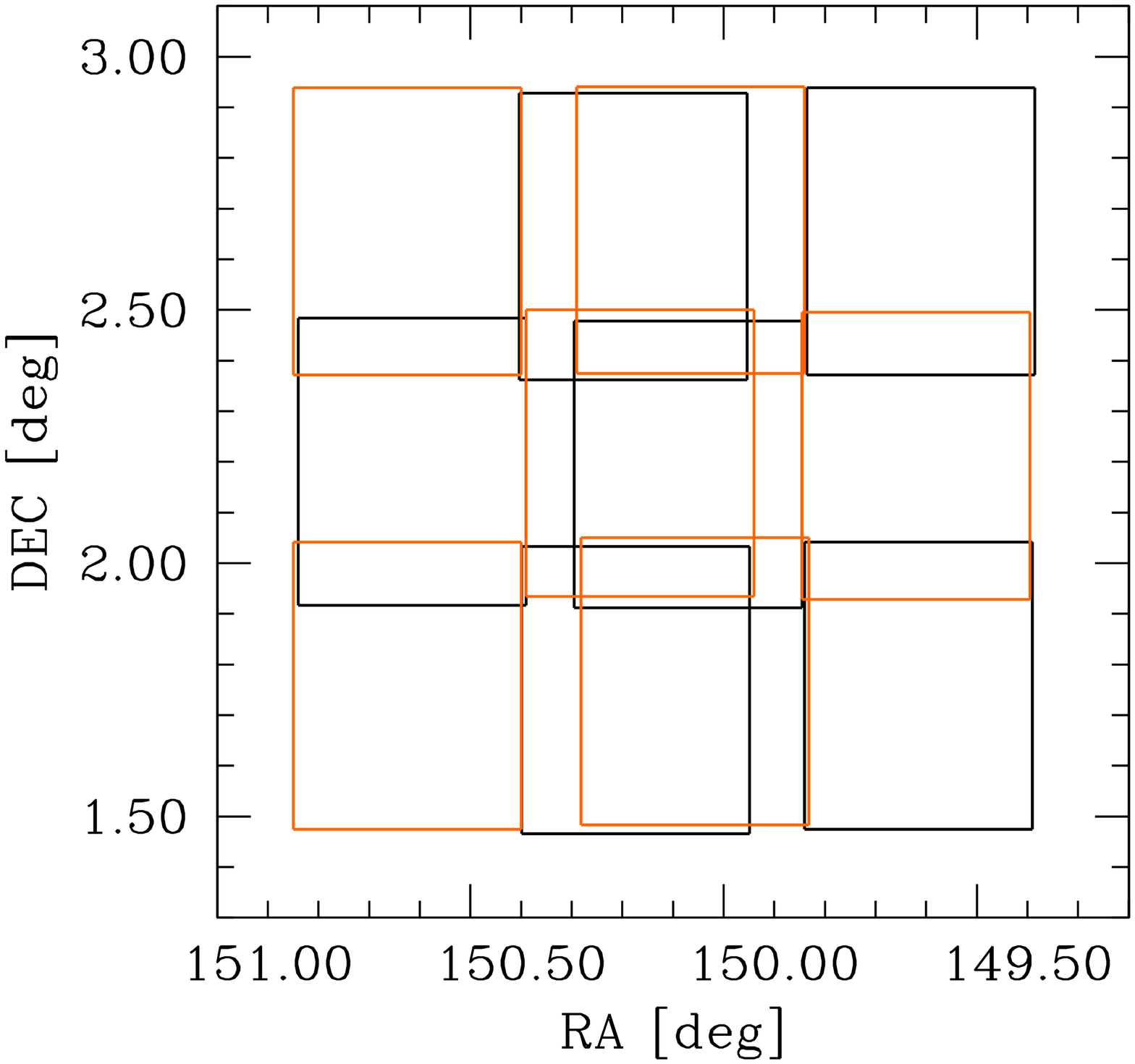}}
\caption{Sketch of the Subaru co-added exposures, totalling 26
  pointings. The total area covered is 1.9 degrees$^2$. The top panel
  shows the pointings for which the camera was not rotated and the
  bottom panel shows the pointings for which the camera was rotated by
  90 degrees. We only use the two pointings represented in blue,
  totalling 0.55 degrees$^2$.\label{pointsketch}}
\end{figure}

The data are co-added using SWarp on a pointing basis. A pointing is
defined according to the rotation of the camera and the dither
pattern, so that only exposures with the same orientation angle and
offset less then $3^\prime$ are stacked together.  A total of 26
pointings is obtained, as shown in Fig.  \ref{pointsketch}. We adopt
this strategy because the Subaru PSF pattern exhibits large variations
across the field-of-view, and by stacking all the data resulted in PSF
pattern that could not be corrected to the level required for the weak
lensing analysis. Due to this fact, each pointing results in a very
shallow final co-added image. There are, however, two pointings with a
higher depth. Therefore, we decide to use only these pointings in our
analysis, yielding an image coverage of 0.55 degrees$^2$. The position
of these two pointings in the field-of-view are shown in blue in the
top panel of Fig. \ref{pointsketch}.

We use the LANCZOS3 kernel to resample the pixels according to the
computed astrometric solution. The co-addition is done using a
weighted mean combination which takes into account the sky-background
noise, the weight maps and the relative photometric
zero-points. During the stacking process the sky background is also
subtracted. The co-added science images have a pixel size of
$0.2^{\prime\prime}$ and are accompanied by weight maps and flag
images containing information on saturated pixels. Masks are created
in a similar way as for CFHT data. Subaru masks cover about $\sim
15\%$ of the total area.

\label{lastpage}

\end{document}